\documentclass[a4paper,11pt]{article}
\pdfoutput=1 

\usepackage{jheppub} 

\usepackage[utf8]{inputenc}
\usepackage[T1]{fontenc}
\usepackage[english]{babel}
\usepackage{hyperref}
\usepackage{xcolor}
\usepackage{amsmath}
\usepackage{amssymb}
\usepackage{amsfonts}
\usepackage{bm}
\usepackage{graphicx}
\usepackage{subcaption}
\usepackage[normalem]{ulem}
\usepackage{physics}

\title{TMDs in the Lens of Generative AI: \\A Pixel-Based Approach to Partonic Imaging}

\author[a]{Marco Zaccheddu,}
\emailAdd{zacch@jlab.org}
\author[b]{Leonard Gamberg,}
\emailAdd{lpg10@psu.edu}
\author[a]{Wally Melnitchouk,}
\emailAdd{wmelnitc@jlab.org}
\author[c]{Daniel Pitonyak,}
\emailAdd{pitonyak@lvc.edu}
\author[a,b]{Alexei~Prokudin,}
\emailAdd{prokudin@psu.edu}
\author[a]{Jian-Wei Qiu,}
\emailAdd{jqiu@jlab.org}
\author[a]{and Nobuo Sato}
\emailAdd{nsato@jlab.org}
\affiliation[a]{Theory Center, Jefferson Lab, 12000 Jefferson Avenue, Newport News, Virginia 23606, USA}
\affiliation[b]{Division of Science, Penn State Berks, Reading, PA 19610, USA }
\affiliation[c]{Department of Physics, Lebanon Valley College, Annville, Pennsylvania 17003, USA}

\preprint{JLAB-THY-26-4691}

\abstract{This work introduces a novel, nonparametric pixel-based framework for the Bayesian inference and imaging of transverse momentum dependent (TMD) parton distributions. The methodology is built upon a fully differentiable framework that integrates TMD evolution with the Collins-Soper-Sterman formalism, enabling the simultaneous extraction of partonic distributions and the nonperturbative evolution kernel. To achieve efficient and exact sampling of the high-dimensional posterior, we leverage generative AI through a hybrid normalizing flow-driven Metropolis-Hastings approach. The framework is validated through multi-scale closure tests of increasing complexity, ranging from basic functional models to convoluted structure functions. Using singular value decomposition (SVD), we rigorously characterize the uncertainty of the reconstructed distributions and reveal the existence of {\it null TMDs}, which are functional components in the null space of the kernel that remain unconstrained by observables. The new framework provides the first integration of pixel-based discretization, generative AI, and SVD within a Bayesian context to solve the TMD inverse problem. This synergy between machine learning and multi-scale data removes inherent degeneracies and enables unbiased 3D partonic imaging.}

\begin{document}
\maketitle

\section{Introduction}
\label{sec:intro}

Achieving a precise and detailed understanding of the nucleon’s partonic structure and dynamics remains a central challenge in high-energy nuclear physics.
While our understanding of collinear parton distribution functions (PDFs) is reaching unprecedented levels of precision, advancing  toward a full three-dimensional (3D) imaging of the nucleon is a primary objective for current and future experimental programs, such as those at Jefferson Lab and the future Electron-Ion Collider (EIC). Central to this 3D imaging program are the transverse momentum dependent (TMD) distributions, which provide a correlated map of partonic longitudinal momentum and transverse position or momentum.

In the perturbative QCD framework, the extraction of TMD PDFs and fragmentation functions (FFs) is governed by factorization theorems~\cite{Collins:1981uk,Collins:1981uw,Collins:1984kg,Idilbi:2004vb,Collins:2004nx,Ji:2004wu,Collins:2011zzd,Echevarria:2011epo,Echevarria:2015usa}. In processes such as Drell-Yan inclusive lepton-pair production, semi-inclusive deep-inelastic scattering (SIDIS), and double hadron production in  $e^+ e^-$ annihilation, cross sections are expressed as convolutions of two TMD distributions. Within the Collins-Soper-Sterman (CSS) formalism \cite{Collins:1984kg,Collins:2011zzd}, this convolution is naturally formulated in the impact parameter ($b_T$) space, related to the observable transverse momentum ($k_T$) via a Bessel (or Hankel) transform~\cite{Piessens:2000}:
\begin{equation}
	W(k_T) =  \frac{1}{2\pi} \int_{0}^{\infty} \dd b_{T}\, b_{T}\, J_0(b_{T} \,k_{T})\, \widetilde{W}(b_{T}) \, .
\label{eq:fredholm}
\end{equation}
Mathematically, the determination of the nonperturbative function $\widetilde{W}(b_{T})$ from experimental data $W(k_T)$ constitutes a Fredholm integral equation of the first kind~\cite{Tikhonov:1977, Groetsch:1984}. Such problems are intrinsically ill-posed because the experimental data are discrete, finite in range, and affected by statistical uncertainties; consequently, there exist infinitely many configurations in $b_T$ space that can reproduce the same observables within uncertainties. Furthermore, the choice of phenomenological prescriptions (e.g., the $b_*$ prescription) has been shown to introduce additional systematic theoretical uncertainties that are difficult to constrain with single-scale data~\cite{Cerutti:2026apy}.

The current landscape of state of the art global TMD analyses comprises an array of extractions and collaborative frameworks. Notable examples include the SV19 fit and subsequent ART series \cite{Scimemi:2019cmh, Moos:2023yma}, the work by the MAP collaboration \cite{Bacchetta:2017gcc, Bacchetta:2022awv}, and the JAM collaboration \cite{Barry:2023qqh}. These and related efforts have collectively pushed the field to N$^4$LL accuracy \cite{Moos:2023yma, Moos:2025art, Camarda:2025lbt} and explored increasingly complex physical features, including flavor dependence \cite{Bacchetta:2024pcn}, longitudinally polarized distributions \cite{Bacchetta:2024hel, Yang:2024hel}, the transversity distribution~\cite{Anselmino:2007fs,Anselmino:2013vqa, Radici:2018pyn, Cammarota:2020qcw, DAlesio:2020vtw, Gamberg:2022kdb, Cocuzza:2023vqs, Cocuzza:2023oam, Boglione:2024dal, Zeng:2024gun}, the Sivers function~\cite{Cammarota:2020qcw, Gamberg:2022kdb, Bury:2021sue, Anselmino:2016uie, Bacchetta:2020gko, Echevarria:2020hpy, Boglione:2024dal, Zeng:2024gun, Boglione:2018dqd}, the internal structure of the pion~\cite{Cerutti:2022vnb, Barry:2023qqh, Rossi:2025pwh}, unpolarized TMD FFs for various hadrons~\cite{Boglione:2020auc, Boglione:2022pyj, Boglione:2023cti}, the Collins FF~\cite{Anselmino:2015sxa, Kang:2015msa, Kang:2017jcy, Cammarota:2020qcw,  Gamberg:2022kdb}, and the polarizing FF for $\Lambda$ hyperons~\cite{DAlesio:2020uaj, Callos:2020qtu, Gamberg:2021lsn, DAlesio:2022brl,Chen:2021hdn}. To handle the inherent ill-posedness of the inverse problem, these groups typically assume rigid analytical functional forms for the nonperturbative distributions. Recently, new directions in methodology have emerged, with some groups now utilizing neural networks to provide more flexible parametrizations of TMDs~\cite{Fernando:2023obn, Bacchetta:2025nn, Fernando:2025xzv}, or using AI agents to explore functional forms for the nonperturbative contributions to the TMD PDF \cite{Kang:2026mod}.

Although these approaches have established a robust baseline for TMD phenomenology, they are primarily designed for point-wise optimization or replica-based uncertainty estimation. However, the inverse problem implies a deeper structural challenge that has yet to be formally addressed:~the existence of what we identify and characterize here as {\it null TMDs}. Building upon analogous concepts that have recently emerged in the study of generalized parton distributions (GPDs)~\cite{Bertone:2021, Moffat:2023}, we provide in this work the first formal mathematical definition and empirical quantification of these distributions within the TMD framework. These null TMDs are functional modes residing in the effective null space of the integral transform which, due to the finite momentum reach of experimental data, remain fundamentally invisible to physical observables.
Whereas traditional functional forms are inherently smooth and stable by construction, and neural networks typically rely on explicit regularization to ensure such properties, both remain insensitive to null-space components. Consequently, the null TMDs remain effectively invisible to these approaches, potentially underestimating the true uncertainty of the 3D map.

In this work, we address these challenges through a nonparametric, Bayesian, pixel-based framework that enables the identification and quantification of the components that remain invisible to the data.  By replacing rigid functional forms with a discrete grid of nodal values, or {\it pixels}, we allow the data to dictate the shape of the distribution with minimal {\it a priori} bias. To handle the resulting high-dimensional posterior distribution, we leverage generative AI through a hybrid normalizing flow-driven Metropolis-Hastings (NF-MH) strategy. The methodology is validated through three closure tests of increasing complexity: a baseline Gaussian model, the $u$-quark TMD PDF, and the unpolarized $F_{UU,T}$ structure function. By utilizing singular value decomposition (SVD), we provide a formal mathematical characterization of the null TMDs that remain unconstrained by observables. This framework represents the first integration of pixel-based discretization, generative AI, and SVD within a Bayesian context to solve the TMD inverse problem, establishing a new path toward high-fidelity 3D partonic imaging driven by the synergy between machine learning and multi-scale data.

The paper is organized as follows: in Section~\ref{sec:gauss_distribution} we introduce the methodological framework through a Gaussian case study, detailing the generation of pseudo-data, the discretization of the Bessel transform, and the Bayesian inference strategy based on normalizing flows (NFs). The numerical performance and consistency of this approach are discussed in Section~\ref{sec:results_performance}, including an ensemble analysis in Section~\ref{sec:NF_replica}. In Section~\ref{sec:svd_analysis} we employ SVD to provide the first formal mathematical characterization of the null TMDs. The impact of energy evolution and the advantages of multi-scale analysis in breaking these degeneracies are explored in Section~\ref{sec:multi_scale}. We then generalize the framework to the full TMD PDF structure in Section~\ref{sec:tmd_distribution}, presenting a nonparametric extraction of the hadronic structure and the evolution kernel. The more complex case of the $F_{UU,T}$ structure function is addressed in Section~\ref{sec:fuu_distribution}. Finally, technical details regarding the local interpolators used in the discretization are provided in Appendix~\ref{sec:local_interp}.

\section{Methodological Framework: The Gaussian Case Study}
\label{sec:gauss_distribution}

In this section we present a simple Gaussian case study to illustrate the core methodology for reconstructing TMDs in impact parameter space ($b_T$) from observables in transverse momentum space ($k_T$), and solving the associated inverse problem. We begin by detailing the generation of pseudo-experimental data, followed by the discretization of the target distribution and the integral transform. Subsequently, we introduce the Bayesian inference framework and the NF sampling strategy used to solve the inverse problem. Finally, we discuss the reconstruction results, focusing on the physical interpretation of the uncertainty quantification in both single-scale and multi-scale scenarios.

\subsection{Generation of Pseudo-experimental Data}
\label{sec:gauss_generation_events}

The data generation process is structured to faithfully emulate the response of an experimental apparatus, starting from a well-defined Gaussian distribution that serves as a proxy for the physical observables. The physical dynamics are encoded through the definition of the width $\sigma^2(\mu^2)$, a parameter that evolves logarithmically with the energy scale $\mu^2$ according to the relation:
\begin{equation}
\sigma^2(\mu^2) = \alpha + g_{\rm evo} \ln\frac{\mu^2}{\mu^2_0},
\label{eq:sigma_evolution}
\end{equation}
where $g_{\rm evo}$ characterizes the energy scale dependence.
For this analysis the reference scale is set at $\mu_0^2 = 1$~GeV$^2$, while the data are initially generated at a scale $\mu^2 = 2$~GeV$^2$, using the phenomenological values $\alpha = 0.84$~GeV$^2$ and $g_{\rm evo} = 1.2$~GeV$^2$. The width not only describes the theoretical shape of the distribution, but also serves as the basis for creating a concrete set of random events.
To transform this theory into a set of discrete ``observables,'' we generate
an ensemble of events via inverse transform sampling (ITS). In this Gaussian
scenario, the inverse of the cumulative distribution function (CDF) is
analytical; starting from a uniform distribution of random numbers $u \in [0,
1]$, each individual event is converted into a transverse momentum value $k_T$
according to the law:
\begin{equation}
k_T = \sqrt{\sigma^2(\mu^2)}\ \text{erfinv}(u),
\label{eq:kt_sampling}
\end{equation}
where the inverse CDF is expressed through the inverse error function, $\text{erfinv}$. This mapping allows for the direct transformation of the uniform
probability space into the target Gaussian distribution, producing an ensemble
whose density reflects the expected profile and simulates the momentum
dispersion typical of high-energy collisions.

Once the events are generated, they are binned to calculate the observed density $f_{\rm obs}$. The sampling grid consists of 100 equally spaced bins covering the range from $0$ to $8$ GeV. To make the dataset realistic, a statistical uncertainty $\sigma_f$ derived from the Poisson distribution is associated with each bin, making the error proportional to the square root of the counts recorded in that particular interval. Consequently, the relative precision of the generated dataset scales with the total number of events $N_{\rm ev}$ as $1/\sqrt{N_{\rm ev}}$, allowing for a systematic study of the reconstruction performance across different statistical regimes.

To analyze the model's sensitivity to the available statistics, three different scenarios were generated, consisting of $200$, $2\,000$, and $20\,000$ events, respectively. The final dataset is then refined through a filtering process: all bins without a signal ($f_{\rm obs} = 0$) are excluded, as well as those characterized by excessive noise, where the relative uncertainty exceeds 100\% ($\sigma_f / f_{\rm obs} > 1.0$).
The result of this procedure is a series of triplets $\{f(k_T), k_T, \sigma_f\}$ suitable for inference. These data follow a distribution in the transverse momentum space defined by: 
\begin{equation}
	f(k_T, \mu^2) = \frac{2}{\sqrt{\pi \sigma^2(\mu^2)}} \exp\left( -\frac{k_T^2}{\sigma^2(\mu^2)} \right).
\label{eq:kt_true_distribution}
\end{equation}
Using the zeroth-order Bessel transform, in the conjugate impact parameter space ($b_T$) this corresponds to the function:
\begin{equation}
	\tilde{f}(b_T, \mu^2) = 2\sqrt{\pi \sigma^2(\mu^2)} \exp\left( -\frac{b_T^2\, \sigma^2(\mu^2)}{4} \right).
\label{eq:bt_true_distribution}
\end{equation}
This distribution in the conjugate domain of impact parameters ($b_T$) is not static, but evolves as the energy scale varies. Mathematically, the function at the scale $\mu^2$ can be directly linked to the one defined at the reference scale $\mu_{0}^2$ through an evolution operator, $U(b_T, \mu^2)$. The evolution operator, which acts as a dynamic scaling factor in the transverse plane, is defined as:
\begin{equation}
	U(b_T, \mu^2) = \sqrt{\frac{\sigma^2(\mu^2)}{ \alpha}} \exp\left(-\frac{b_T^2}{4} \, g_{\rm evo} \ln \frac{\mu^2}{\mu^2_0} \right).
\label{eq:evo_operator}
\end{equation}
In this formulation the evolved distribution $\tilde{f}(b_T, \mu^2)$ is obtained simply by multiplying the intrinsic distribution (at the scale $\mu_{0}^2$) by the evolution operator:
\begin{equation}
\tilde{f}(b_T, \mu^2) = \tilde{f}(b_T, \mu_{0}^2)\,
U(b_T, \mu^2).
\label{eq:bt_true_evolution}
\end{equation}
This approach allows for the isolation of the energy scale dependence from the functional form of the distribution at $\mu_0$, facilitating the simultaneous extraction of both contributions during the fitting phase.


\subsection{Discretization}

In this section we reconstruct the $b_T$ space function from available pseudo-data. 
The core of the numerical inversion lies in the discretization of the integral relationship that connects the two spaces. As discussed in the previous section, the mapping between the impact parameter space ($b_T$) and the transverse momentum space ($k_T$) is governed by the zeroth-order Bessel Transform:
\begin{equation}
\label{eq:bessel_transform}
f(k_{T  j}) = \frac{1}{2\pi} \int_{0}^{\infty} \dd b_{T}\, b_{T}\, J_0(b_{T} k_{T j})\, \tilde{f}(b_{T}),
\end{equation}
where $J_0$ is the zeroth-order Bessel function.
To reconstruct the unknown distribution $\tilde{f}(b_T)$ from discrete and stochastic data, we transition from a continuous analytical form to a discrete numerical representation. This discretization is a foundational element of our numerical inversion methodology. 
We assume that the reconstruction of $\tilde{f}(b_T)$ takes place within a finite interval $b_T \in [b_{T \text{min}}, b_{T \text{max}}]$.

While the upper limit $b_{T \text{max}}$ is chosen such that the distribution is negligible in large-distance regions, the lower limit $b_{T \text{min}}$ is dictated by the finite resolution of the sampled transverse momentum space, as well as the need to ensure the numerical stability of the projection operator. Within this domain, we approximate the function using a basis of $N$ local interpolators $h_i(b_T)$, defined on a grid of nodes $\{b_{T i}\}$. The function is then expanded as:
\begin{equation}
\tilde{f}(b_T) \approx \sum_{i=1}^{N} \tilde{f}_i\, h_i(b_T)\,,
\label{eq:bt_interpolation}
\end{equation}
where the coefficients $\tilde{f}_i \equiv \tilde{f}(b_{Ti})$ represent the values of the distribution at the $N$ grid nodes.

We refer to these nodal values as \textit{pixels}, an analogy inspired by digital imaging. Just as an image is composed of discrete units that collectively represent a continuous scene, our framework treats the distribution as a collection of localized degrees of freedom. In the general multidimensional case, these pixels would form a grid in the $(x, b_T)$ plane, providing a literal 2D image of the nucleon's internal structure. In this work, we focus on the $b_T$ dimension to rigorously isolate and characterize the resolution limits of the inverse problem. This pixel-based representation is particularly well-suited for the generative AI techniques employed here, as models such as NFs are optimized for high-dimensional density estimation and reconstruction tasks. By determining the posterior distribution of these pixels and using local interpolators, or \textit{interpixels}, to recover the continuous function, we achieve a nonparametric imaging of the nucleon that remains agnostic to global functional assumptions.

This approach follows the numerical strategies for discretization and evolution in QCD phenomenology employed in Refs.~\cite{Freese:2024ypk, Salam:2008qg, Bertone:2013vaa, Bertone:2023onx}. The use of these local interpolators allows the continuous integral to be transformed into a discrete summation, ensuring numerical stability in the transition between conjugate spaces. Details about the construction of these functions and their numerical implementation are discussed in Appendix~\ref{sec:local_interp}.
By substituting this expansion into the integral transform, we can express the value of the function at each observed momentum point $k_{Tj}$ as a linear combination of the unknown pixel values:
\begin{equation}
\label{eq:discretized_sum}
f(k_{T j}) = \sum_{i=1}^{N} \mathcal{M}_{ji}\, \tilde{f}_i\, ,
\end{equation}
where the elements of the kernel matrix $\mathcal{M}$ are defined as:
\begin{equation}
\label{eq:kernel_matrix}
\mathcal{M}_{ji} = \frac{1}{2\pi} \int_{b_{T\min}}^{b_{T\max}} \dd b_{T}\, b_{T}\, J_0(b_{T} k_{T j})\, h_i(b_{T}).
\end{equation}
Supposing that we have $M$ momentum points, this procedure leads to a discrete matrix equation of the form:
\begin{equation}
\label{eq:matrix_form}
\boldsymbol{f} = \mathcal{M}\, \tilde{\boldsymbol{f}}.
\end{equation}
In this expression, $\boldsymbol{f} = [f(k_{T\,1}), \dots, f(k_{T\,M})]^T$ represents the vector of the $M$ discrete observations in the transverse momentum space, while $\tilde{\boldsymbol{f}} = \big[\tilde{f}(b_{T\,1}), \dots, \tilde{f}(b_{T\,N})\big]^T$ is the vector of the $N$ nodal values that we aim to reconstruct in the impact parameter space. 

To handle the variability of experimental momentum points {$\{k_{Tj}\}$} while maintaining numerical efficiency, we decouple the observation points from the reconstruction basis. Instead of allowing the computational grid depend on the specific experimental realization, we define a fixed discretization in $b_T$ space consisting of $N$ nodes distributed according to a power-law:

\begin{equation}
\label{eq:bt_grid}
{b_{Ti}} = b_{T\,\text{min}} + (b_{T\,\text{max}} - b_{T\,\text{min}}) \left( \frac{i-1}{N-1} \right)^{\!a}\,,
\end{equation}
where the scaling parameter $a$ controls the density of the nodes.
For this study we have chosen $N=50$ nodes and $a=3$ to cluster the points toward the origin. This configuration allows for high-precision resolution of the distribution near its peak at small $b_T$, where the function varies most rapidly, without burdening the calculation with an excessive number of nodes.
In parallel, we establish a fixed reference grid in the transverse momentum space, $\{k_{T k}^{\text{ref}}\}$, which matches the grid used for binning the pseudo-data. This allows for the pre-computation of a reference kernel matrix $\mathcal{M}_{\text{ref}}$ that projects the nodal values $\tilde{\boldsymbol{f}}$ directly onto the momentum grid:
\begin{equation}
\label{eq:matrix_ref}
\big[\mathcal{M}_{\text{ref}}\big]_{ki} = \frac{1}{2\pi} \int_{b_{T\,\text{min}}}^{b_{T\,\text{max}}} \dd b_{T}\, b_{T}\, J_0(b_{T} k_{T,k}^{\text{ref}})\, h_i(b_{T}).
\end{equation}

To evaluate the theory at the specific momentum points $\{k_{Tj}\}$ of the observations, we introduce an interpolation matrix $\mathcal{M}_{\text{interp}}$ that maps the values from the reference grid to the actual experimental coordinates. Using a basis of local interpolators $h_k(k_T)$ in the momentum space, the elements of this interpolation matrix are given by:
\begin{equation}
\label{eq:matrix_interp}
\big[\mathcal{M}_{\text{interp}}\big]_{jk} = h_k(k_{T j}).
\end{equation}
The final projection operator, or forward mapping, which relates the pixels in the impact parameter space to the observed data, is therefore expressed as the product of these two matrices:
\begin{equation}
\label{eq:matrix_total}
\mathcal{M}_{\text{tot}} = \mathcal{M}_{\text{interp}}\, 
\mathcal{M}_{\text{ref}} \,.
\end{equation}

At this stage it is crucial to specify the physical interpretation of the nodal values $\tilde{f}_i$ within our inference procedure. While the matrices $\mathcal{M}_{\text{ref}}$ and $\mathcal{M}_{\text{interp}}$ operate on the distribution at the scale of the observations ($\mu^2 = 2$~GeV$^2$), our primary objective is the reconstruction of the intrinsic structure at the reference scale $\mu_0^2 = 1$~GeV$^2$.
Following the evolution law introduced in Eq.~\eqref{eq:bt_true_evolution}, the vector $\tilde{\boldsymbol{f}}$ that enters the discrete mapping in Eq.~\eqref{eq:matrix_form} is not treated as a primary set of variables. Instead, it is decomposed as the element-wise product (Hadamard product) of the reference distribution and the evolution operator:
\begin{equation}
\label{eq:bt_decomposition_mapping}
\tilde{\boldsymbol{f}}(\mu^2) = \tilde{\boldsymbol{f}}(\mu_0^2) \odot \mathbf{U}(\mu^2),
\end{equation}
where $\big[\mathbf{U}(\mu^2)\big]_i = U(b_{Ti}, \mu^2)$. In this setup, our sampling algorithm targets the coefficients $\tilde{f}_i(\mu_0^2)$, which represent the fundamental degrees of freedom at the initial scale. This structure ensures that the discretization remains anchored to a universal scale $\mu_0^2$, effectively decoupling the intrinsic functional shape from the energy-dependent broadening described by $g_{\rm evo}$.

This approach ensures that the forward mapping remains efficient across different sampling realizations. However, despite this robust construction, solving Eq.~\eqref{eq:matrix_total} for the reference vector $\tilde{\boldsymbol{f}}(\mu_0^2)$ remains a fundamentally ill-posed inverse problem. In practical terms, the kernel matrix $\mathcal{M}_{\text{tot}}$ is often ill-conditioned, meaning that high-frequency noise in the data $\boldsymbol{f}$ is amplified during any attempt at direct inversion, leading to unphysical oscillations in the reconstructed $b_T$ distribution. This inherent instability prevents a direct solution and motivates the adoption of a probabilistic framework capable of incorporating prior physical knowledge to regularize the inversion.

\subsection{Bayesian Inference Framework}

Classical approaches to addressing this challenge rely on the parametrization of the unknown function, assuming a functional form $\tilde{\boldsymbol{f}} \equiv \tilde{\boldsymbol{f}}(\boldsymbol{\theta})$ governed by a vector of free parameters $\boldsymbol{\theta}$. In this context, the reconstruction is performed by minimizing a figure of merit, typically the $\chi^2$ function, to identify the optimal values of $\boldsymbol{\theta}$ that best describe the observations. This scheme represents the current standard in TMD extractions~%
\cite{Scimemi:2019cmh, Moos:2023yma, Bacchetta:2017gcc, Bacchetta:2022awv, Barry:2023qqh, Moos:2025art, Bacchetta:2024pcn, Bacchetta:2024hel, Yang:2024hel, Anselmino:2013vqa, Radici:2018pyn, Cammarota:2020qcw, DAlesio:2020vtw, Gamberg:2022kdb, Cocuzza:2023vqs, Cocuzza:2023oam, Boglione:2024dal, Zeng:2024gun, Bury:2021sue, Anselmino:2016uie, Anselmino:2007fs, Bacchetta:2020gko, Echevarria:2020hpy, Boglione:2018dqd, Cerutti:2022vnb, Rossi:2025pwh, Boglione:2020auc, Boglione:2022pyj, Boglione:2023cti, Anselmino:2015sxa, Kang:2015msa, Kang:2017jcy, DAlesio:2020uaj, Callos:2020qtu, Gamberg:2021lsn, DAlesio:2022brl, Chen:2021hdn, Fernando:2023obn, Bacchetta:2025nn, Fernando:2025xzv},
whether using theory-inspired analytical forms or flexible neural network parametrizations. While these standard methods successfully estimate uncertainties through error bands or replica techniques, they remain essentially rooted in a local analysis of the cost function around a single ``best-fit'' point.

In this work, we propose a paradigm shift by moving the objective from point-wise optimization to probabilistic inference. Instead of enforcing a specific functional form or seeking a unique set of optimal parameters, we treat each element of the nodal vector $\tilde{\boldsymbol{f}}$ directly as a random variable. By adopting a Bayesian approach, we aim to reconstruct the entire posterior probability distribution of the vector $\tilde{\boldsymbol{f}}$ conditioned on the observed data~$\boldsymbol{f}$.
This allows for a natural mapping of the full structure of correlations and uncertainties, without being limited to a local expansion around a minimum or to {\it a priori} assumptions about the functional form. From this perspective, the reconstruction does not yield a single curve, but a complete probability density in the impact parameter space, enabling an intrinsic and rigorous assessment of the uncertainties and correlations between the different nodes of the distribution.

In this context, the objective shifts toward determining the posterior probability distribution $P(\tilde{\boldsymbol{f}} | \boldsymbol{f})$, which expresses the probability of the configuration in the impact parameter conditioned on the observed data. Formally, the link between {\it a priori} knowledge and experimental information is established by Bayes' theorem:
\begin{equation}
\label{eq:bayes_theorem}
P(\tilde{\boldsymbol{f}} | \boldsymbol{f}) \propto \mathcal{L}(\boldsymbol{f} | \tilde{\boldsymbol{f}})\ 
\pi(\tilde{\boldsymbol{f}}),
\end{equation}
where $\pi(\tilde{\boldsymbol{f}})$ is the prior distribution, which allows for the incorporation of physical constraints or regularizations on the shape of $\tilde{\boldsymbol{f}}$, and $\mathcal{L}(\boldsymbol{f} | \tilde{\boldsymbol{f}})$ represents the likelihood, which encapsulates the compatibility between the discretized theory and the experimental observations. For the sake of notational simplicity, in this and the following sections we use $\tilde{\boldsymbol{f}}$ to denote the nodal values at the reference scale $\mu_0^2$ (as defined in Eq.~\eqref{eq:bt_decomposition_mapping}), unless otherwise specified. It is understood that any comparison with data at a scale $\mu^2 > \mu_0^2$ implicitly involves the evolution operator $U(b_T, \mu^2)$ within the forward mapping $\mathcal{M}$.

Assuming that the uncertainties $\sigma_j$ associated with the data points $f_j$ follow a Gaussian distribution, the likelihood can be expressed in terms of a $\chi^2$ cost function:
\begin{equation}
\label{eq:likelihood}
	\mathcal{L}(\boldsymbol{f} | \tilde{\boldsymbol{f}}) \propto \exp \left[ - \frac{1}{2} \chi^2(\boldsymbol{f} | \tilde{\boldsymbol{f}}) \right],
\end{equation}
where the $\chi^2$ represents the weighted residuals between the observed data and the theoretical prediction obtained via the direct projection $\mathcal{M} \tilde{\boldsymbol{f}}$:
\begin{equation}
\label{eq:chi2}
	\chi^2(\boldsymbol{f} | \tilde{\boldsymbol{f}}) = \sum_{j=1}^{M} \left( \frac{\big[\mathcal{M} \tilde{\boldsymbol{f}}\big]_j - f_j}{\sigma_j} \right)^{\!2}.
\end{equation}
In this context, the term $[\mathcal{M} \tilde{\boldsymbol{f}}]_j$ represents the theoretical value of the distribution in $k_T$ space at the $j$-th momentum point, as predicted by the current configuration of pixels in $b_T$ space.
While the likelihood constrains the reconstruction to reproduce the data, the prior distribution $\pi(\tilde{\boldsymbol{f}})$ provides the necessary regularization to stabilize the inversion by penalizing unphysical configurations. 
In this work, we employ Tikhonov-style regularization up to the second order to constrain the structural properties of the reconstructed distribution. The prior distribution is defined as:
\begin{equation}
\label{eq:tikhonov_prior}
\ln \pi(\tilde{\boldsymbol{f}}) \propto - \sum_{k=0}^{2} \lambda_k\, \Omega_k(\tilde{\boldsymbol{f}}),
\end{equation}
%
where $\lambda_0$, $\lambda_1$, $\lambda_2$ are the regularization coefficients. The penalty terms $\Omega_k(\tilde{\boldsymbol{f}})$ characterize the different structural constraints on the reconstructed distribution:
\begin{subequations}
	\label{eq:regolarizzazioni}
	\begin{align}
	\Omega_0(\tilde{\boldsymbol{f}}) &= \sum_{i=1}^{N} \tilde{f}_{i}^2, \label{eq:omega0} \\[1ex]
	\Omega_1(\tilde{\boldsymbol{f}}) &= \sum_{i=1}^{N-1} (\tilde{f}_{i+1} - \tilde{f}_{i})^2, \label{eq:omega1} \\[1ex]
	\Omega_2(\tilde{\boldsymbol{f}}) &= \sum_{i=2}^{N-1} (\tilde{f}_{i+1} - 2\tilde{f}_{i} + \tilde{f}_{i-1})^2. \label{eq:omega2}
	\end{align}
\end{subequations}
%
%
In particular, $\Omega_0$ prevents unphysically large pixel amplitudes, while $\Omega_1$ and $\Omega_2$ penalize sharp transitions and erratic oscillations, favoring a smooth and stable functional form in $b_T$ space. 
For the study presented in this work, we have adopted a hierarchy of regularization coefficients that prioritizes the structural stability of the distribution: specifically, the values $\lambda_0 = 10^{-3}$, $\lambda_1 = 1$, and $\lambda_2 = 10$ were used. This configuration prioritizes the regularity of the distribution's shape (curvature and slope) over its absolute magnitude. By keeping $\lambda_0$ several orders of magnitude smaller than the others, we ensure that the global normalization remains data-driven, avoiding unjustified bias while effectively filtering out high-frequency numerical noise.

\subsection{Normalizing Flow-driven Metropolis-Hastings}

\begin{figure}[t] 
    \centering
    \includegraphics[width=\textwidth]{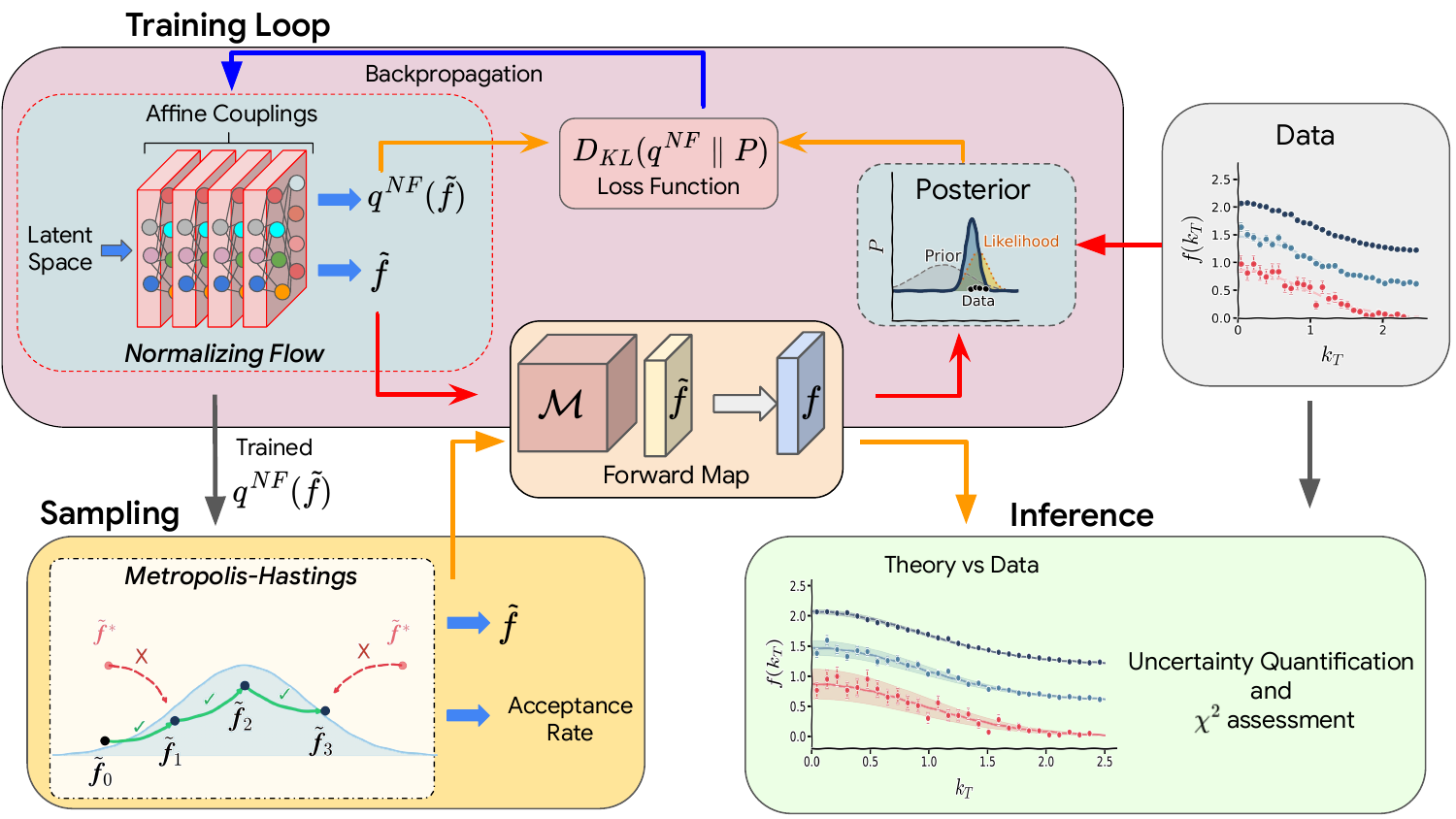}
    \caption{Workflow of the NF-MH framework, illustrating the training loop, the  sampling phase, and the final inference of the TMD distributions.}
    \label{fig:workflow}
\end{figure}

To sample the high-dimensional posterior distribution $P(\tilde{\boldsymbol{f}} | \boldsymbol{f})$, we implement a hybrid strategy that leverages NFs to optimize the efficiency of a Metropolis-Hastings (MH) algorithm. This NF-MH approach is structured into three sequential phases, as illustrated in the workflow of Fig.~\ref{fig:workflow}: (i) \textit{density approximation} via an iterative NF training loop, (ii) \textit{exact sampling} through the MH chain, and (iii) \textit{final inference} of the distributions.

In the first phase, a generative NF model is trained to approximate the target posterior through an iterative \textit{training loop}, as depicted in Fig.~\ref{fig:workflow}. The architecture consists of 12 masked affine coupling layers~\cite{Dinh:2016}, interleaved with ActNorm modules and random permutations. This configuration is specifically designed to capture the complex, nonlocal correlations among the 50 nodes of the $b_T$ grid, effectively mapping the high-dimensional parameter space. 
For each iteration, the NF generates a batch of samples $\tilde{\boldsymbol{f}}$, along with their associated log-probabilities $\ln q^{\text{NF}}(\tilde{\boldsymbol{f}})$. These samples are directly propagated through the discretized forward mapping $\boldsymbol{f} = \mathcal{M} \tilde{\boldsymbol{f}}$ to evaluate the likelihood and the target posterior $P(\tilde{\boldsymbol{f}} | \boldsymbol{f})$. 
By implementing this entire pipeline, from the generative model to the linear projection operator, using a tensor-based framework in PyTorch, we leverage a natively auto-differentiable architecture. This setup allows for the efficient gradient-based optimization of the NF weights and takes full advantage of GPU acceleration for high-throughput batch calculations. 
Optimization is performed by minimizing the Kullback-Leibler (KL) divergence between the NF's parametric distribution $q^{\text{NF}}(\tilde{\boldsymbol{f}})$ and the target posterior, expressed as the expectation value of the log-probability ratio:
\begin{equation}
\label{eq:kl_divergence}
D_{\text{KL}}(q^{\text{NF}} || P) = \mathbb{E}_{\tilde{\boldsymbol{f}} \sim q^{\text{NF}}(\tilde{\boldsymbol{f}})} \left[ \ln q^{\text{NF}}(\tilde{\boldsymbol{f}}) - \ln P(\tilde{\boldsymbol{f}} | \boldsymbol{f}) \right].
\end{equation}

Once trained, the NF serves as a \textit{global proposal distribution} for an \textit{independence sampler}. Unlike standard Markov chain Monte Carlo random walks, new candidates $\tilde{\boldsymbol{f}}_{\text{new}}$ are drawn independently from the learned density $q^{\text{NF}}(\tilde{\boldsymbol{f}})$, bypassing the slow exploration typical of local proposals in high dimensions.
The MH algorithm acts as a corrective filter, compensating for residual biases in the NF approximation by accepting new candidates with probability:

\begin{equation}
\label{eq:mh_acceptance}
\alpha = \min \!\left( 1, \frac{P(\tilde{\boldsymbol{f}}_{\text{new}} | \boldsymbol{f}) }{P(\tilde{\boldsymbol{f}}_{\text{old}} | \boldsymbol{f}) } \frac{q^{\text{NF}}(\tilde{\boldsymbol{f}}_{\text{old}})}{q^{\text{NF}}(\tilde{\boldsymbol{f}}_{\text{new}})} \right)\,.
\end{equation}
The new candidate $\tilde{\boldsymbol{f}}_{\text{new}}$ is accepted if $\alpha \geq u$, where $u \sim \mathcal{U}(0,1)$, a stochastic test that ensures the samples are correctly reweighted. The resulting Markov chain $\tilde{\boldsymbol{f}}$ is then used for the final inference, including data description, uncertainty quantification, and $\chi^2$ evaluation.

\section{Results and Discussion}
\label{sec:results_performance}

In this section we discuss the numerical performance of the NF-MH framework across the different statistical regimes introduced in Section \ref{sec:gauss_generation_events}. We first focus on the single-scale analysis, evaluating the quality of the inversion by comparing the reconstruction against pseudo-data in the momentum space ($k_T$) and analyzing the extracted distribution in the impact parameter space ($b_T$). This dual analysis demonstrates the ability of the sampler to balance experimental observations with the physical constraints imposed by the priors.

\subsection{Momentum Space Analysis: Data Consistency and Fit Stability}

\begin{figure}[t] 
	\centering
	\includegraphics[width=1\textwidth]{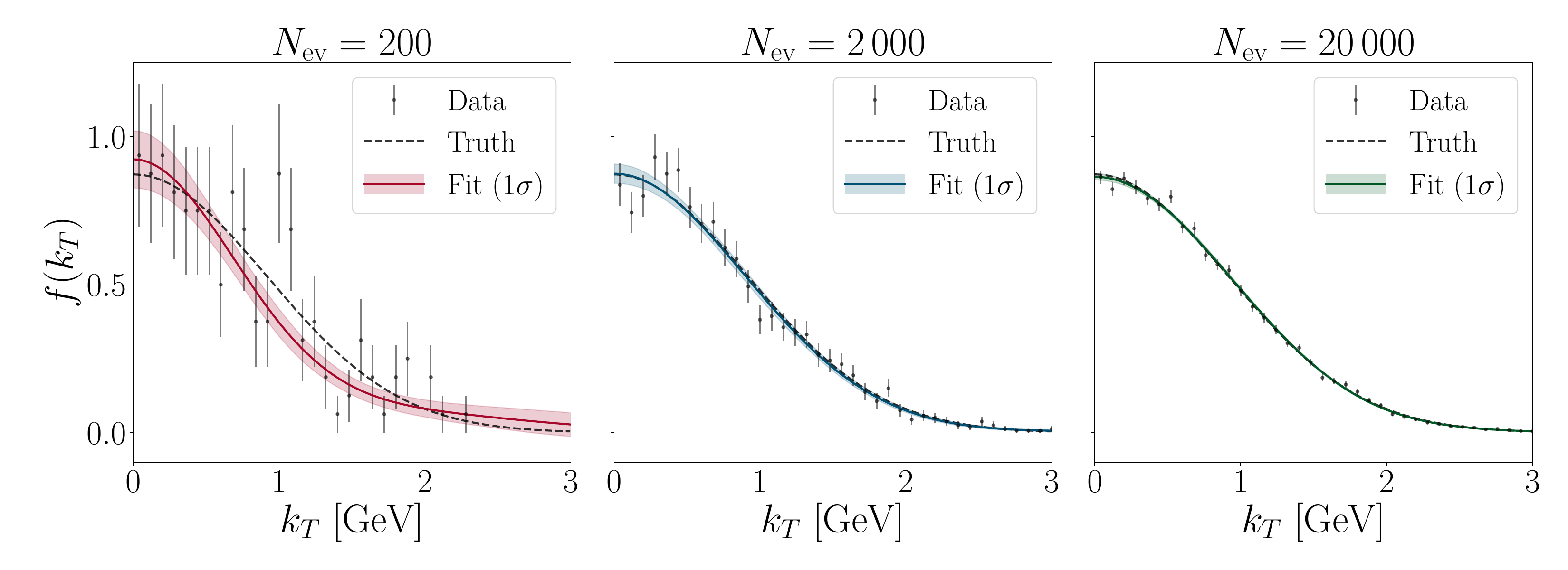}
	\caption{Comparison between pseudo-data and Bayesian fit in $k_T$ space for $N_{\text{ev}} =~200$, 2\,000 and 20\,000. The increase in statistics leads to a more densely sampled spectrum and a significant reduction in fit uncertainty.}
	\label{fig:scaling_fits_kt}
\end{figure}

After validating the sampling procedure, we evaluate the reconstruction by projecting the configurations from the impact parameter space back onto the transverse momentum grid. Figure~\ref{fig:scaling_fits_kt} compares the pseudo-experimental data with the Bayesian posterior for the three statistical scenarios, with $N_{\text{ev}} = 200$, 2\,000 and 20\,000.
The fit performance is summarized in Table~\ref{tab:fit_diagnostics_gauss}. Across all regimes, the framework provides an excellent description of the data; the reduced $\chi^2$ values, consistently below unity, confirm that the reconstructed distributions are statistically compatible with the observations. This stability demonstrates that the combination of Tikhonov-based priors and data likelihood effectively regularizes the inversion without over-fitting the experimental noise.

The table also reports the MH acceptance rates, which range from $61\%$ to $78\%$. These high values indicate that the NF accurately captures the high-dimensional posterior's topology, providing an efficient global proposal even as the likelihood becomes more peaked with higher statistics. While further hyperparameter tuning could potentially refine these rates, the current performance is more than sufficient to ensure a robust and computationally efficient exploration of the 50-dimensional nodal space.

\begin{table}[b] 
    \caption{Performance summary of the Bayesian fit across different statistical scenarios, showing the acceptance rate, the number of momentum bins $N_{\text{pts}}$ surviving the filtering criteria, and the reduced $\chi^2/N_{\text{pts}}$ calculated at the posterior mean.}
    \centering
    \begin{tabular}{rccc}
        \hline
        $N_{\text{ev}}$~ & accept. rate & $N_{\text{pts}}$ & $\chi^2/N_{\text{pts}}$ \\
        \hline
        $200$     & 61\% & 27 & 0.79 \\
        $2\,000$  & 76\% & 38 & 0.66 \\
        $20\,000$ & 78\% & 46 & 0.69 \\
        \hline
    \end{tabular}
    \label{tab:fit_diagnostics_gauss}
\end{table}

In the low-statistics regime ($N_{\text{ev}} = 200$), the sampled points are characterized by sharp fluctuations and significant uncertainties. Consequently, while capturing the general Gaussian trend, the reconstructed fit presents a wider uncertainty band and slight structural irregularities, attributable to the sparse and noisy nature of the data. As the number of events increases, a marked improvement in data precision is observed:~the error bars contract, the points align with greater fidelity to the theoretical ``truth'', and the fit becomes progressively smoother and more tightly constrained.
This behavior underscores the stability of the NF-MH framework:~the model not only converges toward the target solution as the statistics grow but also provides a reliable uncertainty quantification (UQ) when experimental information is limited. The consistency between the posterior projections and the observed data confirms that the combination of Tikhonov regularization and NF-driven sampling effectively manages statistical noise without introducing artificial biases into the functional form of the distribution.

\begin{figure}[t] 
	\centering
	\includegraphics[width=1\textwidth]{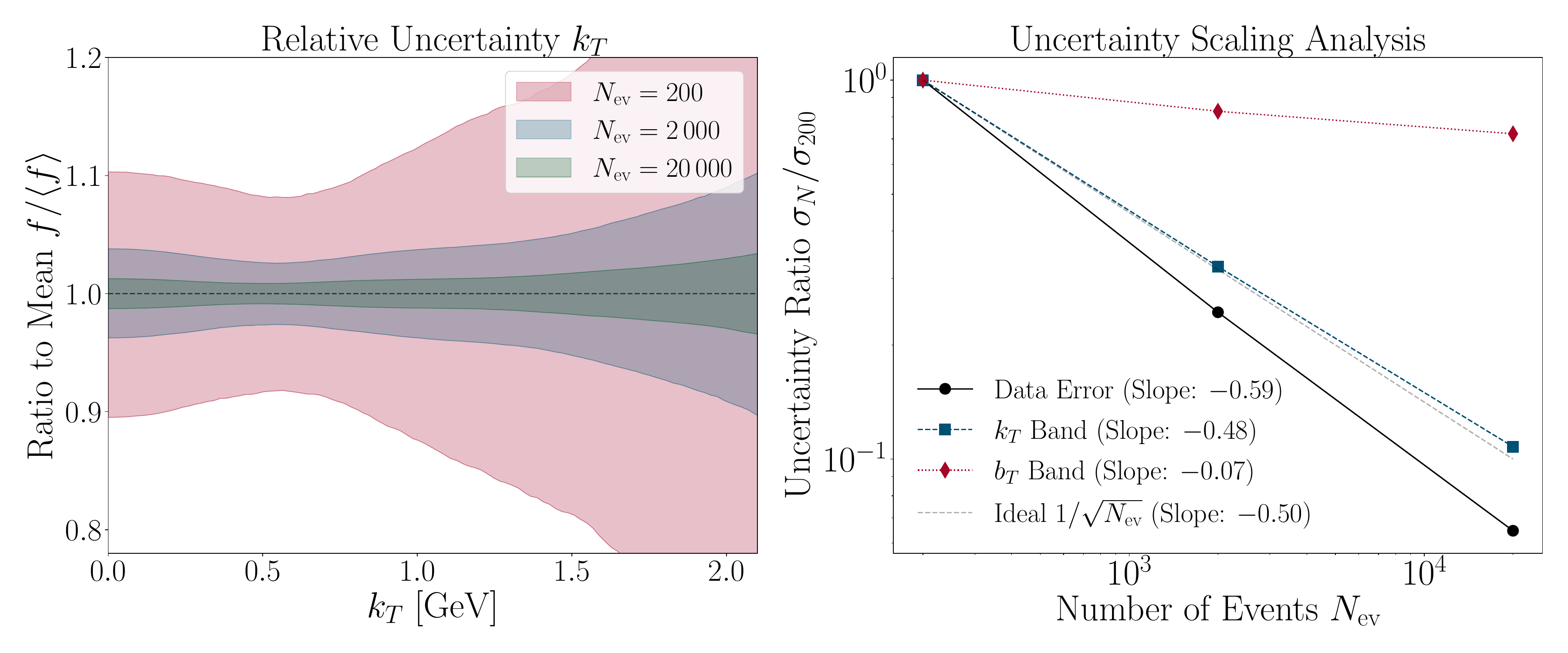}
	\caption{Analysis of the reconstruction precision in $k_T$ space. Both the relative uncertainty (left) and the average scaling (right) follow the ideal $1/\sqrt{N_{\text{ev}}}$ behavior.}
	\label{fig:momentum_precision}
\end{figure}

The effectiveness of the reconstruction is further quantified in Fig.~\ref{fig:momentum_precision}, which illustrates the relative uncertainty and global scaling analysis. The results show a uniform contraction of the relative uncertainty bands across the entire $k_T$ range as $N_{\text{ev}}$ increases, and confirm that this improvement is purely statistical:~both the average data error bars and the fit uncertainty envelopes scale proportionally to $1/\sqrt{N_{\text{ev}}}$. In a log-log representation, the calculated slopes ($-0.59$ for the data and $-0.48$ for the fit) are in excellent agreement with the ideal theoretical value of $-0.5$. This result demonstrates that the NF-MH framework efficiently extracts the available information from momentum-space observations, nearly saturating the statistical limit imposed by the sample size.

Finally, the slight systematic displacement of the posterior mean from the ground truth observed for $N_{\text{ev}} = 200$ is not a failure of the methodology, but a direct consequence of the low information content in the data. A detailed investigation into how these statistical fluctuations propagate through the reconstruction, conducted by analyzing multiple dataset replicas, is presented below in Section~\ref{sec:NF_replica}. This subsequent study highlights the inherent challenges of recovering the underlying distribution from limited observations and provides a more comprehensive assessment of the framework's robustness.

\subsection{Impact Parameter Space Analysis: Resolution Limits and Regularization}

\begin{figure}[ht]
    \centering
    \includegraphics[width=\textwidth]{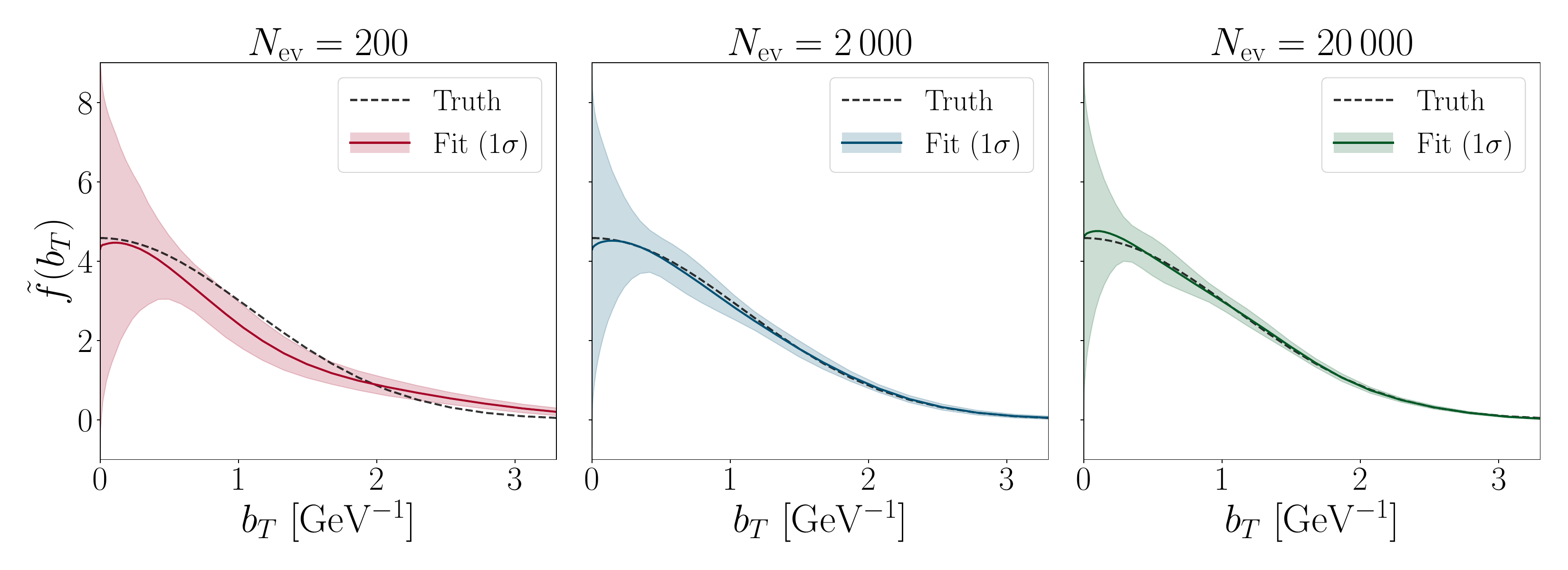}
    \caption{Reconstruction of the impact parameter distribution $\tilde{f}(b_T)$ for $N_{\text{ev}} = 200$, 2\,000, and 20\,000. The $1\sigma$ Bayesian bands consistently encompass the analytical ground truth (dashed lines).}
    \label{fig:scaling_bt_linear}
\end{figure}

The reconstruction results in the impact parameter space ($b_T$), presented in Fig.~\ref{fig:scaling_bt_linear}, allow for a direct evaluation of the framework's ability to invert the Bessel transform and recover the underlying physics from momentum-space observations.

A fundamental validation of the methodology is that the analytical ground truth remains consistently within the $1\sigma$ statistical bands across all investigated event counts. This confirms that the numerical implementation, integrating the Bessel matrix discretization, the interpolation strategy, and the NF-MH sampling, is robust and reconstructs the true distribution without introducing significant systematic biases.

As expected, the posterior mean approaches the ground truth as the number of events increases. However, a detailed examination of the uncertainty bands reveals several nontrivial features. In the intermediate $b_T$ region, we observe a noticeable shrinkage of the bands when transitioning from 200 to $2\,000$ events. Interestingly, moving to $20\,000$ events does not yield a proportional improvement, suggesting that the reconstruction reaches a ``precision floor''. At this stage, the uncertainty is no longer dominated by data statistics but is instead governed by the constraints imposed by the regularization priors.

\begin{figure}[b] 
	\centering
	\includegraphics[width=0.72\textwidth]{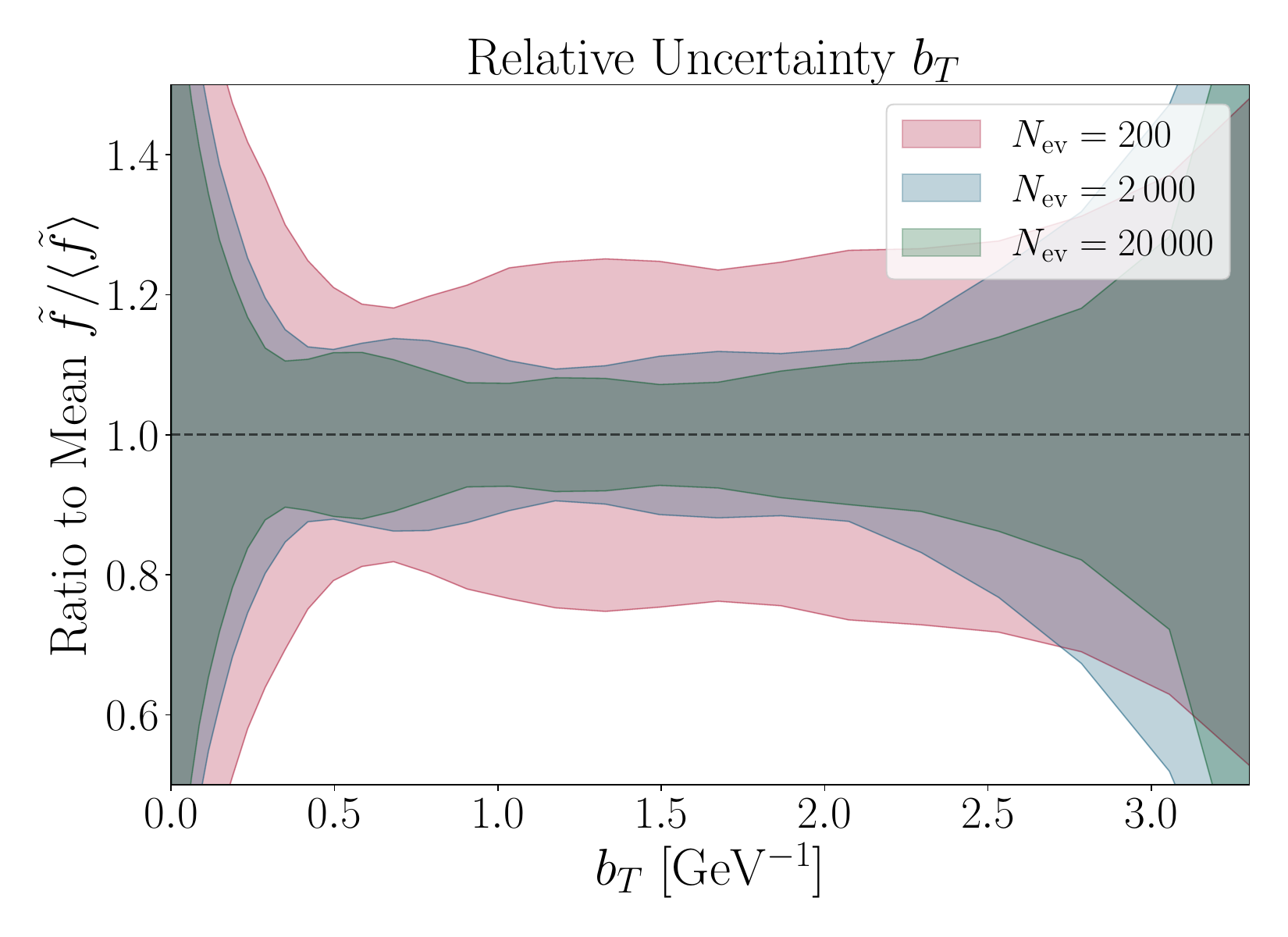}
	\caption{Relative uncertainty in the $b_T$ space, calculated as the ratio of the $1\sigma$ uncertainty bands to the posterior mean. The plot highlights the lack of statistical shrinkage at small distances ($b_T \le 1~\text{GeV}^{-1}$), indicating a clear resolution limit where increasing $N_{\text{ev}}$ does not further constrain the reconstruction.}
	\label{fig:relative_uncertainty_bt}
\end{figure}

This behavior is further quantified in Fig.~\ref{fig:relative_uncertainty_bt}, which illustrates the relative uncertainty across the $b_T$ spectrum. Notably, the description of the distribution at small impact parameters ($b_T \to 0$) shows almost no improvement with increasing statistics, with the relative uncertainty remaining nearly identical for $N_{\text{ev}} = 2\,000$ and $N_{\text{ev}} = 20\,000$. This stands in sharp contrast to the ideal $1/\sqrt{N_{\text{ev}}}$ scaling observed in momentum space (Fig.~\ref{fig:momentum_precision}), indicating that the statistical gain is not uniformly transferred to the conjugate domain and that the average width of the uncertainty bands in $b_T$ does not follow the expected statistical trend.

This result is counter-intuitive, as a ten-fold increase in data would typically suggest a better resolution of the peak. Instead, this behavior indicates that the information carried by the momentum-space data is not equally resolvable across the entire $b_T$ spectrum. This limitation is not a failure of the NF-driven sampling, but an intrinsic property of the Bessel transform itself, which effectively ``filters'' certain components of the distribution, making them inaccessible to the data regardless of the available statistics. This phenomenon, related to the ill-posed nature of the inversion, will be formally explored in Section~\ref{sec:svd_analysis} through SVD.

\subsection{Ensemble Analysis:~Statistical Consistency \& Uncertainty Decomposition}
\label{sec:NF_replica}

Before examining the formal resolution limits of the inversion, it is crucial to disentangle the impact of statistical noise from potential algorithmic deficiencies. To this end, we performed an ensemble analysis of 100 independent pseudo-data replicas for each statistical scenario to verify whether the discrepancies observed in a single reconstruction ({\it e.g.}, at $N_{\mathrm{ev}} = 200$) stem from intrinsic modeling biases or are simply a predictable consequence of the statistical fluctuations inherent in the data. The framework's statistical robustness is confirmed by the distribution of reduced chi-squared values across replicas, typically falling between $0.60$ and $0.90$, with total ranges of $[0.33, 1.39]$, $[0.49, 1.52]$, and $[0.51, 1.59]$ for $N_{\text{ev}} = 200, 2\,000,$ and $20\,000$, respectively. High MH acceptance rates ($50\%$--$85\%$) further demonstrate the NF's ability to accurately approximate the posterior and provide reliable uncertainty estimates.

This methodology allows for a rigorous decomposition of the total uncertainty into two distinct contributions, representing the intrinsic model precision and the statistical stability across different data realizations:
\begin{equation}
\sigma^2_{\text{tot}}(b_T) = \sigma^2_{\text{int}}(b_T) + \sigma^2_{\text{ext}}(b_T) \,.
\end{equation}
The first term, the internal variance $\sigma^2_{\text{int}}$, represents the average model precision and is calculated by averaging the posterior variances obtained within each individual replica, while the second term, the ensemble variance $\sigma^2_{\text{ext}}$, measures the statistical stability of the reconstruction by quantifying the variance of the posterior means across the 100 replicas. This decomposition effectively isolates how much the result fluctuates due to the specific stochastic realization of the data versus the inherent resolution limit of the methodology. 

\begin{figure}[t] 
	\centering
	\includegraphics[width=\textwidth]{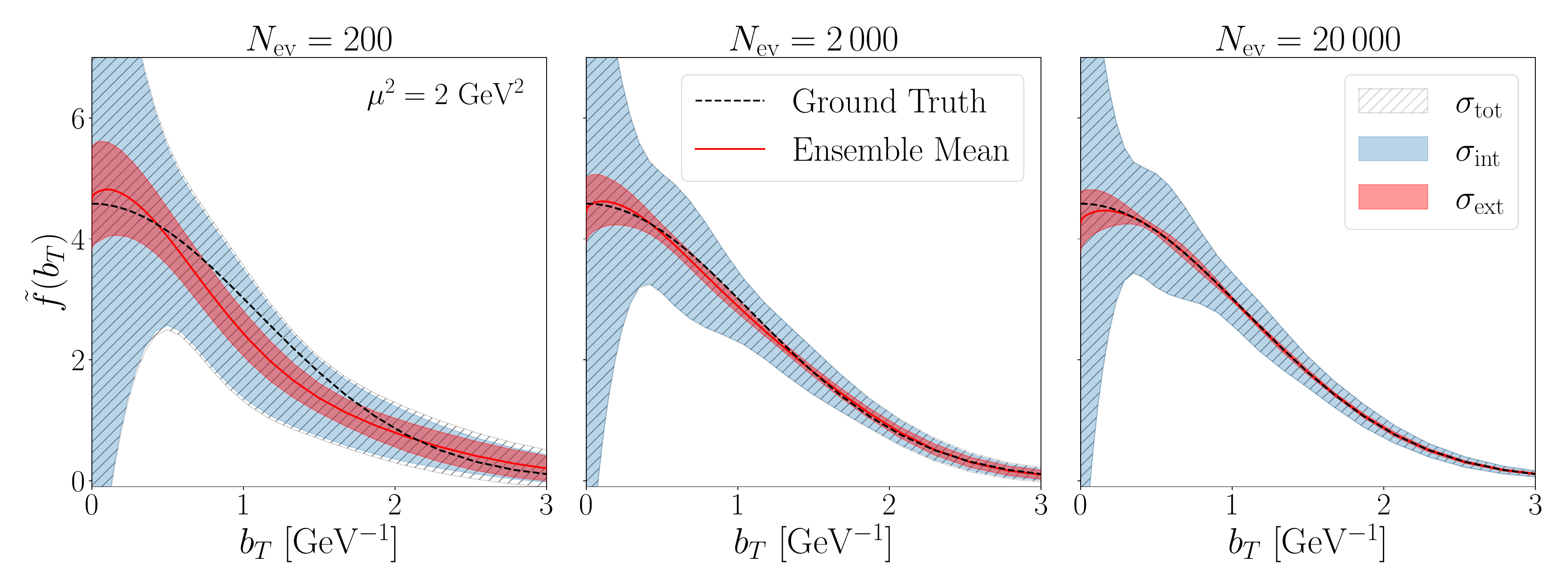}
	\caption{Decomposition of uncertainty in the impact parameter space ($b_T$). The ensemble mean (red line) consistently aligns with the ground truth (black dashed). The total uncertainty (gray hatched band) is the sum of the average model precision (blue band) and the statistical accuracy across replicas (red hatched band).}
	\label{fig:uncertainty_decomposition_bt}
\end{figure}
\begin{figure}[b] 
	\centering
	\includegraphics[width=\textwidth]{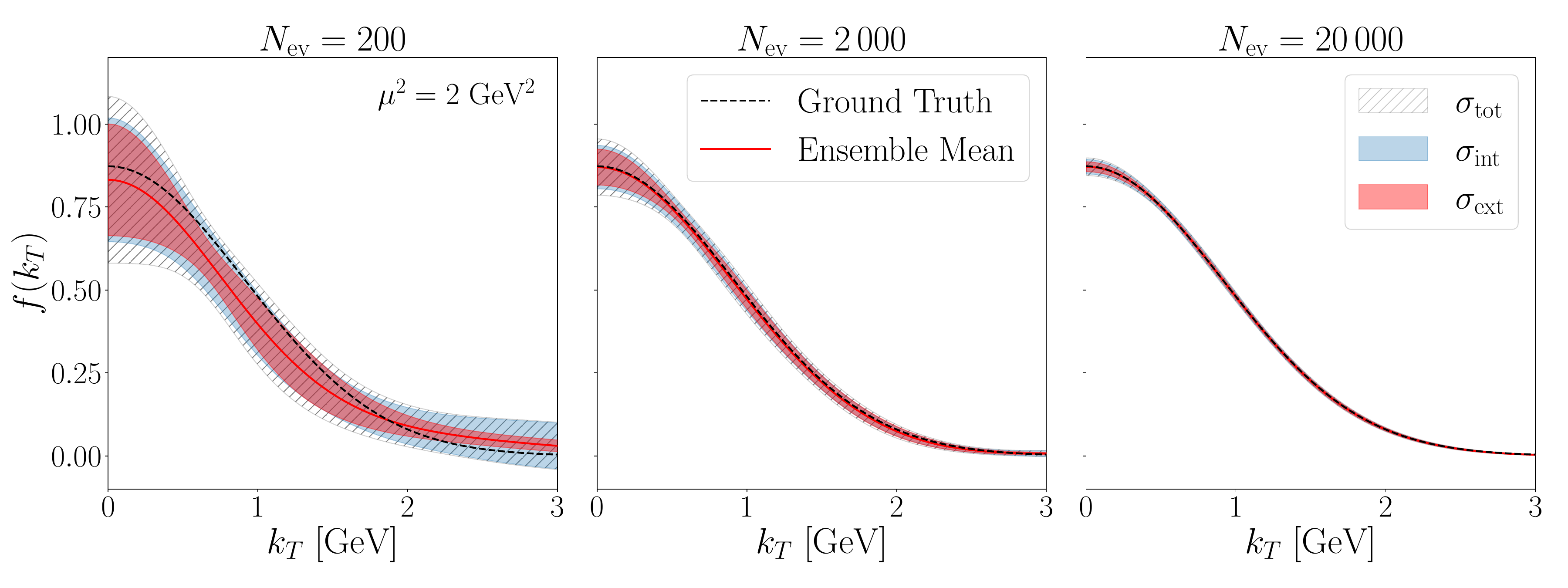}
	\caption{Projection of the uncertainty decomposition in the transverse momentum space ($k_T$). The labels of the lines and bands are as in Fig.~\ref{fig:uncertainty_decomposition_bt}.}
	\label{fig:uncertainty_decomposition_kt}
\end{figure}

The results of this analysis, presented in Fig.~\ref{fig:uncertainty_decomposition_bt} for the impact parameter space and in Fig.~\ref{fig:uncertainty_decomposition_kt} for the transverse momentum projection, reveal several key features of the reconstruction framework. First, the methodology proves to be statistically consistent: even in the low-statistics regime ($N_{\mathrm{ev}} = 200$), where a single reconstruction may deviate from the analytical ground truth, the total uncertainty band 
consistently encompasses the truth. This demonstrates that the framework provides a reliable estimation of its own confidence intervals without introducing systematic biases. Furthermore, the ensemble variance 
narrows significantly as the number of events increases, confirming that the reconstruction converges toward a unique solution as statistical fluctuations average out.

However, this convergence is notably nonuniform across the $b_T$ spectrum. In the large impact parameter region, where the Bessel kernel is most sensitive, the uncertainty bands contract efficiently with increasing statistics. Conversely, at small distances ($b_T~\to~0$), the uncertainty remains substantial even in the high-statistics scenario. This nonuniform scaling confirms that the observed resolution limit is an intrinsic feature of the ill-posed inverse problem: the available data, constrained by a finite $k_T$ range, do not contain sufficient information to resolve the distribution’s behavior at extremely small distances.

In conclusion, the ensemble analysis confirms the robustness of our framework. The model correctly converges in data-sensitive regions, while appropriately reflecting its lack of information through broad, stable uncertainty bands in regions where the data provide no constraining power. This suggests that the ``precision floor'' observed at small distances is not a consequence of insufficient statistics or algorithmic bias, but rather an inescapable limit imposed by the physics of the transformation. To reveal the origin of this behavior, the following section provides a formal characterization of the inversion through a SVD of the mapping operator.

\section{SVD Analysis and the Resolution Limit}
\label{sec:svd_analysis}

To understand the origin of these resolution limits and the nontrivial scaling observed in the impact parameter space, we must analyze the mathematical structure of the forward mapping matrix $\mathcal{M}_{\text{tot}}$. The discrete inverse problem $\boldsymbol{f} = \mathcal{M}_{\text{tot}}\, \tilde{\boldsymbol{f}}$ is inherently ill-posed and rank-deficient. This implies that the data alone are insufficient to uniquely determine the distribution $\tilde{\boldsymbol{f}}$; instead, there exists a manifold of possible configurations in $b_T$ space that can reproduce the observed momentum-space data within their statistical uncertainties.

Importantly, this rank-deficiency is not merely a numerical artifact of the discretization; it is a fundamental property of the Bessel kernel and the finite $k_T$ range sampled. Even with a significantly higher density of observations, the transformation is characterized by an extremely high condition number, meaning that the mapping effectively ``compresses'' large regions of the $b_T$ space into a very narrow subspace in $k_T$.

In this context, the NF-MH framework performs a crucial role:~rather than seeking a single, potentially unstable point estimate, it explores the entire ensemble of solutions compatible with the likelihood and the structural constraints of the prior. The Bayesian credibility bands presented in the previous section are, in fact, a representation of this degenerate space. The lack of shrinkage in the uncertainty bands at small $b_T$ confirms that even a massive increase in data statistics cannot lift this degeneracy if the kernel matrix itself is effectively ``blind'' to those specific degrees of freedom. To formally quantify this loss of information and identify which components of the distribution reside in the effective null-space of the transformation, we employ the SVD as a diagnostic tool.

\subsection{Formalism of the Singular Value Decomposition}

The SVD of the $M \times N$ kernel matrix $\mathcal{M}$ provides a powerful diagnostic tool to analyze the mapping between the impact parameter and transverse momentum spaces~\cite{Aster:2018}. Any matrix $\mathcal{M}$ can be factorized as:
\begin{equation}
\label{eq:svd_decomp}
\mathcal{M} = U \Sigma V^T,
\end{equation}
where $U$ is an $M \times M$ orthogonal matrix whose columns, $\mathbf{u}_i$, form an orthonormal basis for the data space ($k_T$ space), and $V$ is an $N \times N$ orthogonal matrix whose columns, $\mathbf{v}_j$, form an orthonormal basis for the model space ($b_T$ space). The matrix $\Sigma$ is an $M \times N$ pseudo-diagonal matrix containing the singular values $s_1 \ge s_2 \ge \dots \ge s_{\min(M,N)} \ge 0$. 

In practice, the rapid decay of the singular values allows for a compact, low-rank representation of the kernel:
\begin{equation}
\label{eq:svd_compact}
\mathcal{M} \approx U_p\, S_p\, V_p^T = \sum_{i=1}^{p} s_i\, \mathbf{u}_i\, \mathbf{v}_i^T  \,,
\end{equation}
where $p$ represents the effective rank, determined by the number of singular values significantly above the noise floor. In this representation, $U_p$ and $V_p$ are matrices formed by the first $p$ columns of $U$ and $V$, respectively, while $S_p = \text{diag}(s_1, \dots, s_p)$ contains the dominant singular values. Formally, the columns of $V$ and $U$ are the orthonormal eigenvectors of the covariance-like matrices:
\begin{equation}
(\mathcal{M}^T \mathcal{M})\, \mathbf{v}_i = s_i^2\, \mathbf{v}_i\, , \qquad
(\mathcal{M} \mathcal{M}^T)\, \mathbf{u}_i = s_i^2\, \mathbf{u}_i\, ,
\end{equation}
spanning the impact parameter and transverse momentum spaces, respectively. This relationship is manifest in the action of the kernel matrix on the right-singular vectors, $\mathcal{M}\, \mathbf{v}_i = s_i\, \mathbf{u}_i$, which shows that the mapping of any mode $\mathbf{v}_i$ is directly scaled by its singular value.

By partitioning the matrix $V$ into its first $p$ columns and the remaining $N-p$ columns, we can define two fundamental subspaces in the model space. The first $p$ vectors, $V_{\text{obs}} = [\mathbf{v}_1, \dots, \mathbf{v}_p]$, span the observable space, representing the features of the $b_T$ distribution that are effectively mapped to the data space and can thus be constrained by measurements. Conversely, the remaining $N-p$ vectors, $V_{\text{null}} = [\mathbf{v}_{p+1}, \dots, \mathbf{v}_N]$, span the null space of the transformation. Since $s_i \approx 0$ for all $i > p$, it follows immediately that $\mathcal{M}\, \mathbf{v}_i \approx 0$ for these modes. Consequently, any component of the distribution lying within this subspace is ``filtered out'' by the kernel, rendering it mathematically inaccessible regardless of the available statistics.

This partition allows us to define the model resolution through the projection operators onto these subspaces:
\begin{equation}
\label{eq:projection_operators}
P_{\text{obs}} = V_{\text{obs}}\, V_{\text{obs}}^T = \sum_{i=1}^{p} \mathbf{v}_i\, \mathbf{v}_i^T\, , 
\qquad
P_{\text{null}} = V_{\text{null}}\, V_{\text{null}}^T = \sum_{i=p+1}^{N} \mathbf{v}_i \mathbf{v}_i^T\, ,
\end{equation}
where $P_{\text{obs}} + P_{\text{null}} = \mathbb{I}$ is the identity operator in the $N$-dimensional model space. In the literature of inverse problems~\cite{Aster:2018}, $P_{\text{obs}}$ is referred to as the \textit{model resolution matrix} $R_m$. The diagonal elements of $R_m$, known as resolution indices, quantify the degree to which each node in the $b_T$ grid is resolved by the data. A resolution index close to unity indicates that the corresponding pixel is effectively constrained by the observations, whereas values near zero identify nodes dominated by the null space, where the final reconstruction is exclusively determined by the prior information.

Using these operators, any reconstructed distribution $\tilde{\boldsymbol{f}}$ can be uniquely decomposed into its observable and null components:
\begin{equation}
\label{eq:model_decomposition}
\tilde{\boldsymbol{f}} = \tilde{\boldsymbol{f}}_{\text{obs}} + \tilde{\boldsymbol{f}}_{\text{null}} = P_{\text{obs}} \tilde{\boldsymbol{f}} + P_{\text{null}} \tilde{\boldsymbol{f}}  \,,
\end{equation}

This separation has a direct impact on the Bayesian inference process. Since any generic distribution can be uniquely decomposed as $\tilde{\boldsymbol{f}} = \tilde{\boldsymbol{f}}_{\text{obs}} + \tilde{\boldsymbol{f}}_{\text{null}}$, the forward mapping to the data space can be written as:
\begin{equation}
\label{eq:likelihood_insensitivity}
\mathcal{M} \tilde{\boldsymbol{f}} = \mathcal{M} \tilde{\boldsymbol{f}}_{\text{obs}} + \mathcal{M} \tilde{\boldsymbol{f}}_{\text{null}} \approx \mathcal{M} \tilde{\boldsymbol{f}}_{\text{obs}}  \,,
\end{equation}
where we have used the property $\mathcal{M} \mathbf{v}_i \approx 0$ for modes in the null space. Consequently, the likelihood function, which depends exclusively on the residuals between the data and the theoretical prediction $\mathcal{M} \tilde{\boldsymbol{f}}$, is effectively insensitive to any component of the distribution residing in the null space. This invariance implies that the data alone cannot constrain the null-space degrees of freedom.

From a Bayesian perspective, while the prior distribution $\pi(\tilde{\boldsymbol{f}})$ imposes a global structural constraint on the impact parameter distribution, the likelihood acts as a reweighting factor that only affects the components residing in the observable space. Therefore, the posterior distribution for $\tilde{\boldsymbol{f}}_{\text{obs}}$ is shaped by the interplay between data and prior, whereas the null-space component $\tilde{\boldsymbol{f}}_{\text{null}}$ remains exclusively constrained by the prior. This formalizes why the uncertainty bands in regions dominated by the null space of the kernel do not shrink with increasing statistics: in these domains, the NF-MH algorithm is effectively sampling the prior manifold, which remains invariant under changes in the number of events~$N_{\text{ev}}$.

This phenomenon is closely related to the concept of shadow distributions encountered in the study of GPDs~\cite{Bertone:2021, Moffat:2023}. In that context, shadow GPDs represent components that do not contribute to the observables or vanish upon integration. Similarly, in our framework, any term proportional to $\tilde{\boldsymbol{f}}_{\text{null}}$ can be added to the reconstructed distribution without affecting the compatibility with the data. These shadow models occupy the null space of the mapping and are exclusively determined by the theoretical prior. As will be discussed in the following sections, this nonuniqueness persists in realistic TMD extractions, leading to the existence of null TMDs that remain fundamentally unconstrained by momentum-space observables.

\subsection{Spectral Analysis and Resolution Limits}

To analyze the information content resolvable by the kernel, we compare the spectral properties of two distinct discretization matrices at $\mu^2 = 2$~GeV$^2$:~the \textit{reference kernel matrix} $\mathcal{M}_{\text{ref}}$ ($M=100$) and the \textit{total kernel matrix} $\mathcal{M}_{\text{tot}}$ ($M=46$, $N_{\text{ev}}=20\,000$). For both, the impact parameter space is discretized with $N=50$ nodes. The number of nonzero singular values is determined by $\min(M, N)$, resulting in 50 values for the reference case and 46 for the data-driven case. This difference reflects the intrinsic limitation imposed by the experimental sampling on the maximum dimensionality of the observable space.

\begin{figure}[t] 
	\centering
	\includegraphics[width=0.8\textwidth]{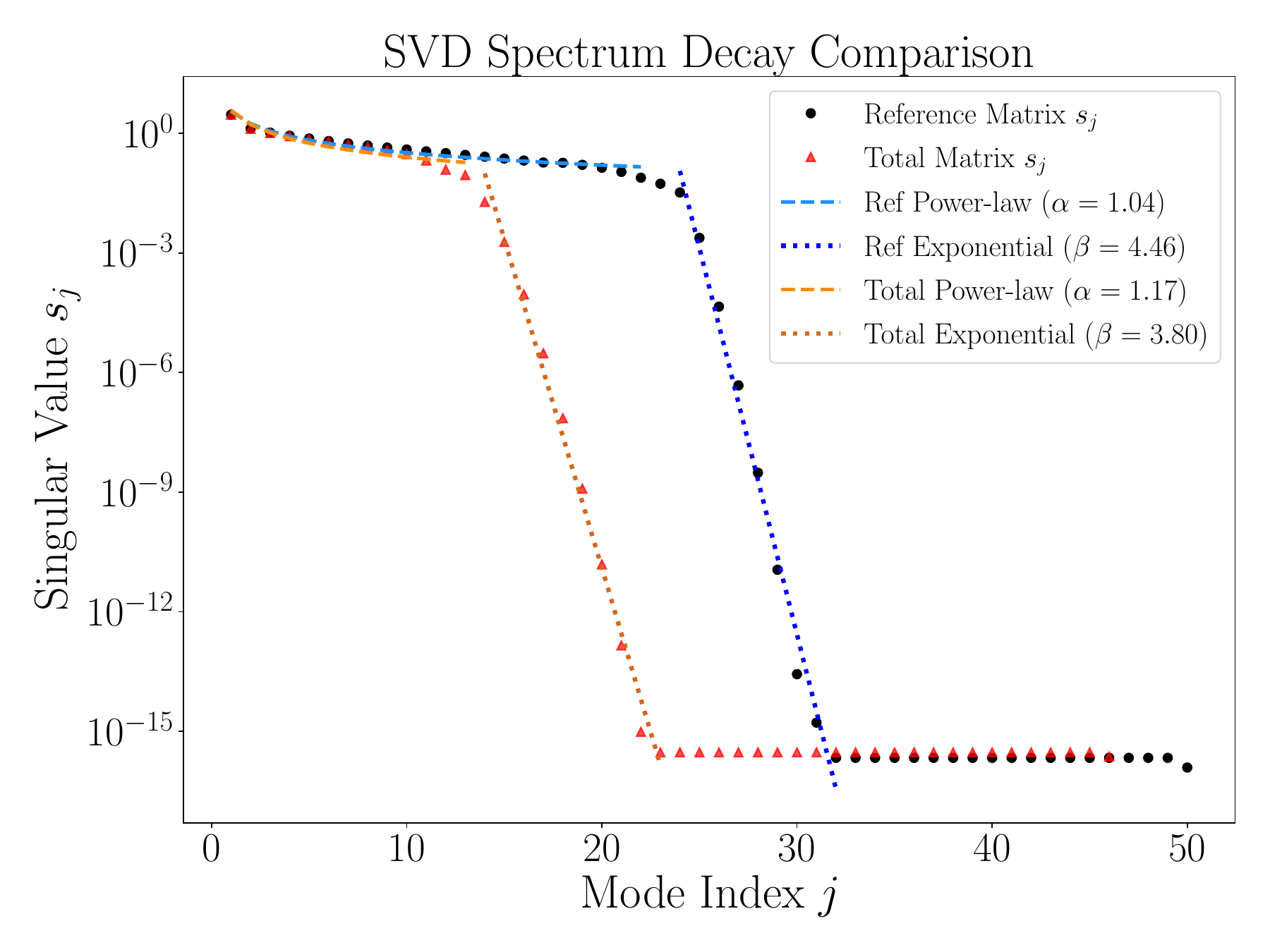}
	\caption{Singular value spectrum $s_j$ for the reference (black) and total (orange) kernels. Both exhibit a transition from moderately to severely ill-posed regimes before reaching the numerical floor.}
	\label{fig:singular_values_decay}
\end{figure}

The singular value spectra $s_j$, shown in Fig.~\ref{fig:singular_values_decay}, confirm the ill-posed nature of the inverse problem. We characterize the decay as mildly ill-posed for $s_j = \mathcal{O}(j^{-\alpha})$ with $\alpha \le 1$, moderately ill-posed for $\alpha > 1$, and severely ill-posed if the decay becomes exponential, $s_j = \mathcal{O}(e^{-\beta j})$.

\begin{figure}[ht]
	\centering
    \vspace*{-0.3cm}
	\includegraphics[width=0.82\textwidth]{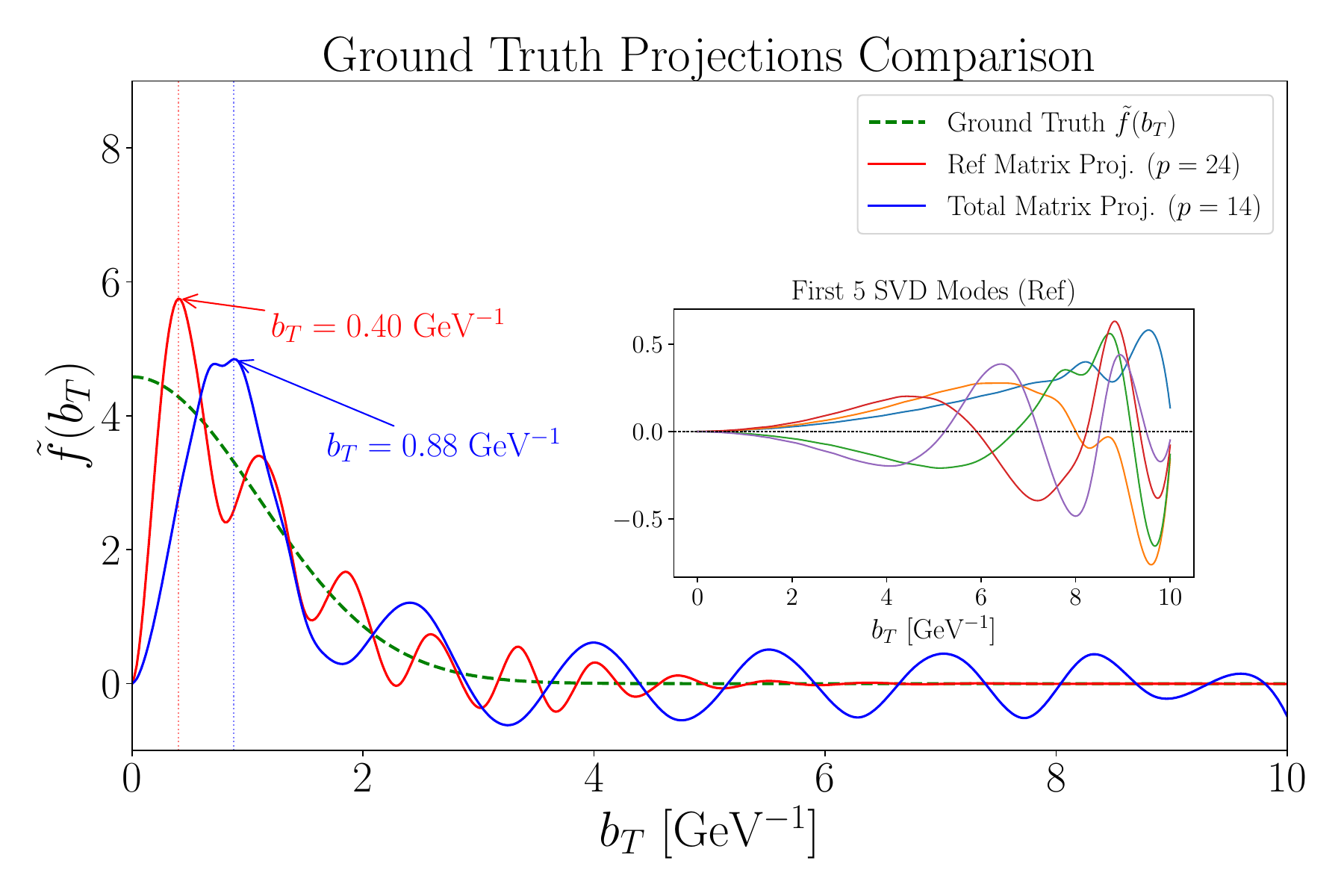}
	\caption{Ground truth distribution and its projections onto the observable spaces spanned by $p=24$ (reference) and $p=14$ (total) modes. The inset shows the first 5 SVD modes. The reference projection peaks at $b_T \approx 0.40~\text{GeV}^{-1}$, while the total kernel projection shifts to $b_T \approx 0.88~\text{GeV}^{-1}$.}
	\label{fig:gt_projection_20k}
\end{figure}

For the reference kernel, the first 24 modes follow a power-law decay ($\alpha \approx 1.04$), identifying a moderately ill-posed regime. In contrast, the total kernel exhibits a more rapid loss of information due to sparser sampling: the power-law regime is restricted to only the first 14 modes ($\alpha \approx 1.17$), followed by a steeper exponential decay. This contraction of the observable subspace, from $p=24$ to $p=14$, quantitatively demonstrates how experimental resolution pushes the information limits toward larger impact parameters.
The physical impact of this spectral decay is illustrated in Fig.~\ref{fig:gt_projection_20k}, which compares the ground truth distribution $\tilde{f}(b_T)$ with its projections onto the observable spaces of both matrices.

\begin{figure}[ht]
	\centering
	\includegraphics[width=0.72\textwidth]{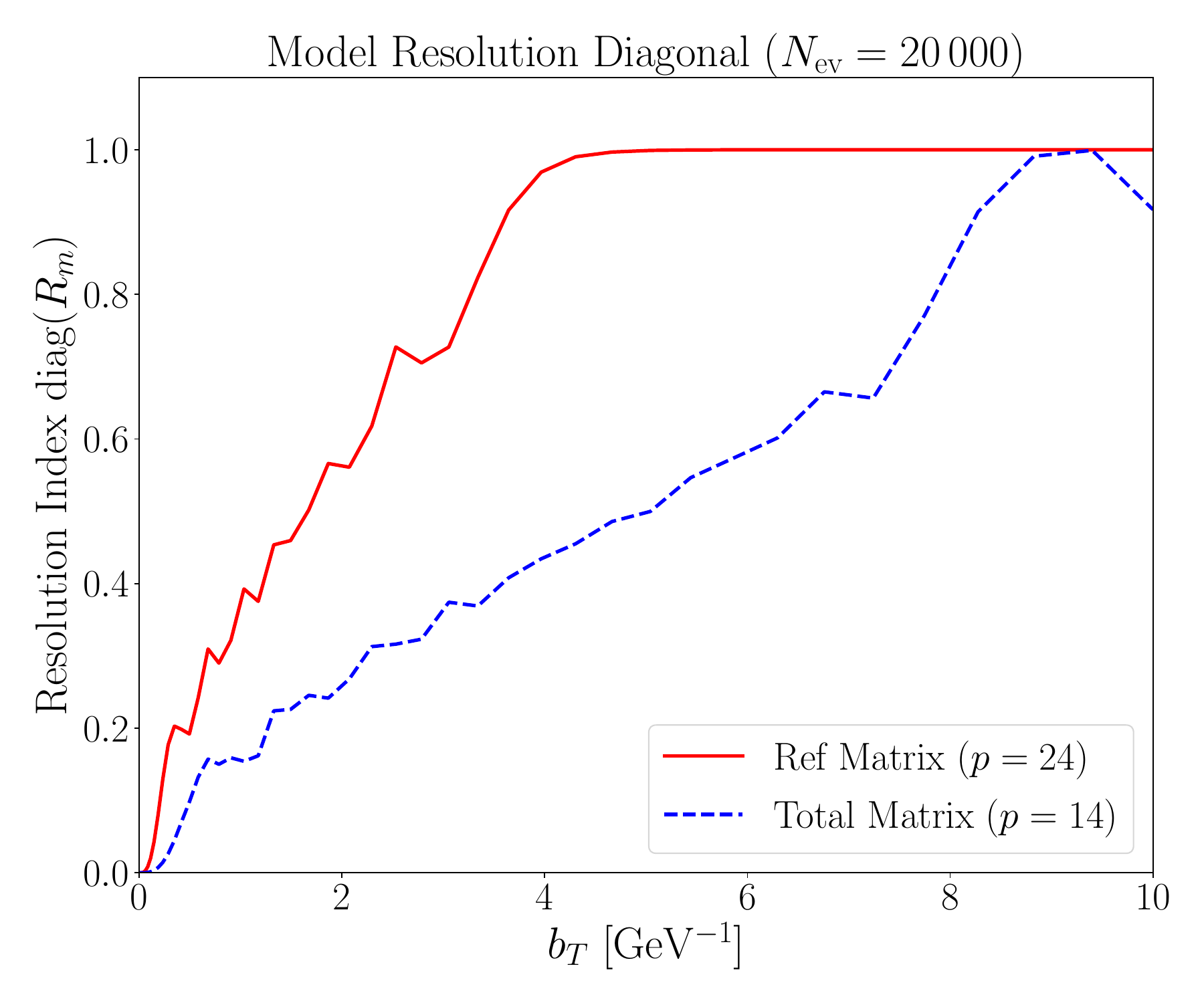}
	\caption{Diagonal of the resolution matrix $R_m$ for $p=14$. The resolution vanishes at small $b_T$, where the reconstruction is dominated by the prior.}
	\label{fig:resolution_diag_20k}
\end{figure}

This is further corroborated by the diagonal elements of the model resolution matrix $R_m$ in Fig.~\ref{fig:resolution_diag_20k}, which drop to zero as $b_T \to 0$. The vanishing of $R_m$ provides a formal justification for the uncertainty bands:~since the data-driven kernel provides zero resolution in this limit, the posterior reverts to the prior, preventing any shrinkage of the credibility intervals regardless of $N_{\text{ev}}$. This region is the natural habitat of the null TMDs, where distinct pixel patterns at the origin remain equally compatible with momentum-space data.

\subsection{Mathematical Origin of the Precision Floor}

The SVD analysis provides a rigorous explanation for the ``precision floor'' observed in our numerical experiments. In the Bayesian framework, the width of the posterior distribution is determined by the interplay between the likelihood and the prior. The lack of shrinkage in the uncertainty bands can be understood through the existence of flat directions in the parameter space. Since $\mathcal{M} \tilde{\boldsymbol{f}}_{\text{null}} \approx 0$, the likelihood function is effectively independent of any component residing in the null space. Consequently, even as experimental uncertainties $\sigma_j \to 0$ (with increasing $N_{\text{ev}}$), the ``curvature'' of the likelihood along these directions remains zero, providing no data-driven force to constrain the null-space modes.

From this perspective, the transition from the prior $\pi(\tilde{\boldsymbol{f}})$ to the posterior is governed by a likelihood reweighting that only affects the observable subspace. For the null-space components, there is no information gain, and the uncertainty bands remain ``frozen,'' purely reflecting the structural constraints of the Tikhonov priors. This sampling of a degenerate manifold explains the emergence of the null TMDs introduced above: 
these are the models that occupy the null space and are exclusively determined by the theoretical prior.

This limitation can be visualized through an optical analogy, where the Bessel transform acts as a diffraction-limited lens. Increasing the statistics is similar to increasing the light intensity on the observed target:~it makes the resolvable details sharper, but it cannot reveal features smaller than the lens's intrinsic resolution limit. Regardless of the precision of the data, the kernel cannot transmit spatial frequencies that belong to its null space.

\begin{figure}[t] 
	\centering
	\includegraphics[width=\textwidth]{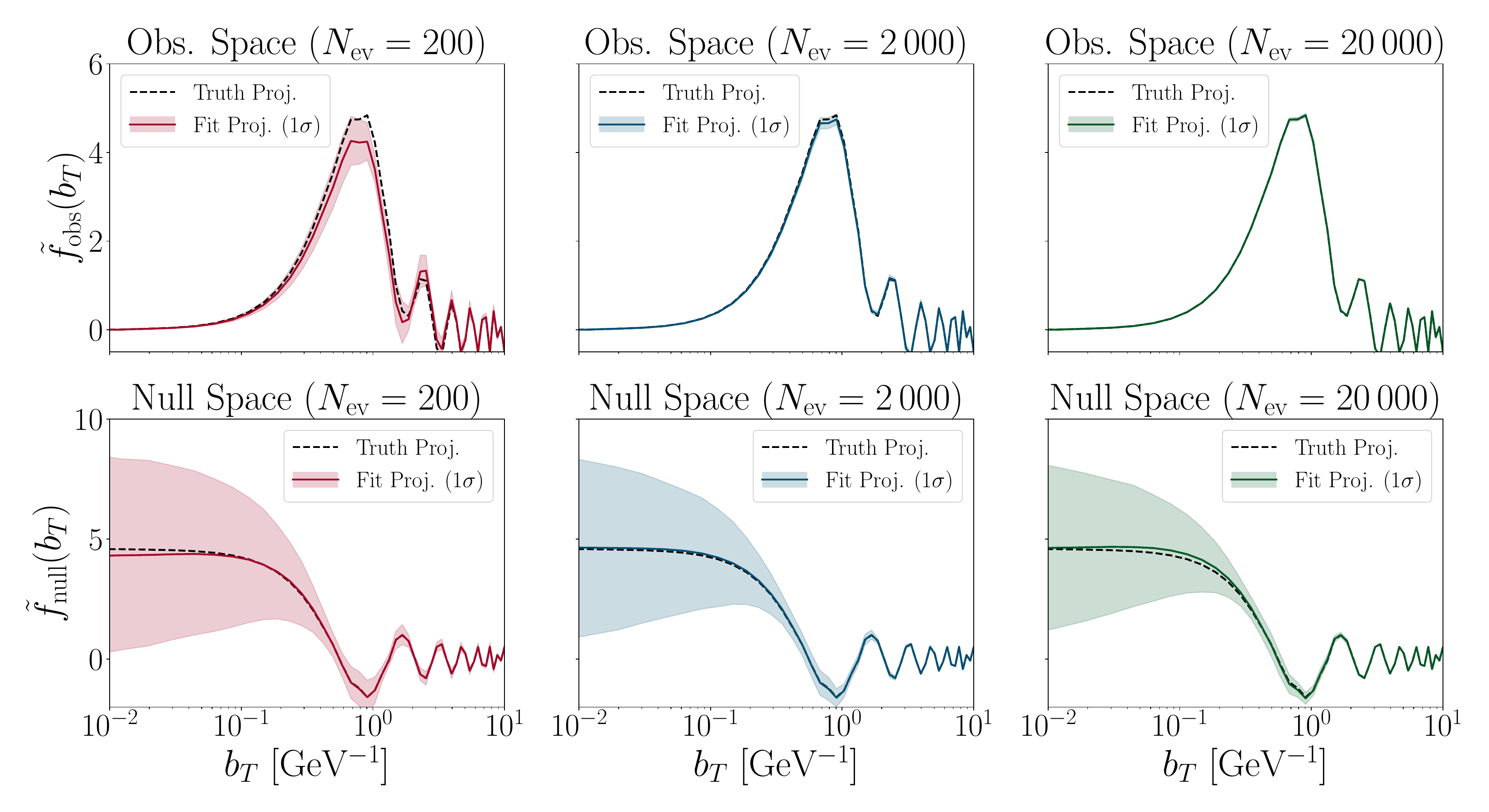}
	\caption{Decomposition of the fit uncertainty into observable (top row) and null (bottom row) subspaces. The statistical gain is only visible in the observable space, while the null space remains unchanged, establishing an irreducible precision floor.}
	\label{fig:projection_scaling}
\end{figure}

The empirical evidence for this mechanism is provided by the scaling analysis shown in Fig.~\ref{fig:projection_scaling}. By decomposing the posterior samples into their observable and null components ($\tilde{\boldsymbol{f}} = \tilde{\boldsymbol{f}}_{\text{obs}} + \tilde{\boldsymbol{f}}_{\text{null}}$), we can track the uncertainty scaling in each subspace independently.  As illustrated, the uncertainty of $\tilde{\boldsymbol{f}}_{\text{obs}}$ shrinks with increasing statistics, following the expected $1/\sqrt{N_{\text{ev}}}$ scaling. Conversely, the uncertainty of $\tilde{\boldsymbol{f}}_{\text{null}}$ remains entirely invariant. Since the total reconstructed distribution is the sum of these components, the null-space uncertainty establishes an irreducible precision floor, explaining why higher statistics do not lead to narrower bands in resolution-limited regions.

While the diffraction limit of the Bessel kernel is a fundamental constraint, the practical resolution limit is strictly dictated by the spectral coverage in momentum space. The comparison between the reference kernel (reaching $k_T = 8~\text{GeV}$) and the data-driven kernel (limited by the sampled Gaussian width) reveals a crucial phenomenological insight. To ``push'' the resolution limit toward smaller impact parameters, it is not sufficient to accumulate more statistics at low transverse momenta.
Instead, one must extend the experimental reach toward the high-$k_T$ tails of the distribution. High-momentum data provide the ``high-frequency'' information necessary to resolve the internal structure of the TMD at small $b_T$. Without this coverage, the core of the distribution remains fundamentally invisible to the data, and its reconstruction will inevitably be shaped by the structural assumptions encoded in the theoretical prior.
This implies that for a single-scale measurement, the resolution frontier is rigidly ``locked'' by the maximum momentum transfer available. To break this impasse, one must either access higher momenta or leverage the dynamical information encoded in the scale dependence of the distribution.

\section{Multi-Scale Analysis and the Impact of Evolution}
\label{sec:multi_scale}

In the previous sections, we established that the resolution of the impact parameter distribution is strictly locked by the maximum transverse momentum $k_T$ effectively probed by the data. To push this frontier toward smaller distances, it is necessary to access information from the high-momentum tails of the distribution. In an experimental context, this is naturally achieved by considering data at different energy scales $\mu^2$: as the scale increases, the broadening of the $k_T$ spectrum provides the high-frequency information required to resolve the short-distance structure of the hadron.

Although the short-distance structure at small $b_T$ is formally constrained by the operator product expansion (OPE) in the context of TMD phenomenology, as discussed later in Section~\ref{sec:tmd_distribution}, exploring the resolution limits in this region remains essential for several reasons. First, it provides a rigorous test of the framework's capability to reconstruct high-frequency functional components from the high-momentum tails of the data. Second, as we will demonstrate, the inclusion of multiple scales does not only provide information at the origin, but effectively reconfigures the null space projection across the entire impact parameter spectrum, leading to a better constrained distribution even at intermediate distances. This data-driven approach allows for a more detailed study of the transition region between perturbative and nonperturbative dynamics and paves the way for future global analyses where TMD and collinear distributions could be extracted simultaneously, as recently demonstrated in Ref.~\cite{Barry:2025glq}.

In this section, we extend our framework to a multi-scale analysis. By simulating a range-extension similar to what is prescribed by the CSS 
evolution equations, we demonstrate how a simultaneous fit across multiple scales provides a superior resolution compared to any single-scale measurement. This approach allows us to ``triangulate'' the distribution, leveraging the increased $k_T$ reach at higher energies to fill the null space that remains inaccessible at lower scales.

\subsection{Modeling the Energy Dependence}

To simulate a dynamic experimental scenario, we consider a set of observations $\{f(k_T, \mu_k^2)\}$ measured at different energy scales. Specifically, we analyze the system over four distinct scales:
\begin{equation}
\label{eq:mu_scales}
\mu_k^2 \in \{2, 5, 20, 50\} \text{ GeV}^2 \,.
\end{equation}
These values are chosen to cover a wide kinematic range, allowing the $k_T$ spectrum to undergo a significant broadening as the scale increases. Following the logic of TMD evolution, the energy-dependent width parameter $\sigma^2(\mu^2)$ is governed by the relation established in Eq.~\eqref{eq:sigma_evolution}, with the evolution parameter fixed at $g_{\text{evo}} = 1.2 \text{ GeV}^2$.
To assess the robustness of the inversion, we generate pseudo-data for each scale across three statistical regimes: $N_{\text{ev}} \in \{200, 2\, 000, 20\, 000\}$ events. This setup allows us to quantify how the reduction of statistical noise interacts with the energy-dependent broadening to resolve the short-distance structure of the hadron.

As previously discussed, the core of our methodology remains the reconstruction of the reference distribution $\tilde{\boldsymbol{f}}$ at the initial scale $\mu_0^2 = 1$~GeV$^2$. For each scale $\mu_k^2$ in Eq.~\eqref{eq:mu_scales}, the theoretical prediction is obtained by applying the evolution operator $U(b_T, \mu_k^2)$ point-wise to the reference nodes, effectively ``rescaling'' the intrinsic structure before projecting it into the observable space.

A crucial aspect of this multi-scale setup is the handling of the discretization matrices. Since each energy scale probes a different portion of the momentum space, extending the effective $k_T$ range as $\mu^2$ grows, the mapping must account for scale-dependent experimental grids. To this end, we define a set of interpolation matrices $\{\mathcal{M}_{\text{interp}}^{(k)}\}$ tailored to the specific $k_T$ coordinates of each dataset. This ensures that the forward mapping:
\begin{equation}
\boldsymbol{f}(\mu_k^2) = \mathcal{M}_{\text{tot}}^{(k)} \cdot \left[ \tilde{\boldsymbol{f}}(\mu_0^2) \odot \mathbf{U}(\mu_k^2) \right]
\end{equation}
remains numerically exact for all energies, leveraging the increased high-momentum sensitivity at $\mu^2 = 50$~GeV$^2$ to constrain the small-$b_T$ behavior of the distribution at the reference scale.

In this multi-scale configuration, we extend our Bayesian framework to include $g_{\text{evo}}$ as a free parameter of the fit, associated with a wide Gaussian prior $\pi(g_{\text{evo}})$. Unlike the single-energy analysis, the global likelihood now couples the four datasets, allowing the \mbox{NF-MH} algorithm to sample the joint posterior $P(\tilde{\boldsymbol{f}}, g_{\text{evo}} | \boldsymbol{f})$. By leveraging the different $k_T$ ranges across scales, this simultaneous inference breaks the degeneracy between the intrinsic functional shape and the evolution rate. This synergy not only provides a data-driven determination of $g_{\text{evo}}$ but also yields a high-resolution reconstruction of the $b_T$ distribution, effectively filling the null space that remains inaccessible in a single-scale measurement.


\begin{figure}[t] 
	\centering
    \vspace*{-0.3cm}
	\includegraphics[width=0.82\textwidth]{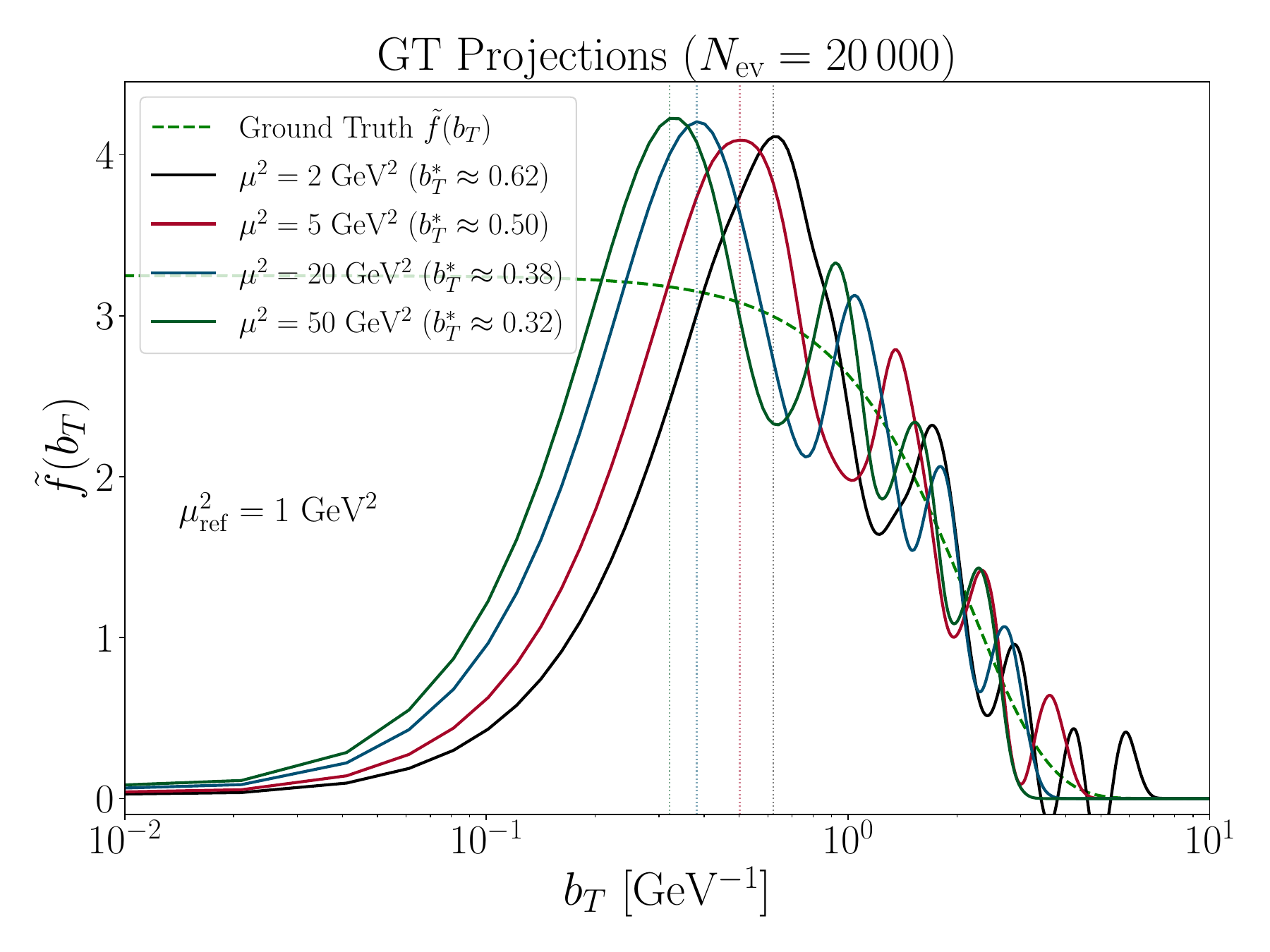}
	\caption{Ground Truth projections into the observable space for different energy scales (using the $N_{\text{ev}}=20\,000$ matrix), where the legend indicates the position of the projection peak, $b_T^*$. As the energy $\mu^2$ increases, the peak of the projection shifts toward smaller $b_T$ values, effectively expanding the observable subspace.}
	\label{fig:gt_projections_multi}
\end{figure}

The core advantage of the multi-scale approach is visualized in Fig.~\ref{fig:gt_projections_multi}, which displays the projections of the ground truth into the observable space for different energy scales. These curves, obtained through the SVD of the total projection matrices $\mathcal{M}_{\text{tot}}^{(k)}$ for each energy set, represent the sensitivity of the experimental $k_T$ spectrum to specific regions of the $b_T$ distribution.
As the energy $\mu^2$ increases, the projection peak shifts toward smaller $b_T$ values, moving from $b_T=0.62$~GeV$^{-1}$ at $\mu^2 = 2$~GeV$^2$ to $b_T=0.32$~GeV$^{-1}$ at $\mu^2 = 50$~GeV$^2$. This systematic shift effectively expands the observable subspace, allowing the framework to resolve shorter distances that remain inaccessible at lower scales.

\subsection{Results and Statistical Performance}

\begin{figure}[t] 
	\centering
    \vspace*{-0.3cm}
	\includegraphics[width=1.02\textwidth]{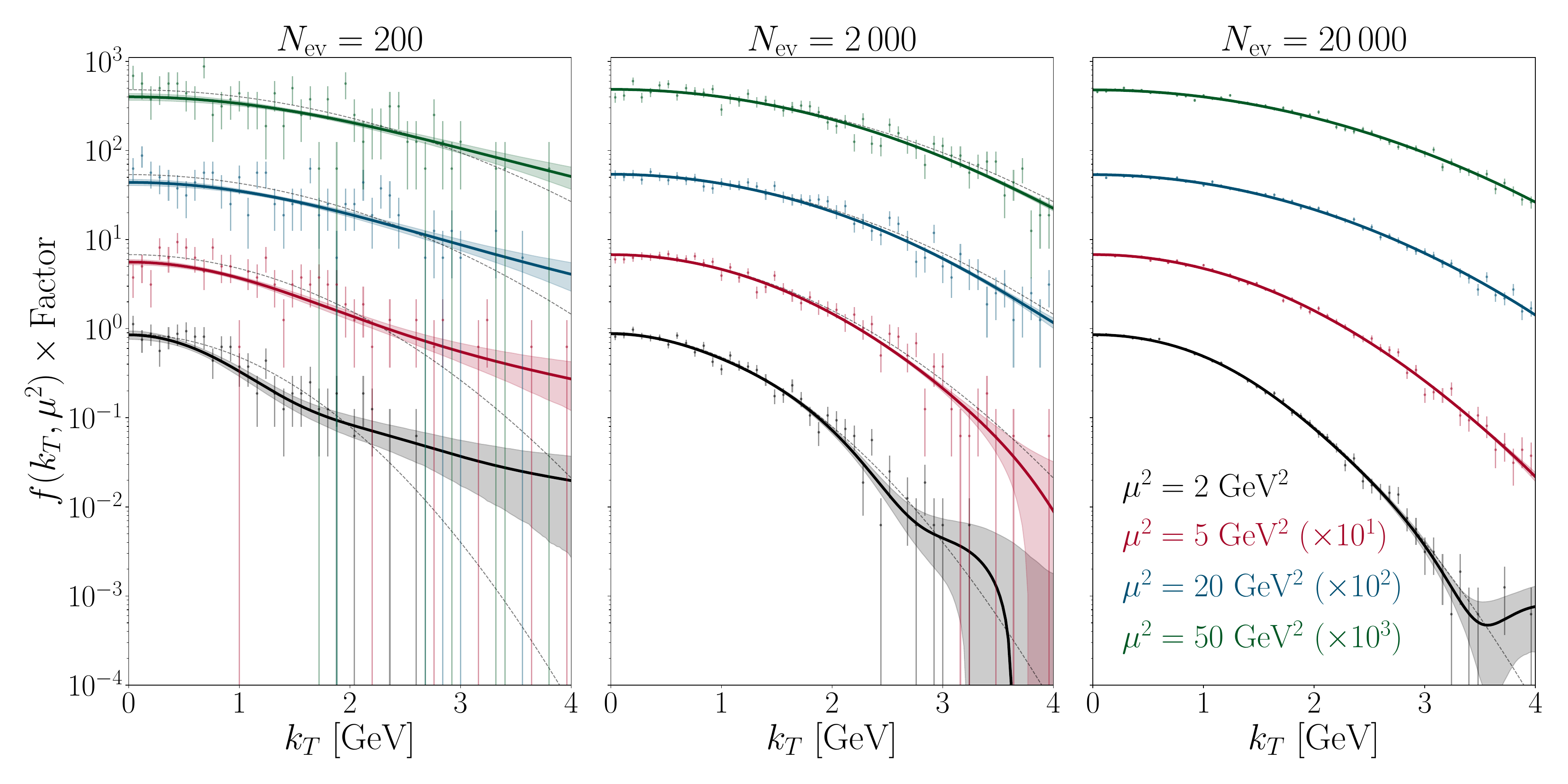}
	\caption{Multi-energy TMD fit versus data. The results are shown for the three event counts ($N_{\text{ev}}$) across four energy scales $\mu^2$. For visual clarity, the curves are shifted by multiplicative offsets as indicated in the labels.}
	\label{fig:multi_energy_shifted}
\end{figure}

The results of the global Bayesian inference are summarized in Fig.~\ref{fig:multi_energy_shifted}, which compares the experimental pseudo-data with the reconstructed TMD distributions across the four energy scales. The framework demonstrates an excellent description of the data in all statistical regimes, correctly capturing the broadening of the $k_T$ spectrum induced by the evolution. Diagnostic metrics, reported in Table~\ref{tab:fit_diagnostics_gauss_multi}, confirm a reduced $\chi^2$ near unity and a stable acceptance rate above 80\% for $N_{\text{ev}} \geq 2\,000$.

\begin{table}[b] 
	\caption{Global fit diagnostics for three statistical regimes for the Gaussian validation case, with $N_{\text{pts}}$ the total number of points across all four scales, and $\chi^2/N_{\text{pts}}$ the reduced chi-squared calculated at the posterior mean. The mean and standard deviation of the extracted $g_{\text{evo}}$ (ground truth 1.2) demonstrate the increased precision 
    with higher statistics.}
	\centering
	\begin{tabular}{rclcc}
		\hline
		$N_{\text{ev}}$~ & accept. rate & $g_{\text{evo}} \, [{\rm GeV}^2]$  & $N_{\text{pts}}$ &  $\chi^2/N_{\text{pts}}$ \\
		\hline
		$200$     & 59\% & ~1.2(3)    & 148 & 0.99 \\
		$2\,000$  & 82\% & ~1.10(5)   & 207 & 1.05 \\
		$20\,000$ & 82\% & ~1.209(15) & 248 & 0.99 \\
		\hline
	\end{tabular}
	\label{tab:fit_diagnostics_gauss_multi}
\end{table}

\begin{figure}[t] 
	\centering
	\includegraphics[width=\textwidth]{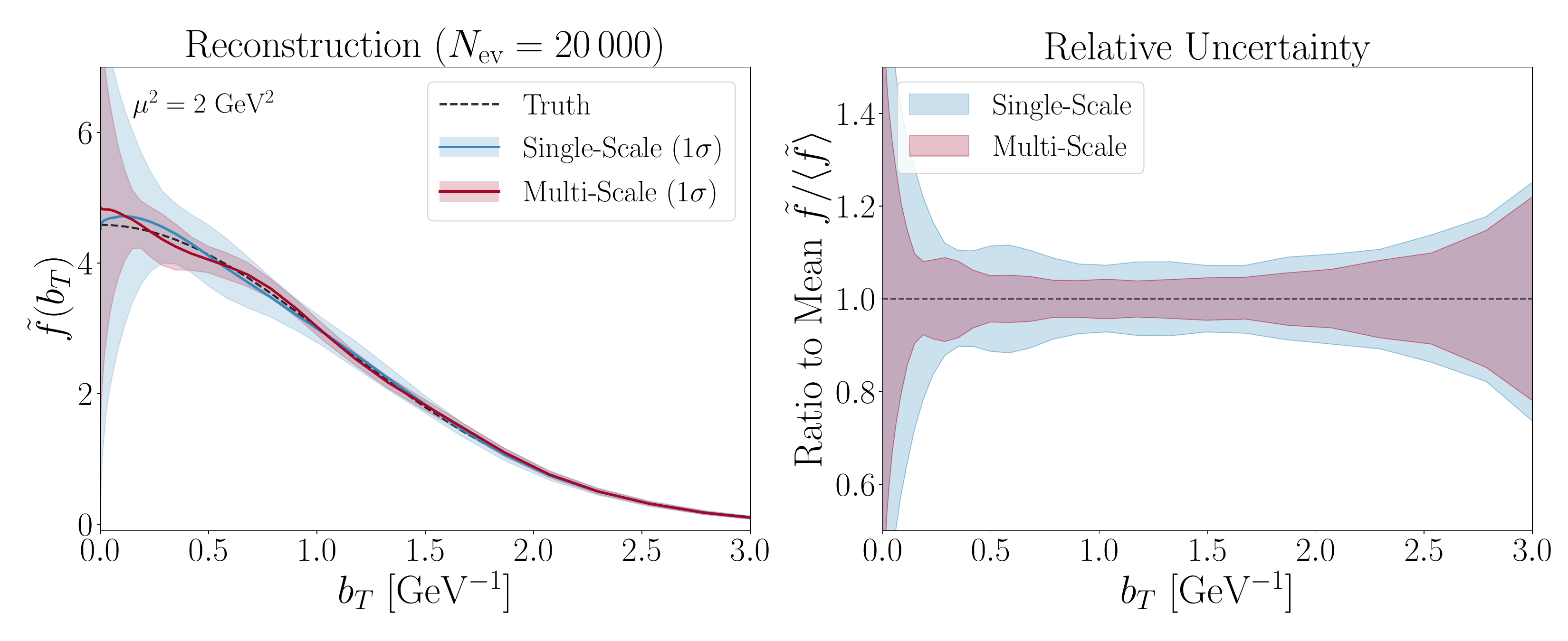}
	\caption{Comparison of the $b_T$ reconstruction for $N_{\text{ev}} = 20\,000$. (Left) Extracted distribution for the single-scale case versus the multi-scale simultaneous fit. (Right) Ratio of the uncertainty bands to the mean value, highlighting the significant precision gain afforded by the inclusion of multiple energy scales.}
	\label{fig:compare_20k}
\end{figure}

The physical advantage of the multi-scale approach is most clearly illustrated in Fig.~\ref{fig:compare_20k}. By comparing the $b_T$ distribution extracted at a single scale versus the multi-scale simultaneous fit (both at $N_{\text{ev}} = 20\,000$), we observe a dramatic improvement in the reconstruction. The ratio of the uncertainty bands to the mean value 
reveals that the multi-energy lever arm suppresses the uncertainty across the entire range, especially in the regions where the single-energy information is intrinsically insufficient to resolve the Bessel inversion.

\begin{figure}[t] 
	\centering
	\includegraphics[width=\textwidth]{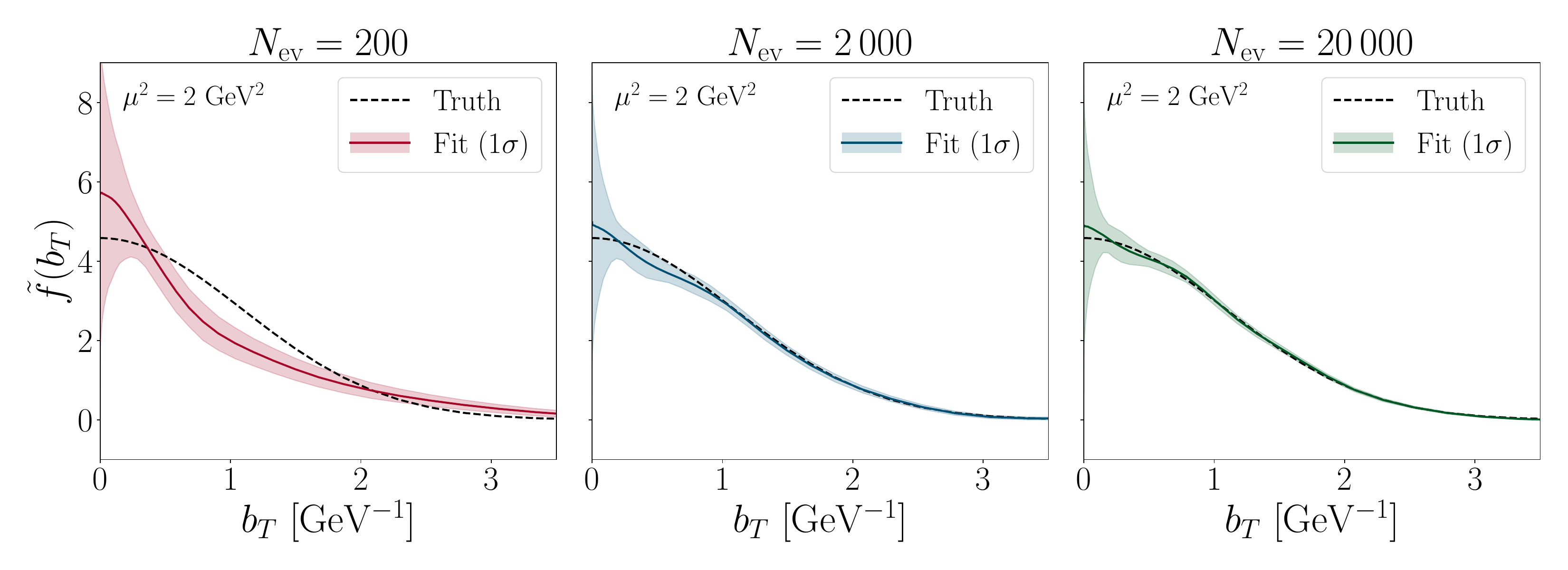}
	\caption{Reconstruction of the impact parameter distribution at the scale $\mu^2 = 2~\text{GeV}^2$ from the multi-energy simultaneous fit. Compared to the single-energy case, the uncertainty bands are significantly narrower, especially in the intermediate and small-$b_T$ regions.}
	\label{fig:bt_comparison_multi}
\end{figure}

The stability of the reconstruction is further detailed in Fig.~\ref{fig:bt_comparison_multi}. Compared to single-energy extractions, the uncertainty bands are significantly narrower across the entire range, particularly in the intermediate and small-$b_T$ regions. While a fundamental resolution limit remains as $b_T \to 0$, the multi-energy precision floor is pushed significantly lower.

The physical driver of this enhanced resolution is the energy-dependent broadening of the $k_T$ spectrum. The evolution stretches the distribution toward higher transverse momenta as $\mu^2$ increases, providing the high-frequency information, and thus the small-distance sensitivity, that is missing at lower scales. By effectively combining these different sensitivity windows, the NF-MH framework leverages the high-energy data to resolve the core of the distribution while the low-energy data anchor the large-distance tail.

\begin{figure}[ht]
	\centering
	\includegraphics[width=\textwidth]{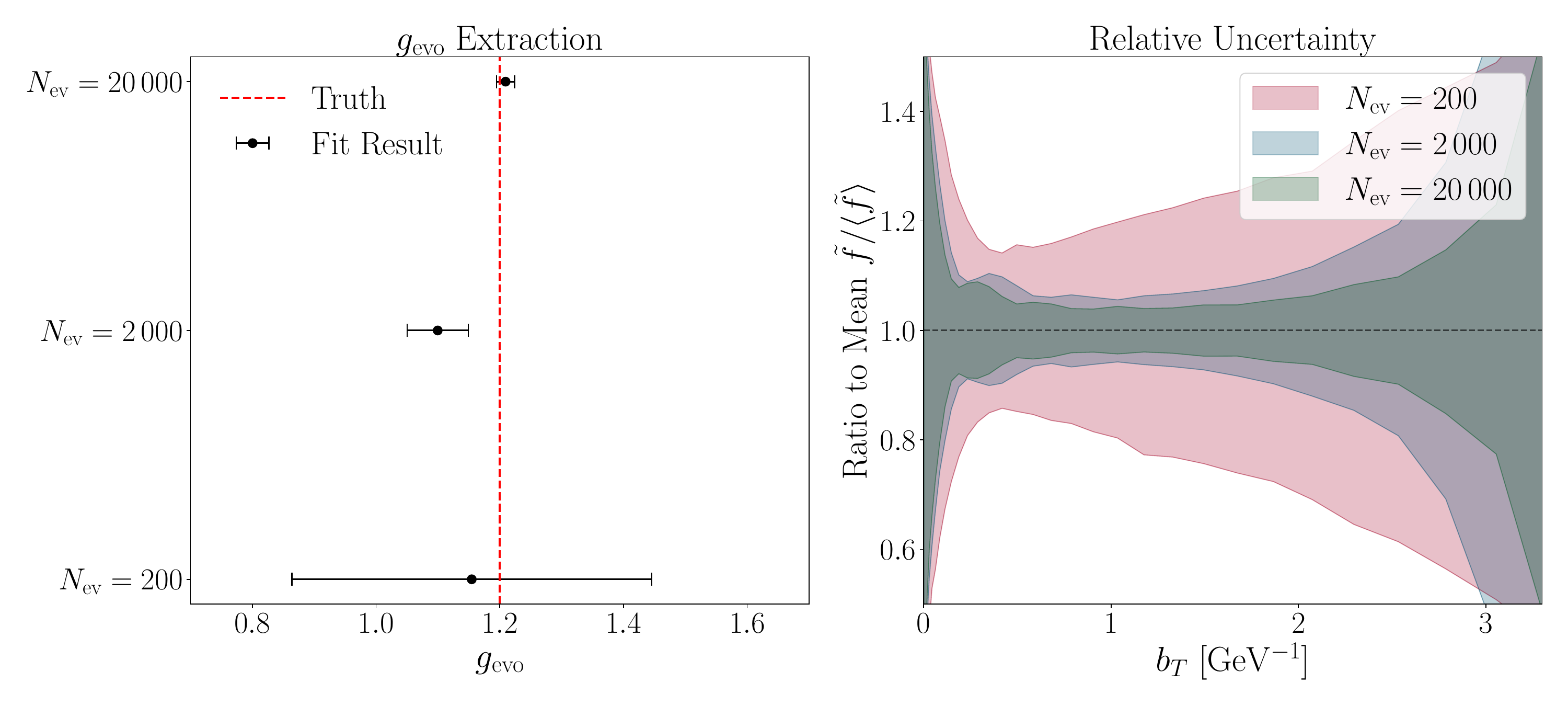}
	\caption{Summary of the multi-energy fit results. (Left) Extraction of the evolution parameter $g_{\text{evo}}$ for different event counts compared to the true value (dashed line). (Right) Relative uncertainty of the $b_T$ reconstruction, showing the scaling of the precision with $N_{\text{ev}}$.}
	\label{fig:combined_gevo_uncertainty}
\end{figure}

Finally, the simultaneous extraction yields a precise determination of the evolution dynamics. As shown in the left panel of Fig.~\ref{fig:combined_gevo_uncertainty}, the uncertainty on $g_{\text{evo}}$ shrinks by nearly a factor of 20 between the $N_{\text{ev}}=200$ and $N_{\text{ev}}=20,000$ regimes. This confirms that the multi-energy information is sufficient to break the degeneracy between the evolution magnitude and the intrinsic functional shape, allowing $g_{\text{evo}}$ to be treated as a free parameter.

\section{TMD Distribution Framework}
\label{sec:tmd_distribution}

In this section, we extend our analysis to the full TMD PDFs. To simplify the validation of the pixelated framework, we focus on the $u$-quark flavor as a representative case study. Following the methodology established for the Gaussian models, we perform a systematic analysis across both single-scale and multi-scale scenarios. For the purpose of this analysis, we treat the TMD PDF as an observable directly measurable in momentum space ($k_T$). However, its formal definition and scale evolution are most naturally expressed in the impact parameter space ($b_T$). The connection between these two representations is established via a Bessel transform:
\begin{equation}
\label{eq:bessel_transform_tmd}
f_{f/P}(x, k_T; Q) = \frac{1}{2\pi} \int_0^\infty \dd b_T \, b_T \, J_0(k_T b_T) \, \tilde{f}_{f/P}(x, b_T; Q) \, ,
\end{equation}
where the $b_T$-space distribution is defined according to the CSS 
formalism as:
\begin{equation}
\label{eq:PDF_func}
\begin{split}
\tilde{f}_{f/P}(x, b_T; Q) &= \sum_j \int_x^1 \frac{\dd\hat{x}}{\hat{x}}\, C_{f/j}\big(x/\hat{x}, b_*; \mu^2_b, \mu_b, g(\mu_b)\big)\, f_{j/P}(\hat{x}, \mu_b) \\
&\quad \times \exp\! \big[ S_{\rm pert}(b_T;\mu_b, Q) \big] M_f(x,b_T) \exp\! \Big[\! - g_K(b_T) \ln \frac{Q}{Q_{0}} \Big]\, .
\end{split}
\end{equation}

The structure of Eq.~\eqref{eq:PDF_func} reflects a rigorous factorization between different physical regimes. The first line represents the OPE at small $b_T$, where the TMD PDF is matched onto the collinear PDFs $f_{j/P}$ where, for this study, we choose to employ the JAM20 set~\cite{Moffat:2021mxy} through the Wilson coefficients $C_{f/j}$, evaluated here at next-to-leading order (NLO)~\cite{Aybat:2011zv}. To avoid the divergence of the perturbative expansion at large $b_T$, we adopt the $b_*$ prescription:
\begin{equation}
b_*(b_T) = \frac{b_T}{\sqrt{1 + (b_T/b_{\text{max}})^2}} \, ,
\end{equation}
with $b_{\text{max}} = 0.8~\text{GeV}^{-1}$. This choice anchors the matching scale to $\mu_b = 2e^{-\gamma_E}/b_*$, ensuring it remains within the perturbative domain.

The second line of Eq.~\eqref{eq:PDF_func} encodes the scale evolution and the intrinsic nonperturbative structure. The term $S_{\rm pert}$ denotes the perturbative Sudakov form factor:
\begin{equation}
\label{eq:perturbative_sudakov}
S_{\rm pert}(b_T;\mu_b,Q) = K(b_*; \mu_b) \ln \frac{Q}{\mu_b} + \int_{\mu_b}^{Q} \frac{\dd \mu'}{\mu'} \Big[ \gamma_D(g(\mu'); 1) - \ln \frac{\sqrt{\zeta}}{\mu'} \gamma_K(g(\mu')) \Big] \, .
\end{equation}
In our implementation, the anomalous dimensions $\gamma_K$ and $\gamma_D$ are considered up to $\mathcal{O}(\alpha_s^2)$ and $\mathcal{O}(\alpha_s^1)$, respectively, providing a consistent NLO accuracy for the evolution kernel. Within the $b_*$ prescription and with the choice of the matching scale $\mu_b$, the term $K(b_*; \mu_b)$ in Eq.~\eqref{eq:perturbative_sudakov} vanishes. 
The long-distance dynamics are then captured by $M_f(x, b_T)$, accounting for the intrinsic nucleon structure, and $g_K(b_T)$, the nonperturbative component of the CSS kernel. As detailed in the following sections, these functions will be extracted using a nonparametric pixel-based approach.

\subsection{Nonperturbative Modeling and Pixel-Based Architecture}

The nonperturbative content of the TMD is encoded in the intrinsic distribution $M_f(b_T)$ and the evolution kernel $g_K(b_T)$. To ensure physical consistency, we implement a constrained architecture that enforces the perturbative limit at small transverse distances while maintaining maximum flexibility at large $b_T$. This is achieved through a logistic masking function:
\begin{equation}
\label{eq:R_mask}
R(b_T) = \left[ 1 + \exp\left( \kappa \frac{b_T - b_{\text{max}}}{w} \right) \right]^{-1} ,
\end{equation}
where the parameters are fixed to $b_{\text{max}} = 0.8 \, \text{GeV}^{-1}$, $w = 0.5 \, \text{GeV}^{-1}$, and $\kappa = 5.0$. This function acts as a transition operator:~as $b_T \to 0$, $R(b_T) \to 1$, effectively anchoring the distribution to its theoretical limit. The intrinsic distribution is then parametrized as:
\begin{equation}
\label{eq:Mf_pixel_model}
M_f(b_T) = R(b_T) + \tilde{p}(b_T) \big[ 1 - R(b_T) \big] ,
\end{equation}
where $\tilde{p}(b_T)$ represents the unknown hadronic structure to be extracted. In this framework, $\tilde{p}$ is not constrained by a fixed functional form but is defined on a discrete grid of nodal points (pixels). For the discretization, we adopt the power-law grid defined in Eq.~\eqref{eq:bt_grid}, configured here with $N=30$ nodes and a scaling parameter $a=4$. In this study, we assume an $x$-independent functional form for the intrinsic component and perform the analysis at a fixed value of $x=0.1$.

To validate the robustness of the NF-MH algorithm, we define a \textit{Ground Truth} scenario where the pixel values are explicitly set to follow a Gaussian profile:
\begin{equation}
\label{eq:p_gaussian_gt}
\tilde{p}_{\text{true}}(b_T) = \exp\left( -\frac{b_T^2\, w_f}{4} \right) ,
\end{equation}
with the ground truth width fixed at $w_f = 0.60 \text{ GeV}^2$. Similarly, the ground truth for the nonperturbative evolution kernel is defined as:
\begin{equation}
\label{eq:gk_model}
g_K(b_T) = \frac{g_2}{2}\, b_T^2 \big[ 1 - R(b_T) \big] ,
\end{equation}
with $g_2 = 0.28 \text{ GeV}^2$. This construction ensures that nonperturbative evolution is suppressed at small $b_T$, where the perturbative Sudakov factor is dominant, and becomes active only in the infrared regime.

While the initial validation (Sec.~\ref{sec:gauss_generation_events}) relied on an analytical Gaussian profile, the global extraction requires a generalized approach to handle the numerical pixel-based parametrization defined in Eq.~\eqref{eq:Mf_pixel_model}. 

The event generation is extended using a numerical ITS technique. Specifically, we compute the cumulative distribution function directly from the theoretical TMD cross-section $f_{\text{th}}(k_T; Q)$, obtained from the $b_T$-space distribution in Eq.~\eqref{eq:PDF_func} via the Bessel transform defined in Eq.~\eqref{eq:bessel_transform_tmd}, as follows:
\begin{equation}
\label{eq:cdf_integral}
\Phi(k_T) = \int_{0}^{k_T} \dd k_T'\, f_{\text{th}}(k_T'; Q)\,  .
\end{equation}
To maintain numerical stability across the nonparametric grid, the inversion $k_{Ti} = \Phi^{-1}(u_i)$ is performed on a fine-grained grid of 2500 points with linear interpolation. 

This approach allows us to generate a set of events that accurately reflects the full theoretical distribution, accounting for the nontrivial interplay between perturbative evolution (Sudakov factor) and nonperturbative dynamics ($M_f$ and $g_K$). By doing so, we ensure that the pseudo-data retain the full complexity of the TMD physics. Once the events are generated, they are binned and subjected to the same filtering criteria established in Sec.~\ref{sec:gauss_generation_events}, ensuring that only bins with a reliable signal-to-noise ratio are considered for the inference. This provides a rigorous benchmark for our extraction framework, confirming that the NF-MH algorithm can reconstruct the underlying structure without being influenced by artifacts of the generation process.

To evaluate how effectively the algorithm constrains the TMD shape, we perform a systematic study across different statistical regimes, generating datasets of $10^3$, $10^4$, and $10^5$ events. This allows us to observe how the reconstruction accuracy improves as the statistical noise in the observable space is reduced. Furthermore, we investigate two distinct physical scenarios:~a static case at a single energy scale without evolution, and a generalized case involving multiple energy scales where the CSS evolution kernel becomes active. This strategy mimics the transition explored in the simpler Gaussian case, aimed at shifting and constraining the observable space to better isolate the nonperturbative structure.
In the following sections, we present the results for both the single-energy and multi-scale scenarios, establishing the baseline performance of the NF-MH algorithm in the numerical TMD framework.

\subsection{Results for Single-Energy Scale Extraction}
\label{sec:results_single_energy}

In this first stage of validation, we focus on the extraction of the TMD PDF at a fixed hard scale $Q = 2$~GeV. To simplify the initial test, we restrict the analysis to the $u$-quark flavor, evaluating the performance of the NF-MH algorithm as a function of the statistical precision of the pseudo-data. The diagnostic results of the inference for different event counts are summarized in Table~\ref{tab:fit_diagnostics}.

\begin{table}[b] 
	\caption{Diagnostic summary of the Bayesian fit across different statistical scenarios for the $u$-quark TMD PDF at $Q=2$~GeV, showing the acceptance rate, number of momentum bins $N_{\text{pts}}$ surviving the filtering criteria, and $\chi^2/N_{\text{pts}}$ calculated at the posterior mean.}
	\centering
	\begin{tabular}{cccc}
		\hline
		$N_{\text{ev}}$ & accept. rate & $N_{\text{pts}}$ & $\chi^2/N_{\text{pts}}$ \\
		\hline
		$10^3$   & 86\% & 37 & 0.83 \\
		$10^4$   & 83\% & 42 & 0.67 \\
		$10^5$   & 69\% & 43 & 0.94 \\
		\hline
	\end{tabular}
	\label{tab:fit_diagnostics}
\end{table}

\begin{figure}[ht]
	\centering
	\includegraphics[width=\textwidth]{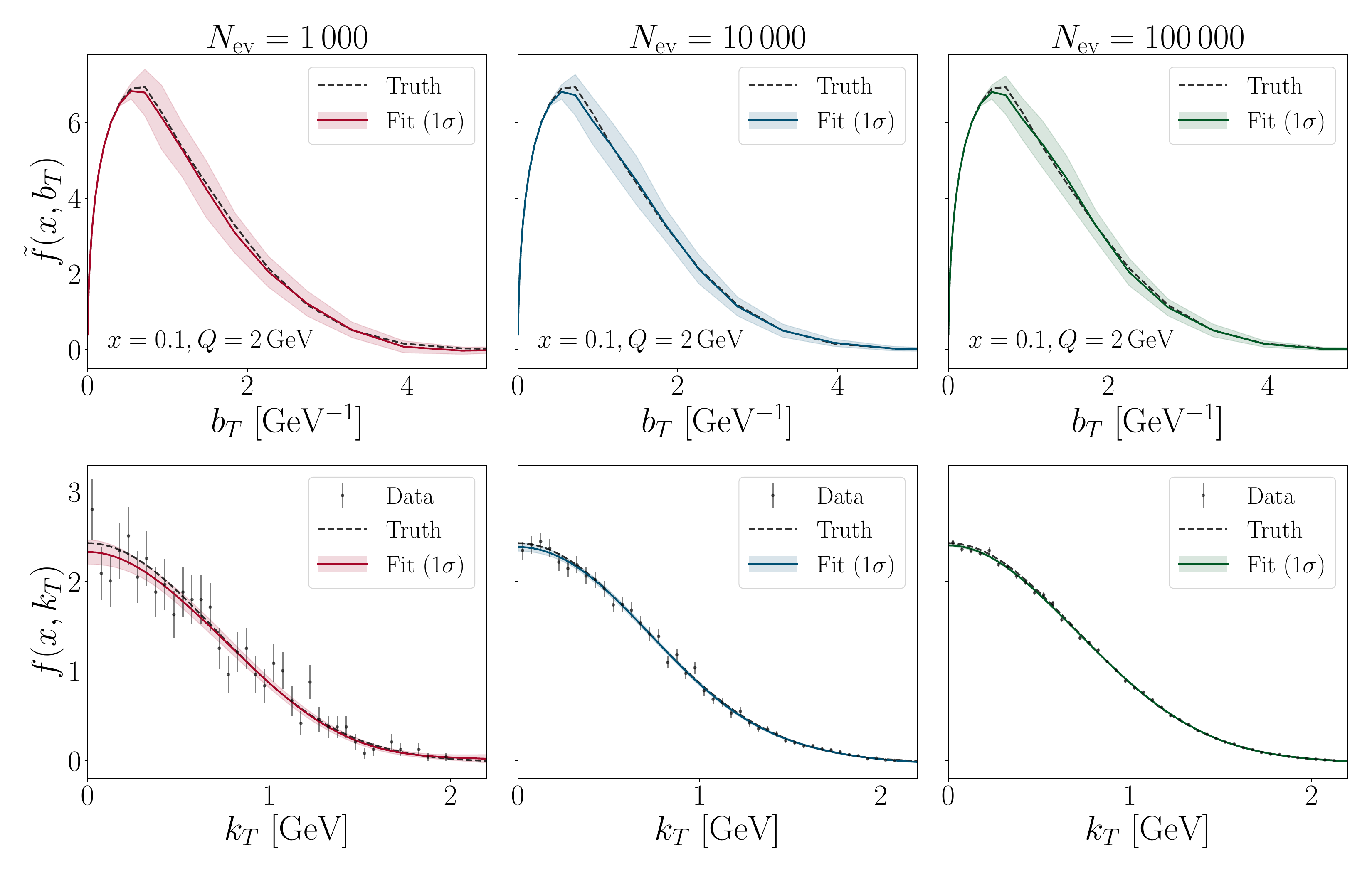}
	\caption{Comparison of the extraction results across three statistical regimes ($N_{\text{ev}} = 10^3$, $10^4$, $10^5$). The top row shows the reconstructed $b_T$-space distribution $\tilde{f}(x, b_T)$ compared to the ground truth. The bottom row displays the corresponding fits to the $k_T$-space pseudo-data.}
	\label{fig:scaling_linear}
\end{figure}

\begin{figure}[ht]
	\centering
	\includegraphics[width=\textwidth]{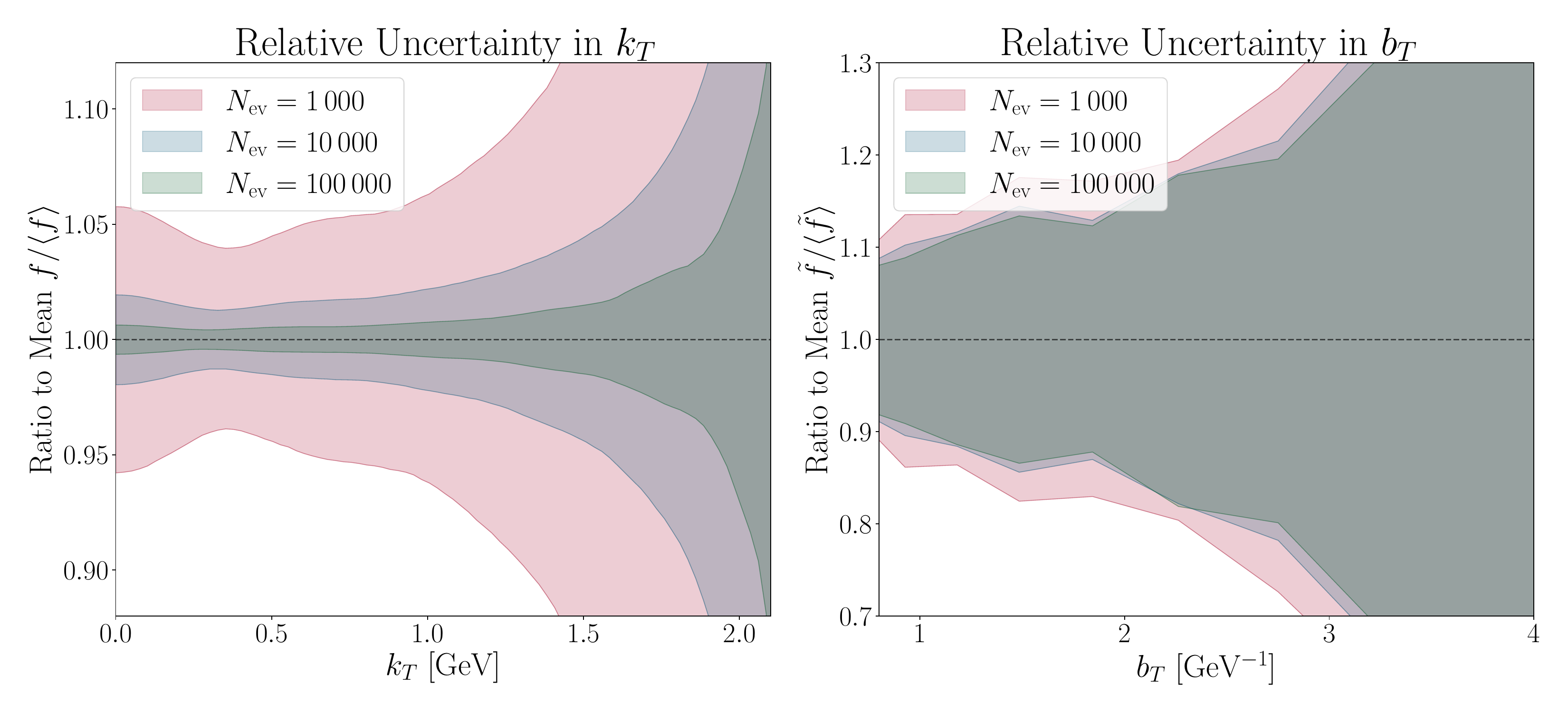}
	\caption{Relative uncertainty (band width divided by the mean value) for the extracted TMD. (Left) Relative uncertainty in momentum space ($k_T$). (Right) Relative uncertainty in impact parameter space ($b_T$). The different colors represent the statistical regimes ($10^3$, $10^4$, and $10^5$ events).}
	\label{fig:relative_uncertainty}
\end{figure}

The algorithm achieves an excellent description of the pseudo-data across all regimes, as evidenced by the $\chi^2/N_{\text{pts}}$ values in Table~\ref{tab:fit_diagnostics} and the visual fits in Fig.~\ref{fig:scaling_linear}. While the uncertainty bands in the top row of Fig.~\ref{fig:scaling_linear} consistently contain the ground truth, the quantitative scaling of the precision is better analyzed through the relative uncertainty, shown in Fig.~\ref{fig:relative_uncertainty}.

\begin{figure}[b] 
	\centering
	\includegraphics[width=\textwidth]{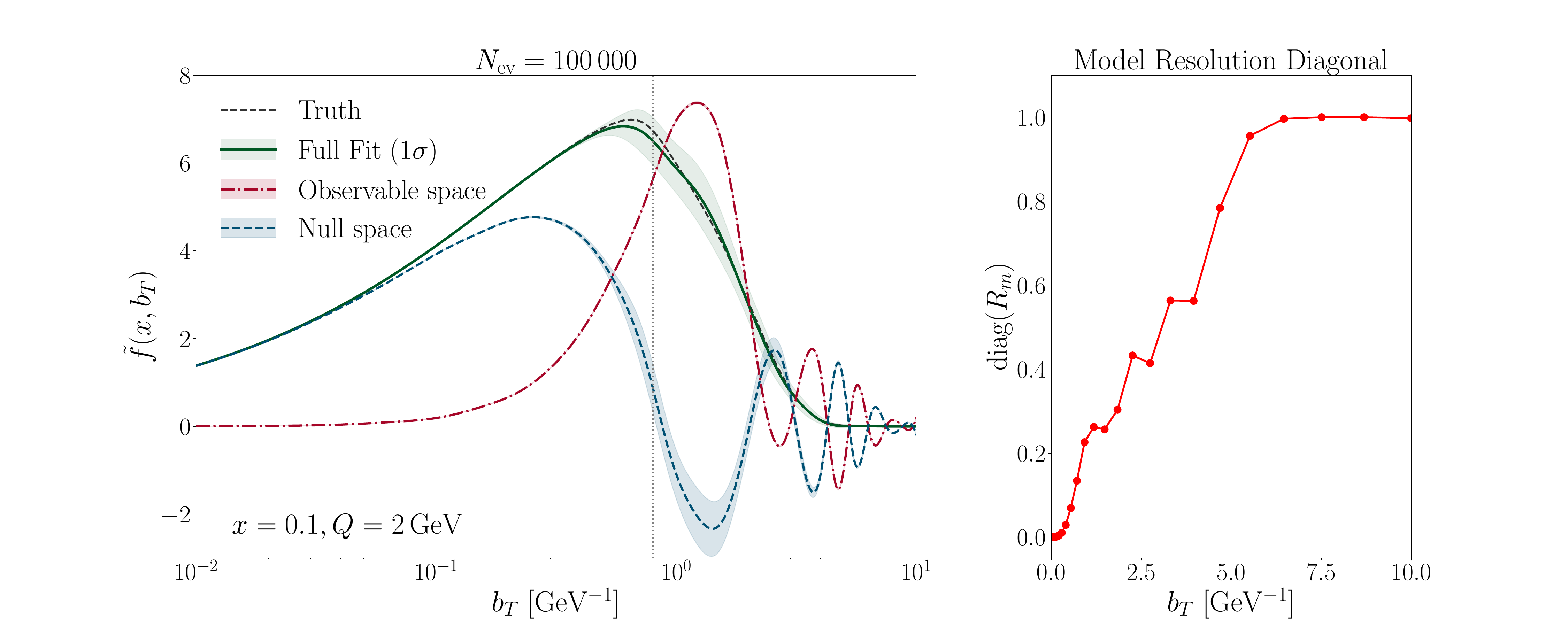}
	\caption{Resolution analysis at $Q=2$~GeV and $10^5$ events. (Left) The total $b_T$-space distribution (green band) decomposed into its projection onto the observable space (red band) and the null space (blue band). (Right) Diagonal elements of the resolution matrix, indicating the sensitivity of the data to different $b_T$ regions.}
	\label{fig:resolution_analysis}
\end{figure}

As shown in the left panel of Fig.~\ref{fig:relative_uncertainty}, the relative uncertainty in $k_T$ space decreases significantly as the statistics increase from $10^3$ to $10^5$ events. This behavior confirms that the NF-MH algorithm effectively leverages higher statistics to provide a more rigid constraint in the observable space. Conversely, the right panel illustrates a different trend for the $b_T$ space: while a slight improvement is visible between $10^3$ and $10^4$ events, the gain becomes almost negligible when moving to $10^5$ events. 
This observation suggests the existence of a ``precision floor'' in $b_T$ space at a fixed energy scale. Despite the increased statistical precision of the $k_T$ bins, the lack of multi-scale information prevents the algorithm from further narrowing the $b_T$ bands, as the inverse Bessel problem remains mathematically under-determined. This highlights the fundamental limitation of single-scale extractions and reinforces the necessity of including CSS evolution to break the precision floor and resolve the nonperturbative structure at large transverse distances.
To formally investigate the origin of the precision floor in $b_T$ space, we perform a resolution analysis on the $10^5$ events case. This is summarized in Fig.~\ref{fig:resolution_analysis}, where the reconstructed TMD distribution is decomposed into its projections onto the observable and null subspaces of the kernel.  

The left panel of Fig.~\ref{fig:resolution_analysis} illustrates a key feature of the inverse problem:~the total uncertainty of the distribution (green) is the result of the combined uncertainties of its projections. Notably, the uncertainty band associated with the observable space (red) is extremely narrow, indicating that the high statistics have almost perfectly constrained the part of the TMD that ``lives'' in the data-accessible domain. Conversely, the null space projection (blue) exhibits a much wider band, which directly drives the total uncertainty of the reconstructed PDF.  A crucial physical insight emerges from the behavior at different $b_T$ values. The null space projection dominates the distribution at small $b_T$, where the observable space contribution is almost negligible. The two components reach a similar magnitude only around $b_T \approx 0.5~\text{GeV}^{-1}$, just below the matching scale $b_{\text{max}} = 0.8~\text{GeV}^{-1}$. 

The right panel, showing the diagonal elements of the resolution matrix, confirms this lack of sensitivity; the resolution indices remain significantly below unity across almost the entire $b_T$ range and drop to zero as $b_T \to 0$. This explains why increasing the statistics in $k_T$ fails to narrow the $b_T$ bands significantly: the data at a single energy scale provide almost zero information about the short-distance (small $b_T$) structure and only a partial constraint on the rest of the distribution, leaving a substantial portion of the functional space to the null space. This mathematically proves that only by including multiple energy scales, which ``scan'' different regions of the resolution matrix through evolution, can we suppress the null space dominance and achieve a full-range reconstruction.

\subsection{Multi-Scale Extraction and TMD Evolution}
\label{sec:results_evolution}

To evaluate the ability of the NF-MH framework to reconstruct the dynamical evolution of the nucleon structure, we extend the extraction to a multi-scale scenario. In this setup, pseudo-data are generated at four distinct hard scales: $Q \in \{ 2, 5, 7, 10 \}$~GeV. This configuration forces the algorithm to simultaneously constrain the intrinsic hadronic structure $M_f(b_T)$ and the nonperturbative evolution kernel $g_K(b_T)$, governed by the $g_2$ parameter. The diagnostic results of the inference for the $u$-quark flavor are presented in Table~\ref{tab:fit_diagnostics_tmd_multi}.

\begin{table}[b] 
	\centering
	\caption{Global fit diagnostics for three statistical regimes for the multi-scale scenario, with $N_{\text{pts}}$ the total number of points across all four scales, and $\chi^2/N_{\text{pts}}$ the reduced chi-squared calculated at the posterior mean. The mean and standard deviation of the extracted $g_2$ (ground truth 0.28~GeV$^2$) demonstrate the increased precision gained with higher statistics. }
	\begin{tabular}{rclcc}
		\hline
		$N_{\text{ev}}$ & accept. rate & $g_{2}\, [{\rm GeV}^2]$ & $N_{\text{pts}}$ & $\chi^2/N_{\text{pts}}$ \\
		\hline
		$10^3$   & 79\% & ~$0.27(2)$  & 217 & 0.90 \\
		$10^4$   & 77\% & ~$0.287(8)$ & 253 & 0.93 \\
		$10^5$   & 61\% & ~$0.280(3)$ & 259 & 1.09 \\
		\hline
	\end{tabular}
	\label{tab:fit_diagnostics_tmd_multi}
\end{table}

\begin{figure}[t] 
	\centering
	\includegraphics[width=\textwidth]{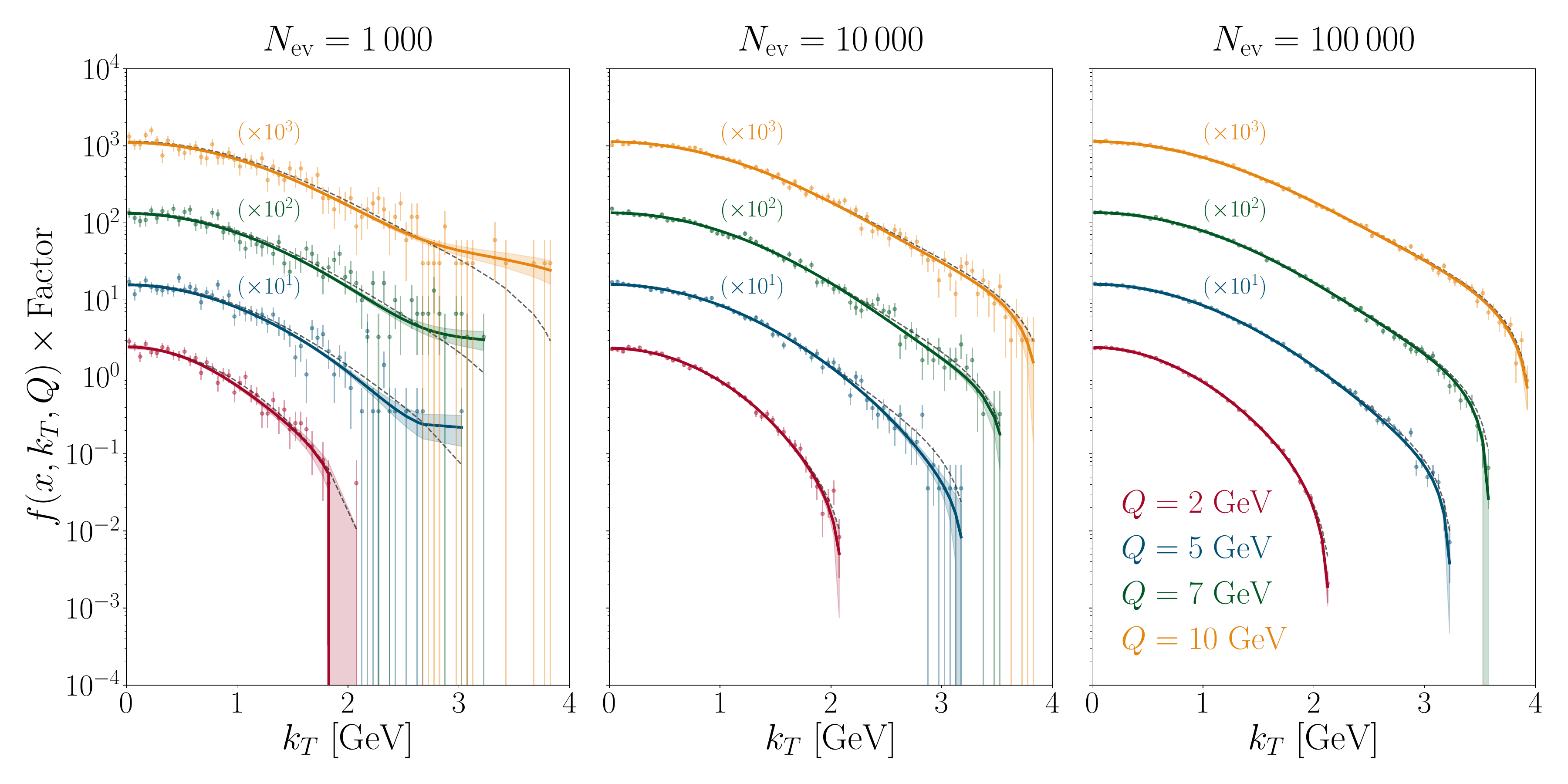}
	\caption{Multi-scale TMD fit results in momentum space. Each column corresponds to a statistical regime ($N_{\text{ev}} = 10^3$, $10^4$, and $10^5$), showing the four data sets ($Q=2$, 5, 7, and 10~GeV) compared to the reconstructed distributions in $k_T$.}
	\label{fig:fit_multi_energy_evo}
\end{figure}

\begin{figure}[t] 
	\centering
	\includegraphics[width=\textwidth]{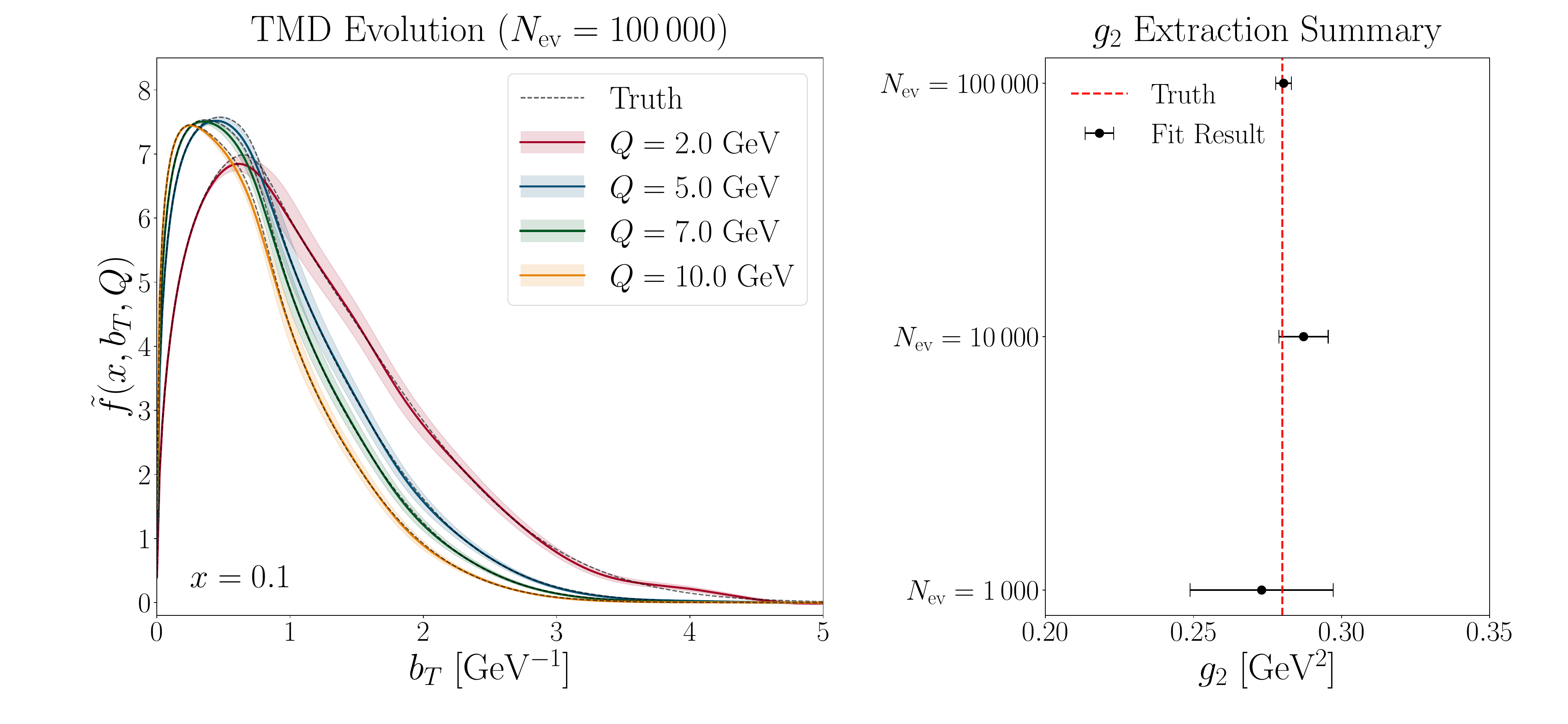}
	\caption{Summary of the multi-scale extraction results. (Left) Reconstructed $b_T$-space distributions $\tilde{f}(x, b_T)$ at the four energy scales for $N_{\text{ev}} = 10^5$, compared to the ground truth. (Right) Extracted values of $g_2$ for the three statistical regimes; the horizontal line indicates the true value ($0.28\,{\rm GeV}^2$) and the error bars represent the $1\sigma$ uncertainty.}
	\label{fig:g2_summary}
\end{figure}

The algorithm maintains excellent performance across all regimes, as shown in Fig.~\ref{fig:fit_multi_energy_evo}. The uncertainty bands of the fit and the error bars of the pseudo-data exhibit a consistent narrowing as the number of events increases, providing an increasingly rigid constraint on the underlying TMD structure. The quantitative details of the extraction and the evolution dynamics are further summarized in Fig.~\ref{fig:g2_summary}.

The left panel of Fig.~\ref{fig:g2_summary} displays the extracted TMD PDF in $b_T$ space for the $10^5$ events case, evaluated at each of the four energy scales. The uncertainty bands consistently enclose the ground truth curves, successfully capturing the scale-dependent suppression at large $b_T$. Although a precision floor remains due to the nonparametric nature of the pixels, the multi-energy information provides a significantly tighter constraint compared to the single-scale case.

The extraction of the evolution dynamics is equally robust, as shown in the right panel of Fig.~\ref{fig:g2_summary}. As the statistics increase, the extracted values of $g_2$ converge toward the true value of $0.28\, {\rm GeV}^2$, with the $1\sigma$ error bars shrinking by an order of magnitude (from 0.02 to 0.003, as detailed in Table~\ref{tab:fit_diagnostics_tmd_multi}). While the mean values may exhibit small statistical offsets from the ground truth, they remain consistently within the expected confidence intervals. This result confirms that the multi-energy lever arm is sufficient to break the degeneracy between the intrinsic shape and the evolution kernel, allowing $g_2$ to be treated as a free and resolvable physical parameter.

\begin{figure}[t] 
	\centering
    \hspace*{-0.2cm}
	\includegraphics[width=1.02\textwidth]{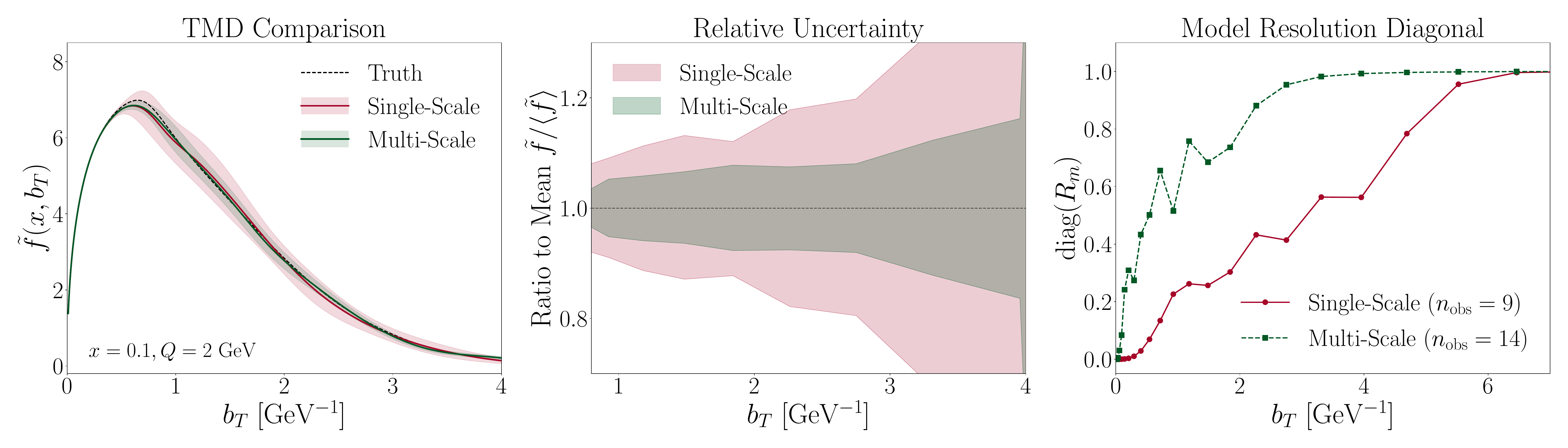}
	\caption{Direct comparison between single-scale ($Q=2$~GeV) and multi-scale extractions for $N_{\text{ev}} = 10^5$. (Left) Reconstructed $b_T$-space distributions. (Center) Comparison of the relative uncertainties, highlighting the precision gain provided by the multi-energy information. (Right) Comparison of the resolution indices for the single-scale and multi-scale scenarios, showing the expanded observable subspace.}
	\label{fig:single_vs_multi}
\end{figure}

The definitive advantage of the multi-scale approach is quantified in Fig.~\ref{fig:single_vs_multi}, which compares the extraction performed at a single scale ($Q=2$~GeV) with the global multi-scale fit, both at $N_{\text{ev}} = 10^5$. As shown in the left panel, while both extractions are consistent with the ground truth, the multi-scale framework yields a significantly narrower uncertainty band.
The center panel, displaying the relative error of the two reconstructions, provides a clear visual proof of the ``lever arm'' effect. The inclusion of multiple energy scales allows the algorithm to access different ranges of transverse momentum $k_T$, effectively shifting the observable space across different regions of the impact parameter $b_T$. 
This improvement is formally quantified in the right panel, which compares the resolution indices of the two scenarios. The multi-scale analysis (green) exhibits a substantial gain in resolution compared to the single-scale case (red), confirming that evolution effectively populates the observable subspace with additional independent information. Consequently, the precision floor observed in the single-scale case is significantly lowered, enabling a much more accurate determination of the nonperturbative hadronic structure.

\section{$F_{UU,T}$ Structure Function Case}
\label{sec:fuu_distribution}

The final and most challenging case considered in this work involves the extraction of TMD distributions from the unpolarized structure function $F_{UU,T}$. Following the validation strategy established in the previous sections, we perform a systematic study across both single- and multi-energy scenarios. In SIDIS, the experimental data is not a direct mapping of the TMD PDF, but rather a convolution in momentum space of the distribution function and the FF. In the impact parameter space, this convolution simplifies to a product of the two functions, as expressed by the factorization theorem~\cite{Bacchetta:2006tn}:
\begin{equation}
\label{eq:unp_conv}
\begin{split}
F_{UU,T}(x, z, q_T, Q^2) &= x \sum_a e_a^2 \frac{1}{2\pi} \int_0^{\infty} \dd b_T \, b_T\, J_0(q_T b_T)\, \tilde{f}_1^a(x, b_T; Q)\, \widetilde{D}_1^a(z, b_T; Q) \, ,
\end{split}
\end{equation}
where the hard-scattering factor $H(Q^2)$ is taken to be unity at the order of this analysis.

It is important to emphasize that while TMD factorization is rigorously valid only in the small transverse momentum regime, $q_T \ll Q$, in this work we extend the analysis to a broader $q_T$ range for the purpose of a closure test. This setup provides a controlled numerical environment where the target distribution is known exactly across the entire spectrum, allowing us to rigorously validate the capability of the pixelated framework to deconvolve the PDF signal from the structure function. In this context, we treat the TMD expression in Eq.~\eqref{eq:unp_conv} as the absolute ground truth, bypassing the added complexity of matching to the collinear region through the so-called $Y$-term. This simplification is intended to isolate and test the performance of the NF-MH sampler and the SVD-based resolution analysis in a high-dimensional inverse problem, while a full phenomenological fit to experimental data would require the standard kinematic cuts and matching procedures.

While $\tilde{f}_1$ encodes the 3D structure of the nucleon, the TMD FF $\widetilde{D}_1$ describes the number density of hadrons $H$ produced from a parton $a$ with a specific transverse momentum and longitudinal momentum fraction. Similar to the PDF case, the TMD FF is decomposed into a perturbative part and a nonperturbative model $M_D(z, b_T)$ as:
\begin{equation}
\label{eq:FF_func}
\begin{split}
\widetilde{D}_{H/ f}(z, b_T; Q) &= \sum_j \int_{z}^1 \frac{\dd\hat{z}}{\hat{z}^{3-2\epsilon}}\, \tilde{C}_{j/f}\big(z/\hat{z}, b_*; \mu_b^2, \mu_b, g(\mu_b)\big)\, d_{H/ j}(\hat{z}; \mu_b) \\
&\quad \times  \exp \!\left\{S_{\text{pert}}(b_T;\mu_b, Q) \right\} M_D(z, b_T) \exp \!\bigg\{\!- g_K(b_T) \ln \frac{Q}{Q_{0}} \bigg\} \, .
\end{split}
\end{equation}
The first line of Eq.~\eqref{eq:FF_func} represents the matching onto the collinear FFs $d_{H/j}$ where, for this analysis, we employ the JAM20 set~\cite{Moffat:2021mxy} at a fixed value of $z=0.2$ through the Wilson coefficients $\tilde{C}_{j/f}$ at NLO~\cite{Aybat:2011zv}. The second line encodes the nonperturbative structure $M_D(z, b_T)$ and the evolution kernel, with the perturbative Sudakov factor $S_{\text{pert}}$ defined as in Eq.~\eqref{eq:perturbative_sudakov}. In this work, we adopt a $z$-independent functional form for the intrinsic nonperturbative component of the TMD-FF, effectively performing the study for a single value of $z$.

Consistent with the modeling of the PDF in Sec.~\ref{sec:tmd_distribution}, the intrinsic nonperturbative component of the FF is defined using the logistic R-mask defined in Eq.~\eqref{eq:R_mask}:
\begin{equation}
\label{eq:MD_model}
M_D(b_T) = R(b_T) + \big[1 - R(b_T)\big] \exp\left( - \frac{b_T^2\, w_{D}}{4} \right) \,,
\end{equation}
where the width parameter is fixed to $w_{D} = 0.20 \text{ GeV}^2$. The nonperturbative evolution kernel $g_K(b_T)$ follows the same functional form as the PDF (Eq.~\eqref{eq:gk_model}). In this analysis, the parameters of the TMD FF are treated as fixed inputs, allowing us to focus on the pixel-based reconstruction of the $u$-quark TMD PDF $M_f(x, b_T)$, which we treat as the sole active flavor to simplify this validation. 
For the $b_T$ grid, we maintain the same nodal configuration established in Sec.~\ref{sec:tmd_distribution}. Furthermore, the generation of pseudo-data follows the numerical ITS method described in that section, where the unpolarized structure function $F_{UU,T}$ from Eq.~\eqref{eq:unp_conv} replaces the TMD PDF in the cumulative distribution function defined in Eq.~\eqref{eq:cdf_integral}. In the multi-scale scenario, while the evolution dynamics are coupled, we only treat the $g_2$ parameter within the TMD PDF as a free physical degree of freedom.
                                                                                    
\subsection{Results for Single-Energy Scale Extraction}
\label{sec:fuu_results_single_energy}

In this section, we evaluate the extraction of the $u$-quark TMD PDF from the unpolarized structure function $F_{UU,T}$ at a fixed scale $Q = 2$~GeV. The transition from a pure TMD PDF extraction to the $F_{UU,T}$ structure function case introduces a significant modification to the kernel of the inverse problem. Since the observable $F_{UU,T}$ involves the product $\tilde{f}_1 \times \widetilde{D}_1$, the fixed FF effectively modulates the sensitivity of the data to the impact parameter space $b_T$. 

The presence of the FF acts as an additional weight within the Bessel integral, which can further suppress or enhance specific $b_T$ regions depending on the $z$-kinematics. This configuration allows us to test the robustness of the NF-MH algorithm in a realistic scenario where the information about the target structure is partially shielded by the fragmentation process. The goal is to verify whether the NF-MH framework can successfully deconvolve the PDF signal from the structure function when the latter is treated as the primary observable. 

To analyze the information transfer in this convoluted case, we first define the impact parameter space term $\widetilde{W}$ as:
\begin{equation}
\label{eq:Wtilde_def}
\widetilde{W}(x, z, b_T; Q) = x \sum_a e_a^2\, \tilde{f}_1^a(x, b_T; Q)\, \widetilde{D}_1^a(z, b_T; Q) \, ,
\end{equation}
which relates to the observed structure function via:
\begin{equation}
\label{eq:fuu_wtilde}
F_{UU,T}(x, z, q_T, Q^2) = \frac{1}{2\pi} \int_0^{\infty} \dd b_T \, b_T\, J_0(q_T b_T)\, \widetilde{W}(x, z, b_T; Q) \, .
\end{equation}
By discretizing the unknown TMD PDF $\tilde{f}_1(b_T)$ using the basis of local interpolators $h_i(b_T)$ defined in Appendix \ref{sec:local_interp}, we can express the forward mapping in a compact matrix form:
\begin{equation}
\label{eq:fuu_matrix}
\mathbf{F}_{UU,T} 
= \mathcal{M}_{\text{eff}}\, \tilde{\boldsymbol{f}}_1 
= \mathcal{M}_{\text{bridge}} \cdot \left[ x \sum_a e_a^2\, \widetilde{\mathbf{D}}_1^a(z, \mathbf{b}_T; Q) \right] \odot \tilde{\boldsymbol{f}}_1 \, ,
\end{equation}
where $\tilde{\boldsymbol{f}}_1$ is the vector of nodal values and $\mathcal{M}_{\text{bridge}}$ represents the discrete Bessel transform operator mapping $b_T$ space to $q_T$ data. In this notation, the FF $\widetilde{\mathbf{D}}_1^a$ is treated as a weight vector acting on the impact parameter pixels. The total \textit{effective kernel matrix} $\mathcal{M}_{\text{eff}}$ that drives the Bayesian inference can therefore be defined as:
\begin{equation}
\label{eq:M_eff_def_revised}
\mathcal{M}_{\text{eff}} = \mathcal{M}_{\text{bridge}} \cdot \text{diag}\left( x \sum_a e_a^2\, \widetilde{\mathbf{D}}_1^a \right) \, ,
\end{equation}
absorbing the kinematic factors and the fragmentation weight. This bridge-based construction demonstrates that the FF acts as an additional nonperturbative layer that modulates the transfer of information from the nucleon's 3D structure to the experimental $q_T$ spectrum. Consequently, any degeneracy in the reconstruction of $\tilde{\boldsymbol{f}}_1$ will depend on the interplay between the Bessel transform's null space and the zeroes (or the rapid decay) of the fragmentation kernel.

\begin{table}[b] 
	\caption{Diagnostic summary of the Bayesian fit for the $F_{UU,T}$ structure function case at $Q=2$~GeV. $N_{\text{pts}}$ represents the number of $q_T$ bins surviving the filtering criteria, and $\chi^2/N_{\text{pts}}$ is the reduced chi-squared calculated at the posterior mean.}
	\centering
	\begin{tabular}{cccc}
		\hline
		$N_{\text{ev}}$ & accept. rate & $N_{\text{pts}}$ & $\chi^2/N_{\text{pts}}$ \\
		\hline
		$10^3$   & 89\% & 45 & 0.83 \\
		$10^4$   & 89\% & 48 & 0.92 \\
		$10^5$   & 84\% & 48 & 1.29 \\
		\hline
	\end{tabular}
	\label{tab:fit_fuu_single}
\end{table}

The diagnostic results for the Bayesian inference across different statistical regimes are summarized in Table~\ref{tab:fit_fuu_single}.
The algorithm achieves a remarkably stable performance, with the $\chi^2/N_{\text{pts}}$ consistently remaining near unity across all statistical regimes. 
Notably, the acceptance rate remains high (above 84\%) even for the $10^5$ events case, indicating that the NF-MH effectively maps the posterior distribution of the pixelated PDF despite the additional numerical complexity introduced by the fragmentation kernel.

\begin{figure}[ht]
	\centering
	\includegraphics[width=\textwidth]{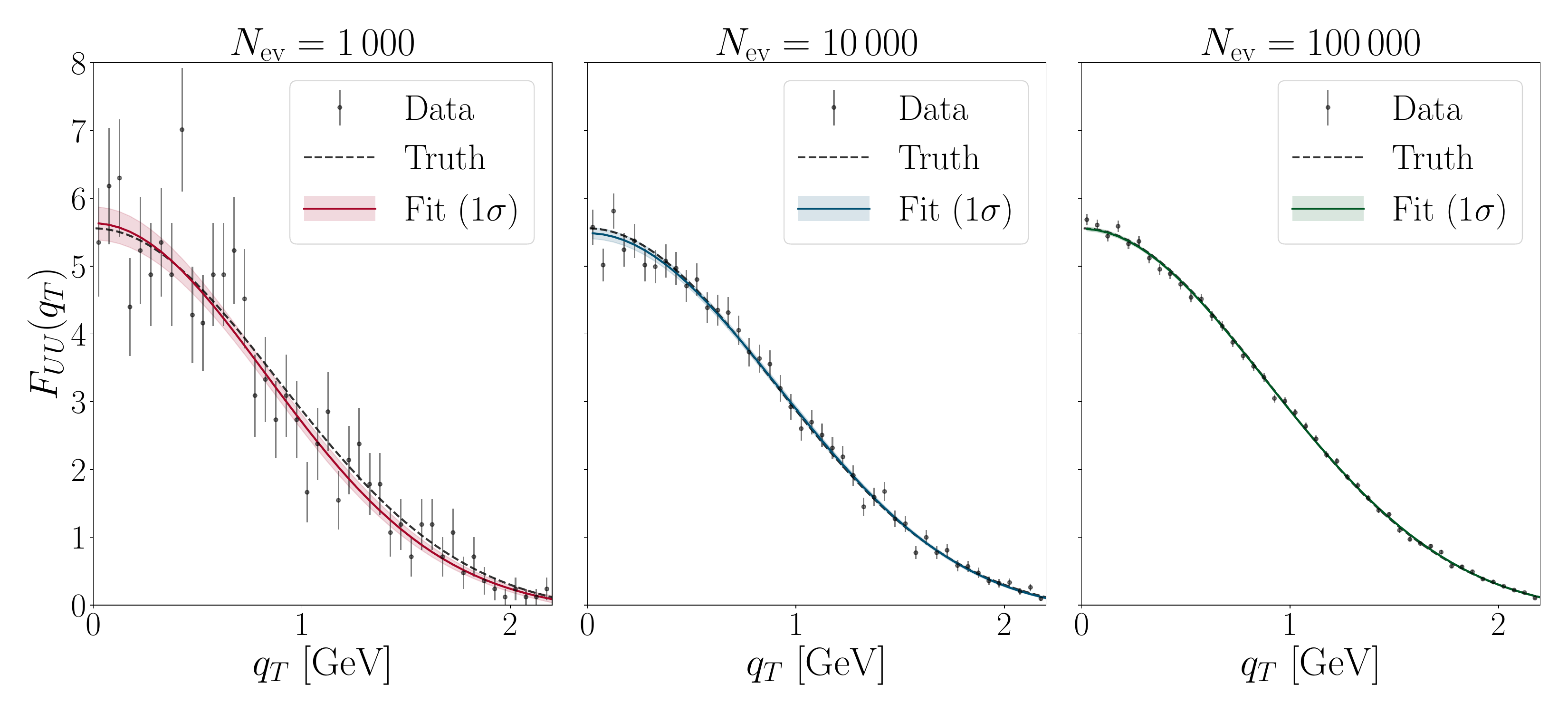}
	\caption{Comparison between $F_{UU,T}$ pseudo-data and the Bayesian fit in $q_T$ space for $N_{\text{ev}} = 10^3$, $10^4$, and $10^5$. The framework successfully deconvolves the signal and describes the structure function across all statistical regimes.}
	\label{fig:fuu_qt_fit}
\end{figure}

As illustrated in Fig.~\ref{fig:fuu_qt_fit}, the NF-MH framework provides a high-fidelity description of the unpolarized structure function. The mean of the posterior distribution, representing the central fit, remains perfectly centered on the pseudo-data, with the uncertainty bands shrinking consistently as $N_{\text{ev}}$ increases due to the suppression of statistical noise. This demonstrates that the pixelated PDF representation is sufficiently flexible to capture the underlying structure without being biased by the convolution. 

\begin{figure}[ht]
	\centering
	\includegraphics[width=\textwidth]{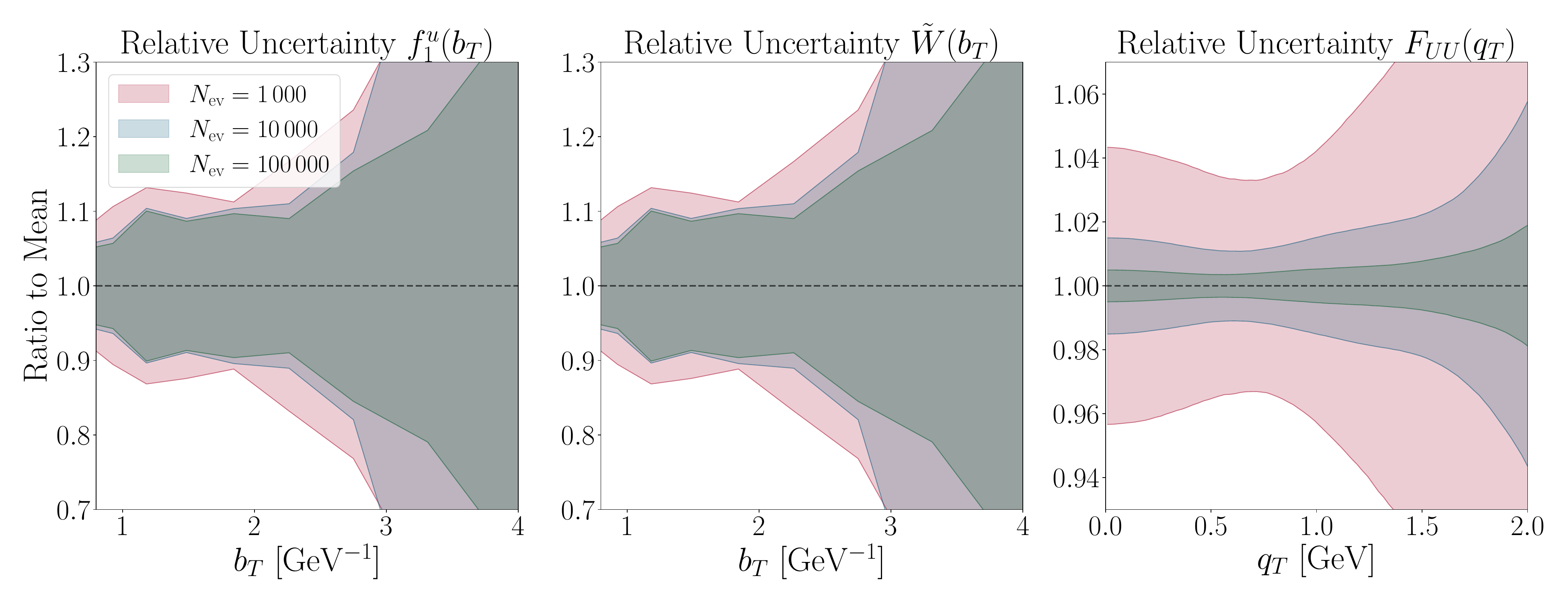}
	\caption{Analysis of relative uncertainties (band width divided by the mean value) for $F_{UU,T}$. The three columns represent the relative uncertainty for the TMD PDF (left), the impact-parameter space term $\widetilde{W}$ (center), and the structure function $F_{UU,T}$ in $q_T$ (right).}
	\label{fig:fuu_relative_uncertainties}
\end{figure}

The information transfer and its statistical scaling are further clarified in Fig.~\ref{fig:fuu_relative_uncertainties}, which highlights a distinct hierarchy between the observable and conjugate spaces. In the rightmost column, the relative uncertainty of $F_{UU,T}$ in the momentum space $q_T$ exhibits a consistent and significant contraction as the number of events increases. 

In contrast, the reconstruction of the TMD PDF (left column) and the convoluted term $\widetilde{W}$ (center column) in the impact parameter space reveals the emergence of a precision floor. While a noticeable improvement is achieved when moving from $10^3$ to $10^4$ events, the transition to $10^5$ events yields only marginal gains. Consistent with our observations in the previous sections, this behavior confirms that the resolution limit at a single energy scale is an intrinsic property of the inverse problem that cannot be overcome by simply increasing the statistical precision. Even with the modulation provided by the FF, the short-distance structure of the nucleon remains partially hidden in the null space of the transformation, as further explored in the resolution analysis.

\begin{figure}[b] 
	\centering
	\includegraphics[width=0.96\textwidth]{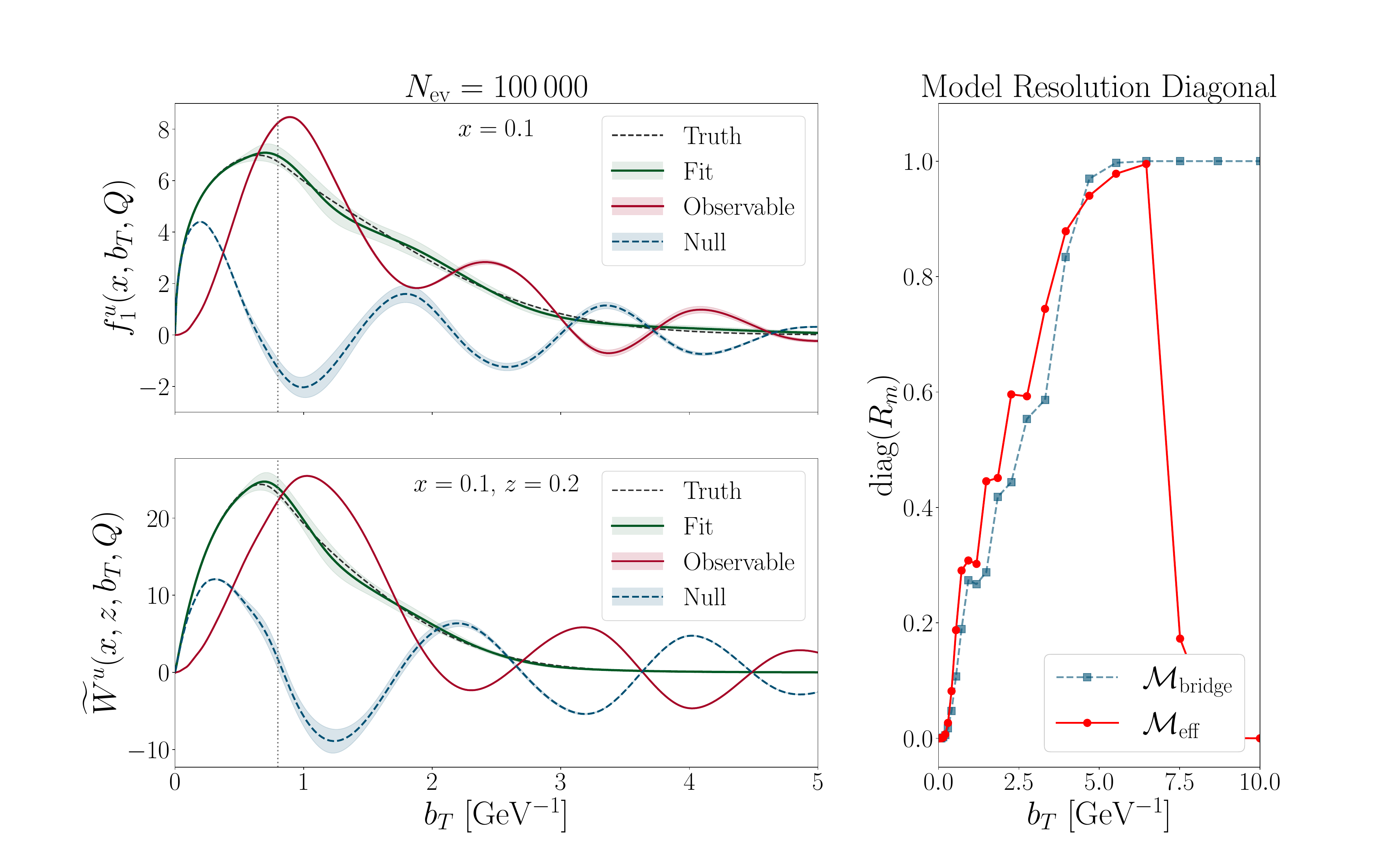}
	\caption{Resolution analysis for the $F_{UU,T}$ case at $N_{\text{ev}} = 10^5$. (Top-Left) Reconstructed TMD PDF with projections onto the observable space (red) and null space (blue) using the effective kernel $\mathcal{M}_{\text{eff}}$. (Bottom-Left) Reconstructed $\widetilde{W}$ with projections using the bridge matrix $\mathcal{M}_{\text{bridge}}$. (Right) Resolution indices (diagonal elements of the resolution matrix) for $\mathcal{M}_{\text{eff}}$ (red) and $\mathcal{M}_{\text{bridge}}$ (blue).}
	\label{fig:fuu_resolution}
\end{figure}

To formally characterize the impact of the fragmentation kernel on the resolvability of the nucleon structure, we perform a detailed resolution analysis at $10^5$ events, as shown in Fig.~\ref{fig:fuu_resolution}. This analysis highlights how the information is partitioned between the observable and null spaces when dealing with convoluted observables.

The top-left panel displays the reconstruction of the TMD PDF. When using the effective kernel $\mathcal{M}_{\text{eff}}$, which includes the weight of the FF, the projection onto the observable space (red) is well-constrained up to $b_T \approx 3~\text{GeV}^{-1}$. Beyond this range, the observable uncertainty band exhibits a slight widening, indicating a loss of sensitivity. In contrast, the reconstruction of the convoluted term $\widetilde{W}$ (bottom-left panel), projected via the pure bridge matrix $\mathcal{M}_{\text{bridge}}$, shows an observable component that remains narrow and perfectly resolved across the entire $b_T$ range.

The origin of this difference is revealed by the resolution indices shown in the right panel. Interestingly, for impact parameters up to $b_T \approx 4~\text{GeV}^{-1}$, the effective kernel (red) provides a slightly higher resolution compared to the pure bridge matrix (blue), suggesting that the fragmentation weight can locally enhance the sensitivity of the observables. However, this trend reverses at larger distances, where the resolution for the effective kernel drops sharply after $b_T \approx 6~\text{GeV}^{-1}$. This ``resolution collapse'' is a direct consequence of the
FF's behavior:~since the TMD FF vanishes at large impact parameters, it effectively acts as a limiting aperture that masks the large-distance structure of the TMD PDF. 
Consequently, the nucleon's structure at high $b_T$ becomes part of the null space of the convolution, explaining why the observable projection of the PDF is not uniformly fine across the whole spectrum. This confirms that in SIDIS-like processes, the FF not only provides a scale-dependent weight but also sets a fundamental limit on the spatial resolution of the 3D partonic map.

\subsection{Multi-Scale Extraction and Evolution from $F_{UU,T}$}
\label{sec:fuu_results_evolution}

To further test the framework's ability to deconvolve the PDF signal and the evolution kernel from convoluted data, we perform a global multi-scale fit to the $F_{UU,T}$ structure function. In this scenario, we use pseudo-data generated at four hard scales $Q \in \{2, 5, 7, 10\}$~GeV, simultaneously extracting the pixelated TMD PDF $M_f(x, b_T)$ and the evolution parameter~$g_2$. The performance of the global fit is summarized in Table~\ref{tab:fit_fuu_multi}.

\begin{table}[b] 
	\caption{Global fit diagnostics for the multi-scale $F_{UU,T}$ case. The extraction successfully recovers the evolution parameter $g_2$ (ground truth $0.28\,{\rm GeV^2}$) while accounting for the fragmentation convolution. $N_{\text{pts}}$ represents the total number of points across all four scales, and $\chi^2/N_{\text{pts}}$ is the reduced chi-squared calculated at the posterior mean.}
	\centering
	\begin{tabular}{lclcc}
		\hline
		$N_{\text{ev}}$ & accept. rate & $g_2\,[{\rm GeV^2}]$  & $N_{\text{pts}}$ & $\chi^2/N_{\text{pts}}$ \\
		\hline
		$10^3$   & 84\% & ~$0.22(5)$ & 251 & 1.02 \\
		$10^4$   & 85\% & ~$0.28(2)$ & 276 & 0.94 \\
		$10^5$   & 68\% & ~$0.272(5)$ & 284 & 1.07 \\
		\hline
	\end{tabular}
	\label{tab:fit_fuu_multi}
\end{table}

The global fit maintains a high degree of statistical consistency, with $\chi^2$ remaining close to 1 even with increasing complexity of the kernel. As observed above, the multi-energy lever arm is critical for breaking the degeneracies of the inverse problem. By scanning different $q_T$ ranges through evolution, the algorithm is able to resolve the intrinsic hadronic structure with significantly higher precision than in the single-scale case, effectively deconvolving the PDF from both the fragmentation kernel and the evolution effects.

\begin{figure}[t] 
	\centering
	\hspace*{-0.3cm}\includegraphics[width=1.01\textwidth]{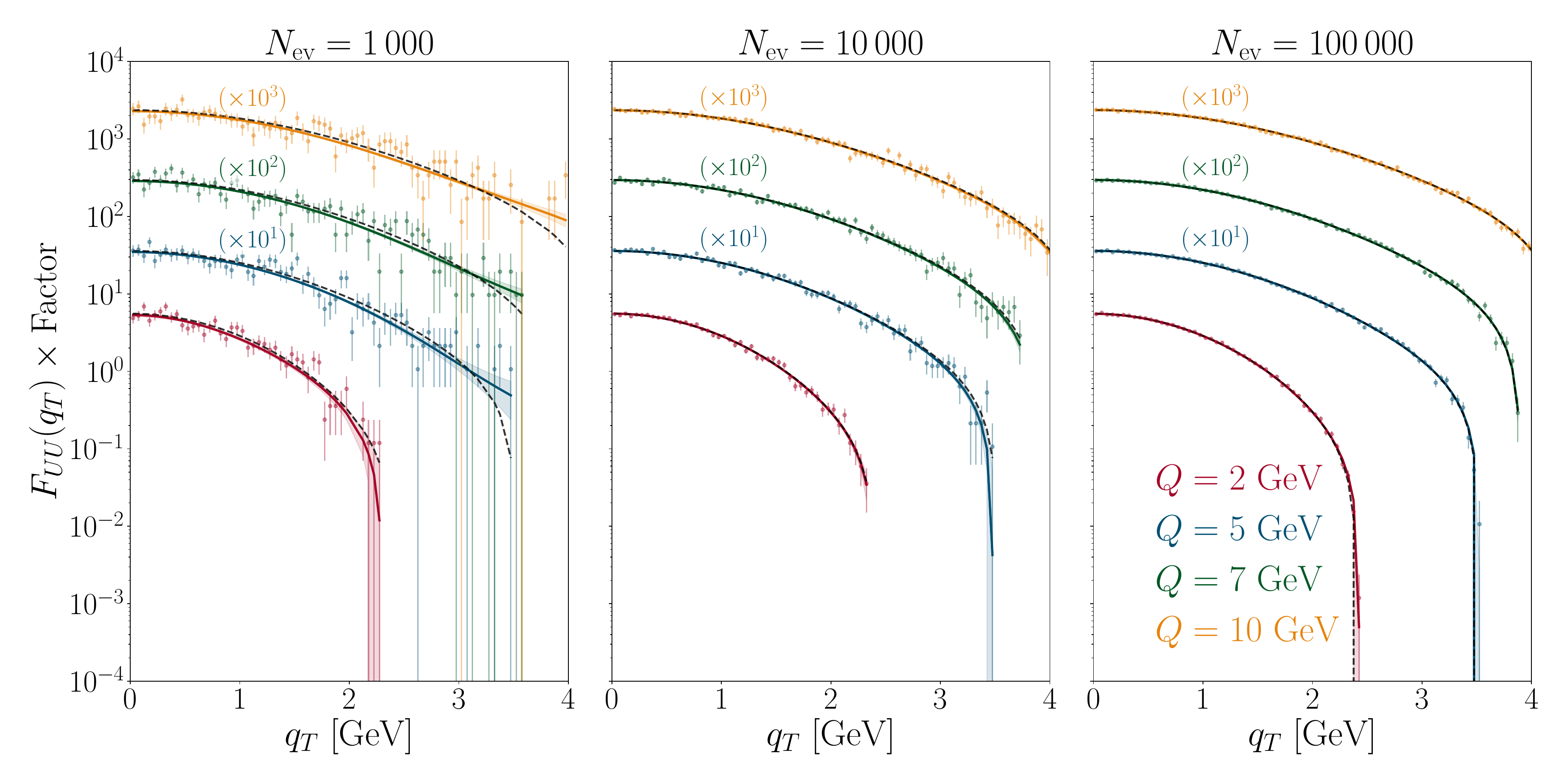}
    \vspace*{-0.2cm}
	\caption{Multi-scale $F_{UU,T}$ fit results. The plot shows the comparison between pseudo-data and the Bayesian fit across three statistical regimes ($N_{\text{ev}} = 10^3$, $10^4$, and $10^5$) and four energy scales $Q$. The excellent agreement is reflected in the $\chi^2$ values and the visual alignment of the distributions.}
	\label{fig:fuu_multi_fit}
\end{figure}

As shown in Fig.~\ref{fig:fuu_multi_fit}, the framework accurately describes the unpolarized structure function across all scales and statistical regimes. This confirms that the NF-MH algorithm successfully captures the energy-dependent broadening, demonstrating that the pixel-based approach can simultaneously extract evolution parameters and nonparametric distributions even from convoluted observables.

\begin{figure}[t] 
	\centering
	\hspace*{-0.55cm}\includegraphics[width=1.06\textwidth]{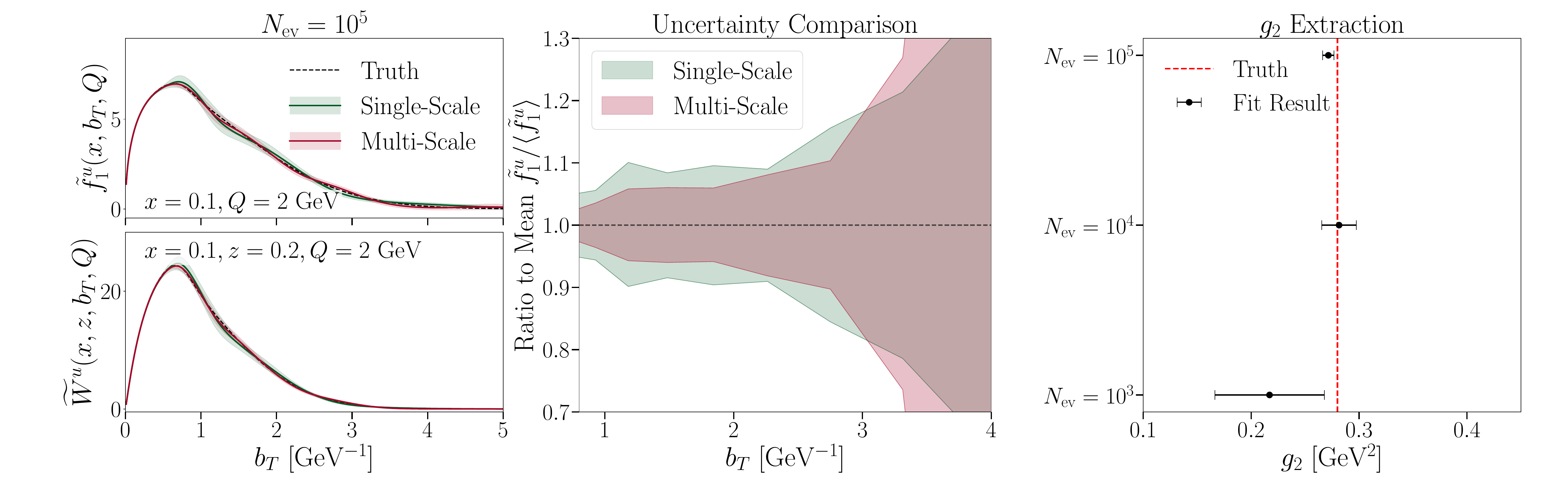}
	\caption{Summary of the multi-scale $F_{UU,T}$ extraction results.  Reconstructed TMD PDF (top-left) and $\widetilde{W}$ (bottom-left) for $N_{\text{ev}} = 10^5$ in single-scale versus multi-scale fits. (Center) Comparison of the relative TMD PDF uncertainties. (Right) Extracted values of the evolution parameter $g_2$ across the three statistical regimes.}
	\label{fig:fuu_final_summary}
\end{figure}

The final multi-scale results are summarized in Fig.~\ref{fig:fuu_final_summary}. Both the reconstructed TMD PDF and the convoluted term $\widetilde{W}$ at $N_{\text{ev}} = 10^5$ (left column) show excellent agreement with the ground truth, confirming that the framework correctly recovers the hadronic structure even from convoluted data. The multi-energy lever arm (center panel) effectively reconfigures the resolution profile: as the dynamical information from evolution suppresses the null-space dominance at small and intermediate distances, the precision floor is pushed toward shorter distances. However, as shown in the relative uncertainty plot, the multi-scale bands can become larger than the single-scale ones for $b_T > 3~\text{GeV}^{-1}$.

\begin{figure}[b] 
	\centering
	\includegraphics[width=\textwidth]{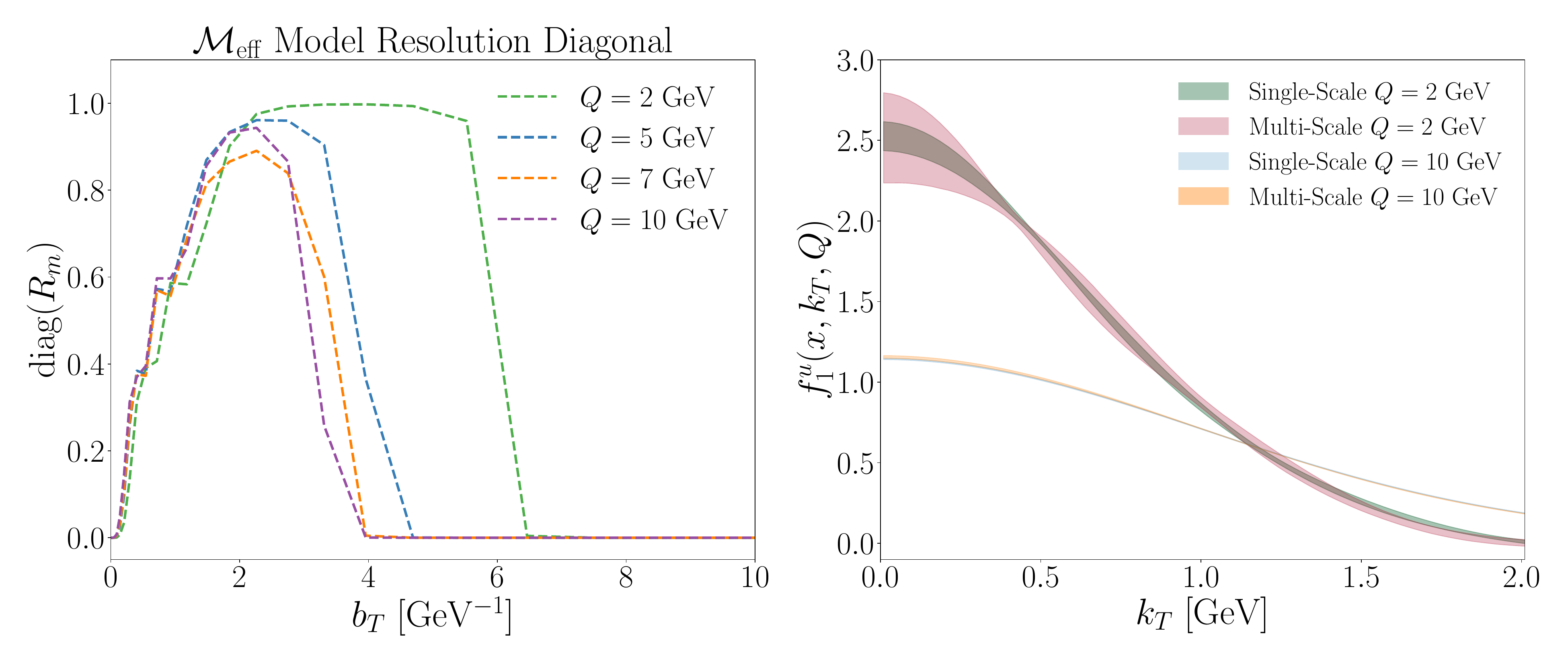}
	\caption{Summary of the resolution and precision analysis for the $F_{UU,T}$ case. The left column shows the diagonal elements of the effective resolution matrices at different energy scales, illustrating the shift of the resolution frontier. The right column displays the extracted TMD PDF in $k_T$ space at $Q=2$ and $10~\text{GeV}$ for both single-scale and multi-scale extractions, highlighting how evolution affects the uncertainty bands.}
	\label{fig:fuu_kt_precision}
\end{figure}

This behavior is driven by two combined factors. First, the inclusion of the evolution parameter $g_2$ as a free degree of freedom in the global fit introduces an additional source of uncertainty that propagates through the impact parameter space. Second, the multi-scale analysis shifts the focus of the inverse problem toward the small-$b_T$ region. This is corroborated by the diagonal elements of the effective resolution matrices at different energies, shown in the left column of Fig.~\ref{fig:fuu_kt_precision}, where we observe a higher sensitivity below $b_T \approx 3-4~\text{GeV}^{-1}$. While a more rigid constraint at large distances would require a higher density of data points at very low transverse momentum $q_T$, the current setup successfully demonstrates the intended relocation of the resolution frontier.

Furthermore, at higher energy scales, the evolution kernel strongly suppresses the large-$b_T$ contribution, damping any oscillations in that region. This leads to a behavior in momentum space that may appear counter-intuitive when observing the TMD PDF in $k_T$ space via a Bessel transform. As illustrated in the right column of Fig.~\ref{fig:fuu_kt_precision}, the uncertainty band of the TMD PDF extracted from the multi-scale fit at $Q=2~\text{GeV}$ is larger than that of the single-scale extraction; however, this effect significantly diminishes at $Q=10~\text{GeV}$, where the evolution dynamics dominate.

The extracted evolution dynamics (right panel of Fig.~\ref{fig:fuu_final_summary}) show $g_2$ converging to the true value with uncertainties shrinking systematically as statistics increase. This confirms that the NF-MH algorithm successfully breaks the degeneracy between the intrinsic shape and the evolution rate, even with convoluted observables, proving the framework's robustness for future global analyses of SIDIS and Drell-Yan data.

\section{Conclusions}
\label{sec:conclusions}

In this work, we introduced a nonparametric pixel-based framework for TMD imaging, replacing rigid functional forms with a discrete nodal representation. Using a hybrid NF-MH algorithm, we performed exact Bayesian sampling to reconstruct the posterior distribution across three case studies: a baseline Gaussian model, the $u$-quark TMD PDF, and the unpolarized $F_{UU,T}$ structure function, demonstrating successful deconvolution of the hadronic signal even from convoluted experimental data.

A key finding of this study is the formal characterization of null TMDs, functional modes residing in the null space of the Bessel kernel. Our SVD analysis demonstrates that single-scale data leave the small-$b_T$ structure fundamentally unconstrained, establishing an irreducible precision limit driven by the theoretical prior. We showed that this resolution barrier is effectively broken by multi-scale data, where energy evolution acts as a kinematic lever arm to resolve the hadronic core and overcome the resolution limits inherent to both the integral transform and the fragmentation process.

The methodology developed here establishes a robust foundation for future 3D imaging programs at facilities like Jefferson Lab and the Electron-Ion Collider. Future work will focus on applying this framework to experimental data, with particular emphasis on the simultaneous extraction of TMD PDFs and FFs. Understanding the correlation between these distributions and achieving a rigorous quantification of their uncertainties will be essential for delivering unbiased partonic maps driven by empirical evidence. \\ \\

\acknowledgments

This work was supported by  the U.S. Department of Energy contract No.~DE-AC05-06OR23177, under which Jefferson Science Associates, LLC operates Jefferson Lab (W.M., A.P., J.Q., N.S., and M.Z.), contract  No.~DE-SC0026320 (L.G.), the National Science Foundation under Grants No.~PHY-2308567 (D.P.), and No.~PHY-2310031, No.~PHY-2335114 (A.P.).
The work of N.S. was supported by the DOE, Office of Science, Office of Nuclear Physics in the Early Career Program.
The work of M.Z. was supported and conducted in part under the Laboratory Directed Research and Development Program at Thomas Jefferson National Accelerator Facility for the U.S. Department of Energy.
This work has benefited by interactions within the Quark-Gluon Tomography (QGT) Topical Collaboration funded by the U.S. Department of Energy, Office of Science, Office of Nuclear Physics with Award DE-SC0023646. \\ \\

\appendix
\section{The Interpixel Framework}
\label{sec:local_interp}

In this appendix we describe the mathematical foundations of the local interpolators used in our pixel-based discretization, which we refer to as the \textit{interpixel framework}. As well as being employed in the present work, the method has also recently been used in Refs.~\cite{Freese:2024ypk, Salam:2008qg, Bertone:2013vaa, Bertone:2023onx}.

\subsection{General Formalism}

We begin by firstly reviewing the basics of Lagrange interpolation in 1D. Given two arrays $\boldsymbol{X,Y}$ of size $N$, one constructs a Lagrange basis polynomial of the form:
\begin{equation}
    l_j(x;\boldsymbol{X})\, = \prod_{0\leq m\leq N,~m\neq j} \frac{x-x_m}{x_j-x_m}\,,
    \label{eq:lagrange_interpolation_app}
\end{equation}
where $x_j, x_m \in \boldsymbol{X}$. The polynomial $l_j$ is by construction of degree $N-1$. The Lagrange interpolation can be written as:
\begin{equation}
    L(x;\boldsymbol{X};\boldsymbol{Y})\, =\, \sum_j y_j~l_j(x;\boldsymbol{X}),
\end{equation}
where $y_j\in \boldsymbol{Y}$. When the size of the arrays $\boldsymbol{X,Y}$ becomes too large, the global Lagrange interpolation exhibits Runge's phenomenon and its accuracy degrades. Instead, a local interpolation strategy here is more appropriate. Specifically, for a given value of $x$, we identify a slice of $\boldsymbol{X}$ around which the $x$ value is contained. We write this slice as $\boldsymbol{X}_x$ and the corresponding slice for $\boldsymbol{Y}$ as $\boldsymbol{Y}_x$. The local Lagrange interpolation can then be written as:
\begin{equation}
    L(x;\boldsymbol{X}_x;\boldsymbol{Y}_x)\, =\, \sum_{j} y_{x,j}~l_j(x;\boldsymbol{X}_x)\, .
    \label{eq.local-lagrange_app}
\end{equation}
In this notation, $\boldsymbol{X}_x$ and $\boldsymbol{Y}_x$ are the images of a \emph{slicing} map acting on the arrays $\boldsymbol{X}$, $\boldsymbol{Y}$ around a given value of $x$:
\begin{subequations}
\begin{align}
(x,\boldsymbol{X})&\, \to\, \boldsymbol{X}_x \;, \\
(x,\boldsymbol{Y})&\, \to\, \boldsymbol{Y}_x \;.
\end{align}
\end{subequations}
For example, given the arrays $\boldsymbol{X} = [x_0,x_1,x_2,x_3,x_4]$ and $\boldsymbol{Y} = [y_0,y_1,y_2,y_3,y_4]$, and a value of $x$ close to $x_2$, a slicing map with a window of size 3 yields $\boldsymbol{X}_x = [x_1,x_2,x_3]$ and $\boldsymbol{Y}_x = [y_1,y_2,y_3]$. Provided the slice size is small, local Lagrange interpolation avoids Runge's phenomenon regardless of the total size of $\boldsymbol{X}, \boldsymbol{Y}$.

We now introduce the concept of the \textit{interpixel function}. Let $\boldsymbol{\delta}^{\alpha}$ be an array of zeros except at the $\alpha$-th entry, where its value is 1. The interpixel function is defined as the local Lagrange interpolation for the $\boldsymbol{\delta}^\alpha$ arrays:
\begin{equation}
    \Phi^{\alpha}(x) \equiv L(x;\boldsymbol{X}_x;\boldsymbol{\delta}^{\alpha}_x) = \sum_j \delta^{\alpha}_{x,j}~l_j(x;\boldsymbol{X}_x)\,.
\end{equation}
This allows us to write the interpolation in terms of the full arrays:
\begin{equation}    
    L(x;\boldsymbol{X};\boldsymbol{Y}) = \sum_{\alpha} y_{\alpha}\, \Phi^{\alpha}(x)\, ,
    \label{eq.local-lagrange2_app}
\end{equation}
where we have effectively pushed the local interpolation logic into the basis functions $\Phi^{\alpha}(x)$.

The power of this representation lies in the decoupling of linear operators. Consider a generic operator $\Box$ acting on Eq.~\eqref{eq.local-lagrange2_app}:
\begin{equation}   
    \Box\big(L(x;\boldsymbol{X};\boldsymbol{Y})\big) = \sum_{\alpha} y_{\alpha}\, \Box\big(\Phi^{\alpha}(x)\big).
    \label{eq.box-local_app}
\end{equation}
Typically in QCD phenomenology, the array $\boldsymbol{Y}$ is unknown and needs to be inferred from data. On the other hand $\Box\big(\Phi^{\alpha}(x)\big)$ is calculable independently of $\boldsymbol{Y}$. This means that when performing regression on $\boldsymbol{Y}$, the quantity $\Box\big(\Phi^{\alpha}(x)\big)$ can be precalculated and stored as a look-up table.

\subsection{Application to TMD Imaging}

In the context of this work, we apply the interpixel framework to the discretization of TMD distributions in both impact parameter ($b_T$) space and in momentum ($k_T$) space. 
The mapping from the general formalism to the physical problem is given by following:
\begin{itemize}
    \item The variable $x$ corresponds to the impact parameter $b_T$.
    \item The array $\boldsymbol{X}$ corresponds to the fixed grid of nodes $\{b_{Ti}\}$ defined in Eq.~\eqref{eq:bt_grid}.
    \item The array $\boldsymbol{Y}$ corresponds to the vector of unknown pixel values $\tilde{\boldsymbol{f}} = [\tilde{f}_1, \dots, \tilde{f}_N]^T$.
    \item The interpixel functions $\Phi^\alpha(x)$ correspond to the basis functions $h_i(b_T)$ introduced in Section~\ref{sec:gauss_distribution}.
\end{itemize}
Specifically, the kernel matrix $\mathcal{M}$ in Eq.~\eqref{eq:kernel_matrix} is constructed by applying the Bessel transform operator $\mathcal{B}$ to the interpixel functions:
\begin{equation}
    \mathcal{M}_{ji} = \mathcal{B}[\Phi^i](k_{Tj}) = \frac{1}{2\pi} \int_{b_{T\,\text{min}}}^{b_{T\,\text{max}}} \dd b_{T}\, b_{T} J_0(b_{T} k_{T j})\, \Phi^i(b_{T}).
\end{equation}
By precalculating these integrals, the forward mapping $\boldsymbol{f} = \mathcal{M} \tilde{\boldsymbol{f}}$ becomes a simple matrix multiplication, enabling the efficient use of the NF-MH sampler for high-dimensional inference.

\clearpage
\bibliographystyle{JHEP}
\bibliography{references}

\end{document}